\documentclass[journal=ancac3,manuscript=article]{achemso}
\usepackage[english]{babel}
\usepackage{notes2bib}
\usepackage{amsfonts}
\usepackage[psamsfonts]{amssymb}
\usepackage{graphicx}
\usepackage[version=3]{mhchem} 
\usepackage{color}

\author{Joan Camunas-Soler}
\affiliation{Small Biosystems Lab, Departament de F\'isica Fonamental, Universitat de Barcelona, Avda. Diagonal 647, 08028 Barcelona, Spain} 
\alsoaffiliation{CIBER de Bioingenier\'ia, Biomateriales y Nanomedicina, Instituto de Salud Carlos III, Madrid, Spain}

\author{Silvia Frutos}
\affiliation{Small Biosystems Lab, Departament de F\'isica Fonamental, Universitat de Barcelona, Avda. Diagonal 647, 08028 Barcelona, Spain} 
\alsoaffiliation{CIBER de Bioingenier\'ia, Biomateriales y Nanomedicina, Instituto de Salud Carlos III, Madrid, Spain}

\author{Cristiano V. Bizarro}
\affiliation{Small Biosystems Lab, Departament de F\'isica Fonamental, Universitat de Barcelona, Avda. Diagonal 647, 08028 Barcelona, Spain} 
\alsoaffiliation{CIBER de Bioingenier\'ia, Biomateriales y Nanomedicina, Instituto de Salud Carlos III, Madrid, Spain}
\altaffiliation{Current Address: Centro de Pesquisas em Biologia Molecular e Funcional/PUCRS Avenida Ipiranga 6681, Tecnopuc, Partenon 90619-900, Porto Alegre, RS, Brazil}

\author{Sara de Lorenzo}
\affiliation{Small Biosystems Lab, Departament de F\'isica Fonamental, Universitat de Barcelona, Avda. Diagonal 647, 08028 Barcelona, Spain} 
\alsoaffiliation{CIBER de Bioingenier\'ia, Biomateriales y Nanomedicina, Instituto de Salud Carlos III, Madrid, Spain}

\author{Maria Eugenia Fuentes-Perez}
\affiliation{Centro Nacional de Biotecnolog\'ia, CSIC, 28049 Cantoblanco, Madrid, Spain}

\author{Roland Ramsch}
\affiliation{Institut de Qu\'imica Avan\c cada de Catalunya, Consejo Superior de Investigaciones Cient\'ificas (IQAC-CSIC), 08034 Barcelona, Spain}
\alsoaffiliation{CIBER de Bioingenier\'ia, Biomateriales y Nanomedicina, Instituto de Salud Carlos III, Madrid, Spain}

\author{Susana Vilchez}
\affiliation{Institut de Qu\'imica Avan\c cada de Catalunya, Consejo Superior de Investigaciones Cient\'ificas (IQAC-CSIC), 08034 Barcelona, Spain}
\alsoaffiliation{CIBER de Bioingenier\'ia, Biomateriales y Nanomedicina, Instituto de Salud Carlos III, Madrid, Spain}

\author{Conxita Solans}
\affiliation{Institut de Qu\'imica Avan\c cada de Catalunya, Consejo Superior de Investigaciones Cient\'ificas (IQAC-CSIC), 08034 Barcelona, Spain}
\alsoaffiliation{CIBER de Bioingenier\'ia, Biomateriales y Nanomedicina, Instituto de Salud Carlos III, Madrid, Spain}

\author{Fernando Moreno-Herrero}
\affiliation{Centro Nacional de Biotecnolog\'ia, CSIC, 28049 Cantoblanco, Madrid, Spain}

\author{Fernando Albericio}
\affiliation{Institute for Research in Biomedicine (IRB Barcelona), Barcelona Science Park, Baldiri Reixac 10-12, 08028 Barcelona, Spain} 
\alsoaffiliation{CIBER de Bioingenier\'ia, Biomateriales y Nanomedicina, Instituto de Salud Carlos III, Madrid, Spain}
\alsoaffiliation{Department of Organic Chemistry, University of Barcelona, 08028. Barcelona, Spain}
\alsoaffiliation{School of Chemsitry and Physics, University of KwaZulu-Natal, 4001-Durban, South Africa}

\author{Ram\'on Eritja}
\affiliation{Institut de Qu\'imica Avan\c cada de Catalunya, Consejo Superior de Investigaciones Cient\'ificas (IQAC-CSIC), 08034 Barcelona, Spain}
\alsoaffiliation{Institute for Research in Biomedicine (IRB Barcelona), Barcelona Science Park, Baldiri Reixac 10-12, 08028 Barcelona, Spain} 
\alsoaffiliation{CIBER de Bioingenier\'ia, Biomateriales y Nanomedicina, Instituto de Salud Carlos III, Madrid, Spain}

\author{Ernest Giralt}
\affiliation{Institute for Research in Biomedicine (IRB Barcelona), Barcelona Science Park, Baldiri Reixac 10-12, 08028 Barcelona, Spain} 
\alsoaffiliation{CIBER de Bioingenier\'ia, Biomateriales y Nanomedicina, Instituto de Salud Carlos III, Madrid, Spain}

\author{Sukhendu B. Dev}
\affiliation{Institute for Research in Biomedicine (IRB Barcelona), Barcelona Science Park, Baldiri Reixac 10-12, 08028 Barcelona, Spain} 

\author{Felix Ritort}
\email{fritort@gmail.com}
\affiliation{Small Biosystems Lab, Departament de F\'isica Fonamental, Universitat de Barcelona, Avda. Diagonal 647, 08028 Barcelona, Spain} 
\alsoaffiliation{CIBER de Bioingenier\'ia, Biomateriales y Nanomedicina, Instituto de Salud Carlos III, Madrid, Spain}


\title[\texttt{achemso} demonstration]
{Electrostatic Binding and Hydrophobic Collapse of Peptide-Nucleic Acid Aggregates Quantified Using Force Spectroscopy}

\begin{document}

\begin{abstract}
Knowledge of the mechanisms of interaction between self-aggregating peptides and nucleic acids or other polyanions is key to the understanding of many aggregation processes underlying several human diseases ({\em e.g.} Alzheimer's and Parkinson's diseases). Determining the affinity and kinetic steps of such interactions is challenging due to the competition between hydrophobic self-aggregating forces and electrostatic binding forces. Kahalalide F (KF) is an anticancer hydrophobic peptide which contains a single positive charge that confers strong aggregative properties with polyanions. This makes KF an ideal model to elucidate the mechanisms by which self-aggregation competes with binding to a strongly charged polyelectrolyte such as DNA. We use optical tweezers to apply mechanical forces to single DNA molecules and show that KF and DNA interact in a two-step kinetic process promoted by the electrostatic binding of DNA to the aggregate surface followed by the stabilization of the complex due to hydrophobic interactions. From the measured pulling curves we determine the spectrum of binding affinities, kinetic barriers and lengths of DNA segments sequestered within the KF-DNA complex. We find there is a capture distance beyond which the complex collapses into compact aggregates stabilized by strong hydrophobic forces, and discuss how the bending rigidity of the nucleic acid affects such process. We hypothesize that within an {\em in vivo} context, the enhanced electrostatic interaction of KF due to its aggregation might mediate the binding to other polyanions. The proposed methodology should be useful to quantitatively characterize other compounds or proteins in which the formation of aggregates is relevant.\newline

\end{abstract}

[keywords: DNA condensation | aggregation | single-molecule | force spectroscopy | DNA-peptide complex | optical tweezers]

\maketitle
\newpage

Understanding the driving forces by which self-aggregating molecules bind their targets
inside the cell is of the utmost importance to elucidate their
mechanisms of action.\cite{calamai2006nature, bucciantini2002inherent, gsponer2006theoretical, coan2008stoichiometry, puchalla2008burst, feng2008small, coan2009promiscuous} Self-aggregating peptides bearing a definite charge are able to establish strong electrostatic interactions with oppositely charged polymers. In particular, recent bulk studies have shown that amyloid peptides with positive charges ({\em e.g.} human lysozyme, A$\beta$40, $\alpha$-synuclein, histidine-leucine peptides) have a strong binding affinity  to negatively charged polymers ({\em e.g.} nucleic acids, polysaccharides, polylysines) stimulating aggregation and fibril formation.\cite{calamai2006nature, braun2011amyloid, di2012binding, macedo2012nonspecific, motamedi2012rapid, cherny2004double,cohlberg2002heparin} Such ubiquituous interaction has triggered discussion on its connection to neurodegenerative diseases, and on the hypothetical role of aggregating peptides as scaffolds for polynucleotide assembly in early prebiotic life.\cite{calamai2006nature, dale2006protein} Despite their prevalence, and even if much progress has happened within the last few years, the complex and nonspecific nature of these interactions makes difficult to quantitatively determine key parameters such as their binding affinities and the kinetic steps during the interaction process.

\begin{figure}
\includegraphics{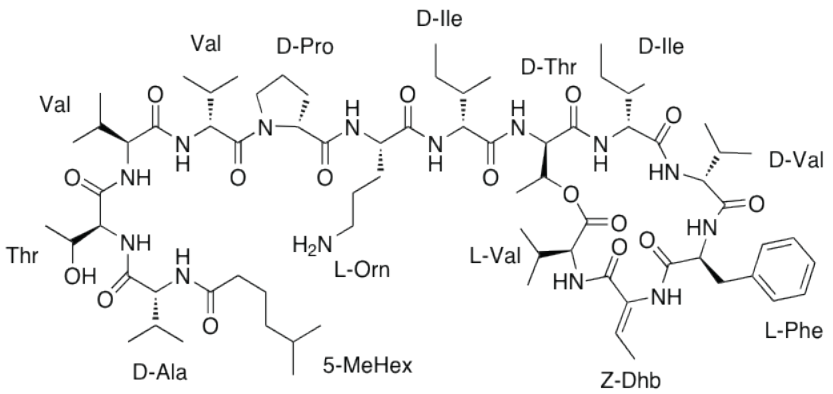}
\caption{{\bf Kahalalide F structure.} }
\label{fig:1}
\end{figure}

Indeed, a full characterization of the interaction between hydrophobic peptides and polyanions is challenging due to the competition between peptide-peptide self-aggregating interactions and peptide-substrate
  electrostatic binding forces, as well as due to the transient and heterogenous nature of the formed complexes.\cite{motamedi2012rapid} An excellent model to address such
  questions is the anticancer drug Kahalalide F (KF), a 14-residue
  cyclic depsipeptide originally isolated from the Hawaiian mollusk
  \emph{Elysia rufescens}.\cite{hamann1993kahalalide} KF is a low solubility compound with a
  highly hydrophobic structure and a single positive residue (L-ornithine) (Figure \ref{fig:1}),\cite{lopez2001synthesis} which exhibits a potent cytotoxic activity
  against several tumor cell lines \cite{garcia1996antitumoral,
    suarez2003kahalalide, sewell2005mechanism} causing the disruption of
  the plasma membrane due to the accumulation of peptide aggregates.\cite{molina2011irvalec} Although KF is a molecule with a strong
  tendency to aggregate it also has a single positive charge capable
  of establishing electrostatic interactions with negatively charged
  substrates such as DNA. Here we use optical tweezers, AFM imaging, and
  dynamic light scattering (DLS) to fully characterize the interaction
  of KF aggregates binding to DNA.  Although the formation of KF-DNA
  complexes can be directly observed in AFM and DLS
  measurements, only force spectroscopy methods make possible to quantitatively
  determine the driving thermodynamic forces. In addition, by applying mechanical force to the ends of the DNA it is possible to control and gain insight into the kinetic steps involved in the formation of the complex.

We have found that binding of DNA to KF occurs in two kinetic
steps: First, DNA binds KF particles due to the electrostatic attraction between the negatively charged DNA and the
positively charged groups exposed on their surface (L-Orn). Electrostatic binding compacts DNA by sequestering DNA segments along the surface of the aggregate in a way reminiscent of a
condensation process. This is followed by a slow remodeling of hydrophobic contacts and
the irreversible entrapping of DNA within the KF-DNA complex. Modeling of the stretching curves
yielded characteristic parameters of the interaction,
such as the average length of DNA segments electrostatically bound to
the aggregate, their affinity of binding and the barrier to unpeel them. DNA unzipping experiments show that KF also forms complexes with
ssDNA. However the different  
mechanical (bending rigidity) and chemical (hydrophobicity) properties of
the polyelectrolyte determine the kinetics of formation of the complex.

\section{Results}
\label{sec:Results}

\subsection{KF compacts dsDNA}
\label{sec:dsDNA}
\begin{figure}
\begin{center}
\includegraphics{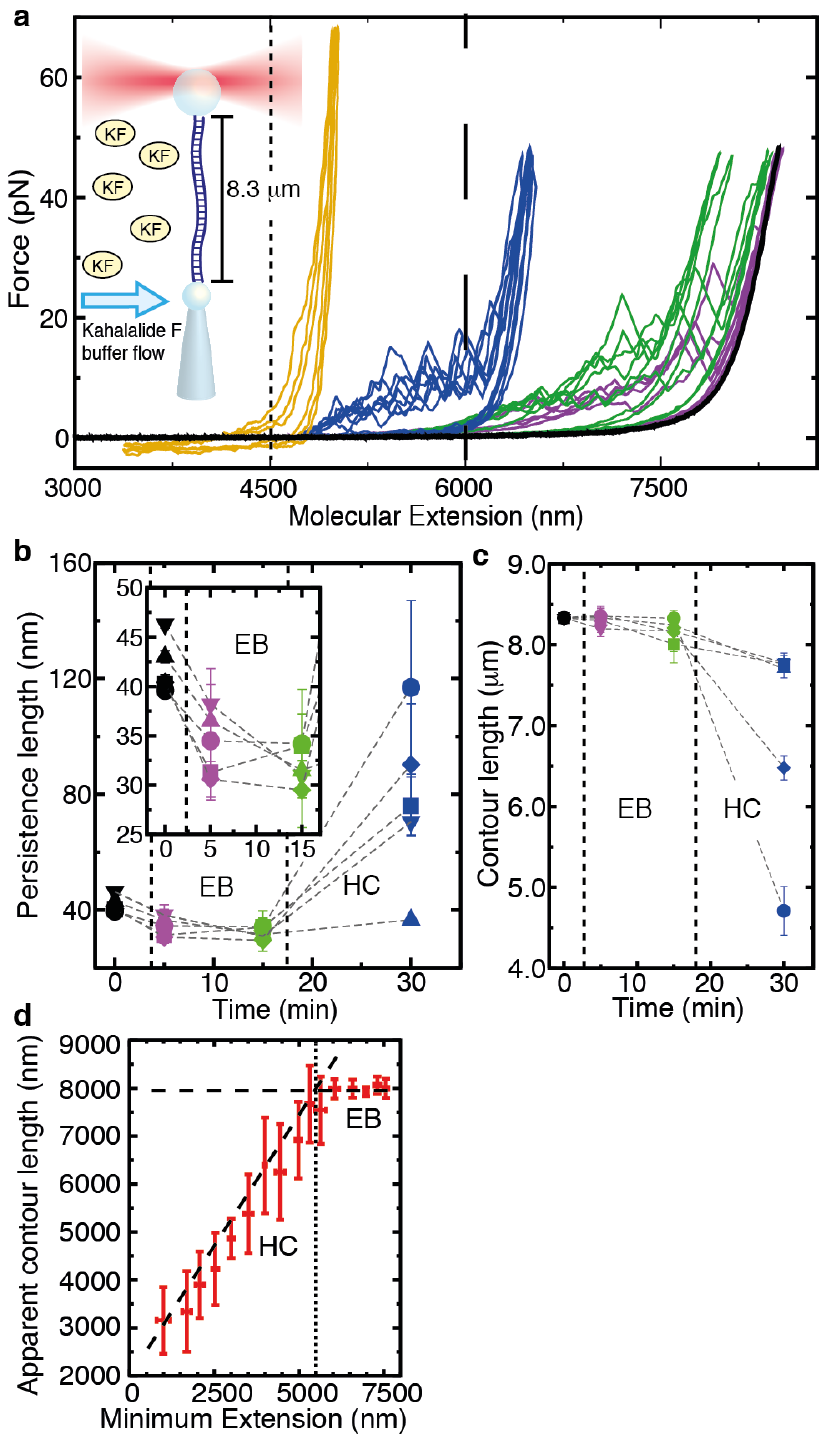}
\end{center}
\caption{{\bf KF binds to dsDNA.} {\bf (a)} DNA pulling curves before (black) and after flowing KF at different waiting times: 5 min (purple), 15 min (green) and 30 min (blue). The molecule is maintained at an extension of 6 $\mu$m (vertical dashed line), and the flow is temporarily stopped to perform pulling cycles between a minimum extension of 4.5 $\mu$m (dotted line) and a maximum force of 45 pN. The sawtooth pattern observed indicates that KF induces the compaction of DNA. Pulling cycles reaching end-to-end distances lower than 4 $\mu$m are shown in yellow. Data is filtered at 10 Hz bandwidth, v=500 nm/s. {\bf (Inset)} Experimental set-up. {\bf (b,c)} Persistence and contour length of five DNA molecules after flowing KF (black corresponds to naked DNA).  The changes in the elastic parameters are a signature of the two regimes observed in KF-DNA complex formation: electrostatic binding and hydrophobic collapse.  {\bf (d)} Apparent contour length of a DNA molecule repeatedly pulled between a maximum force of 40 pN and a minimum extension that decreases in steps of 500 nm per pulling cycle (mean$\pm$SD, N=10). }
\label{fig:2}
\end{figure}

To study how KF binds DNA we stretched a single half $\lambda$-DNA (24-kb) in the presence of KF in the optical tweezers set-up (Figure \ref{fig:2}{\bf a}, Inset). First a DNA molecule was tethered between two beads and its elastic properties measured using the Worm-Like Chain (WLC) model (see Methods). Next the DNA molecule was rinsed with 40 $\mu$M KF while it was maintained at an end-to-end distance of 6 $\mu$m. In this configuration the DNA can explore bended conformations due to thermal fluctuations, as the force remains below 0.4 pN at this extension. The flow was temporarily stopped after 5, 15 and 30 min in order to record a series of force-extension curves (Figure \ref{fig:2}{\bf a}). We collected measurements for at least ten different molecules finding a reproducible pattern (Figure S1, Supporting Information).

After flowing KF for 5 min, DNA maintained at low tension was compacted by the peptide. In order to stretch the compacted molecule, the KF-DNA complex must be unraveled, and therefore a sawtooth pattern with many force rips was observed (Figure \ref{fig:2}{\bf a}, purple). This suggests that KF behaves as a DNA condensing agent, inducing kinks and loops on the DNA. The relaxation curves however, remained similar to those obtained for naked DNA (Figure \ref{fig:2}{\bf a}, black) indicating that DNA compaction took place after the extension of the molecule was reduced. The relaxation curves were well described by the WLC model and showed a decrease of $25\%$ in the persistence length (Figure \ref{fig:2}{\bf b}, Inset). In experiments where KF was flowed for 5 or 15 min, the whole contour length of the DNA could be recovered after pulling up to 40 pN (Figure \ref{fig:2}{\bf c}). This reduction of the persistence length is likely due to the positively charged L-Orn residue that decreases the self-repulsion of the DNA phosphate-backbone.

However, after 15 to 30 min, many interactions could not be disrupted leading to an apparent shorter contour length, that was correlated to an increase of the persistence length. This phenomenology suggests that the KF-DNA complex started to collapse into a more stable and stiffer structure after 15 min. Remarkably, a repulsive negative force was detected after 30 min in most of the experiments if end-to-end distances lower than 4 $\mu$m were allowed (Figure \ref{fig:2}{\bf a}, yellow), suggesting the formation of a thick KF-DNA aggregate of 1-3 $\mu$m of length. It was not possible to remove the bound KF by rinsing the molecule with peptide-free buffer for more than 45 min, reflecting the high stability of the final complex.

Binding of KF to DNA  exhibits two regimes. First, there is a weak and fast regime apparently determined by the electrostatic attraction between the positively charged residues of the KF particles and the negatively charged backbone of DNA. According to our interpretation in this regime DNA binds to hydrophilic spots exposed on the surface of KF particles. Unpeeling of DNA segments requires forces typically lower than 20 pN. We will refer to this mode of binding as electrostatic binding (EB). This regime is observed in the first 15 min of the experiments shown in Figure \ref{fig:2}{\bf a}, and is characterized by a constant contour length, the presence of force rips associated to unpeeling events and a reduced persistence length; the increased flexibility of the filament indicates a charge compensation that reduces self-repulsion of phosphates along the DNA backbone. There is a second stronger binding regime that occurs over longer timescales, that we attribute to the formation of an increasing number of stable hydrophobic contacts within the growing KF-DNA complex. Our interpretation is that in this regime DNA gets buried inside the bulk of the aggregate after being recruited by the hydrophilic spots exposed on its surface creating a stiff filament (as suggested by its increased persistence length). The slower timescales observed in this regime suggest that this mode of binding requires the remodeling and growth of a strongly hydrophobic complex. We will therefore refer to this second regime as hydrophobic collapse (HC). It is characterized by an irreversible decrease in contour length, an increase in persistence length, and the final collapse of the KF-DNA complex that is eventually compressed by pushing the two beads closer than 4 $\mu$m. The force required to disrupt such HC structure is above those accessible with our set-up ($\sim$ 100 pN), as suggested from previous AFM pulling experiments of single hydrophobically collapsed polymers.\cite{li2011signature}

What is the parameter that controls the prevalence of each binding regime? We expect that DNA bending fluctuations determine its binding to KF and the subsequent stabilization of the complex. Therefore, the molecular extension -or distance between beads- should be the parameter controlling the transition between both regimes. To verify this hypothesis we carried out experiments where the DNA was repeatedly pulled in the presence of KF between a maximum force of 40 pN and a minimum extension that progressively decreased from 8 $\mu$m to 2 $\mu$m in steps of 500 nm per pulling cycle. Such minimum extension controls the degree of compaction reached by the complex. For each cycle we then measured the apparent contour length at the maximum force (40 pN). The results (Figure \ref{fig:2}{\bf d}) confirm the presence of the aforementioned regimes, which are separated by a threshold capture distance of 5.5 $\mu$m (corresponding to 66\% of the contour length of the molecule). In the EB regime (relative extension $\ge$ 66\%) the apparent contour length of the DNA fiber does not change whereas in the HC regime (relative extension $\le$ 66\%) it decreases linearly with the minimum distance between the two beads. A KF analog in which the ornithine residue was replaced by a negatively charged glutamic acid was investigated. No interaction between the KF analog and DNA was observed (Figure S2, Supporting Information), confirming that the positive charge of the ornithine residue is essential for electrostatic binding and providing further evidence that electrostatic interactions are key for the initial binding of the peptide to DNA. As well, a salt titration showed that the initial binding of KF to DNA and the subsequent DNA compaction is highly dependent on the ionic strength of the buffer, in agreement with the proposed mechanism (Figure S3, Supporting Information).

\begin{figure}
\begin{center}
\includegraphics{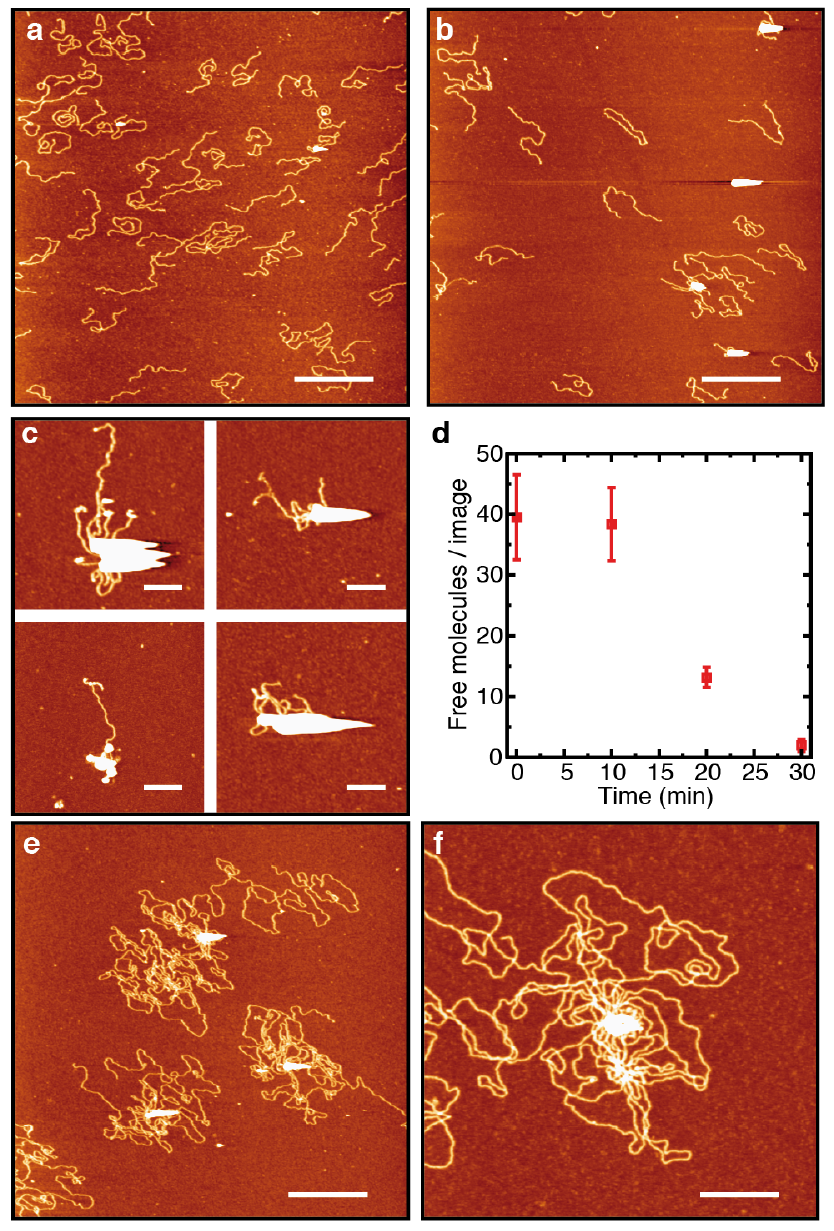}
\end{center}
\caption{{\bf AFM images of KF-DNA complexes.} {\bf(a-c)} AFM images of reactions of 1.65 ng linearized pGEM plasmid (2743-bp) and 100 $\mu$M KF obtained at 0, 20 and 30 min incubation times at room temperature, respectively. The number of free DNA molecules decreases with incubation time and large compaction blobs are observed. {\bf(d)} DNA surface density at different incubation times, determined as the average number of free individual molecules per image of 9 $\mu$m$^2$ (mean$\pm$SD, N$\ge$6). {\bf(e, f)} Full $\lambda$-DNA (48-kb) incubated with 100 $\mu$M KF for 30 min. DNA condensation and formation of blobs are also seen for this larger DNA substrate. Bar scale is 600 nm (a, b, e) and 200 nm (c, f). Color scale (from dark to bright) is 0-2 nm in all AFM images.}
\label{fig:3}
\end{figure}

The formation of KF-DNA aggregates was directly observed by AFM imaging using a 2743-bp DNA fragment (Figure \ref{fig:3}{\bf a-c}). Formation of blobs was observed at the initial time of mixing, their average size and number increased with time. Notably, after 20 min, a sharp decrease in the number of individual molecules bound to the mica surface was observed (Figure \ref{fig:3}{\bf d}). We attribute this to the formation of intermolecular complexes in which several DNA molecules are recruited into a single aggregate. As a consequence, no free DNA was observed after 30 minutes incubation time. These results were further confirmed with the use of longer DNA molecules ($\lambda$-DNA, 48-kb, Figure \ref{fig:3}{\bf e, f}). In the absence of DNA, large aggregates were not found on the images (Figure S4, Supporting Information). The observed aggregation of KF and binding to DNA was also characterized with DLS (Section S2, Supporting Information). The hydrodynamic radius of KF aggregates increased with time and remained constant after addition of DNA.

\subsection{KF-DNA affinity measurements in the EB regime}
\label{sec:dsDNA2}

\begin{figure}
\includegraphics{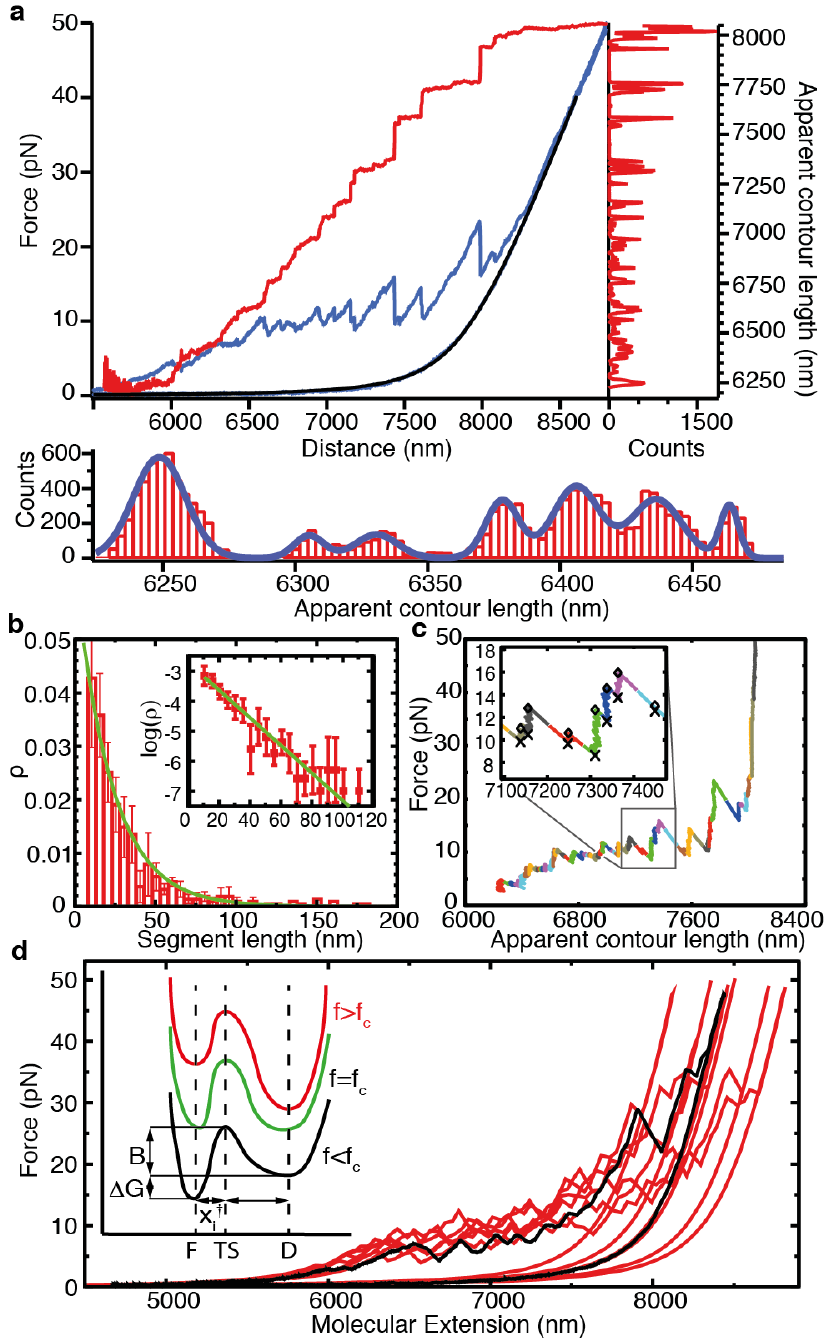}
\caption{{\bf Modeling of DNA stretching experiments.} {\bf (a)} (Top, left) Blue curve shows a typical force-distance curve in a KF-DNA pulling experiment (20 Hz bandwidth, v=30 nm/s). Red curve shows the apparent contour length $l_{0}$ (right axis) for each experimental data point. The release cycle is fitted to the WLC model (black). (Top, right) Histogram of  $l_{0}$ values. (Bottom) Detailed view of the histogram (red) and fit to a sum of Gaussians (blue). {\bf (b)} Histogram of unpeeling segment lengths $\Delta l_{0}$ and fit to an exponential distribution (green). Inset shows a log-normal plot (mean$\pm$SD, N=435 events, 3 molecules). {\bf (c)} Force {\em vs.} apparent contour length representation of the pulling experiment. Each color identifies a state (apparent contour length) temporally stabilized by KF-DNA contacts during the unpeeling process. {\bf (Inset)} Sharp transitions between states are observed. Minimum and maximum forces of every state are indicated with crosses and diamonds respectively. {\bf (d)} The black curve is an experimental pulling curve after 15 min interaction with KF (v=500 nm/s). A set of six simulations of the theoretical model is shown in red. {\bf (Inset)}  Scheme of the free-energy landscape of a two-states system at different forces (f). The main parameters describing the system are: the free energy difference ($\Delta$G) between the formed (F) and dissociated (D) conformations, the height of the barrier (B), the distance ($x_i$) separating the two conformations, and the distance ($x_i^{\dagger}$) from the transition state (TS) to the formed conformation. As the force is increased the free-energy landscape is tilted favoring the dissociated conformation once the critical force ($f_c$) is reached.}
\label{fig:4}
\end{figure}

Pulling experiments of DNA in the presence of KF show a force-distance curve pattern with force rips and hysteresis even if the pulling is performed at very low speeds (Figure \ref{fig:4}{\bf a}, blue). Low pulling speeds are particularly useful to characterize the affinity of DNA binding to KF aggregates during the EB regime. In these experiments, the slope between two consecutive force rips reflects the elastic response of DNA with a given apparent contour length  $l_{0}$. Each force rip is due to the unpeeling of a DNA segment that was electrostatically bound to the KF particle. A statistical analysis of force rips was used to determine the length of the DNA segments released during the unpeeling process (see Methods).\cite{huguet2009statistical} In this way, each experimental data point was associated to an apparent contour length $l_{0}$ (Figure \ref{fig:4}{\bf a}, red left). A histogram of all the $l_{0}$ values showed a series of peaks that identify states that are stabilized by KF-DNA contacts (Figure \ref{fig:4}{\bf a}, red right). The distance between two consecutive peaks is the length of the DNA segment released at every unpeeling event. The histogram was then fitted to a sum of Gaussians (Figure \ref{fig:4}{\bf a}, bottom), and the distance between the mean of consecutive peaks was calculated. The experimental distribution of unpeeling events is broad (from a few nm to $\sim$400 nm) and follows an exponential distribution with  mean size $\Delta l_{0}^{*}=31\pm6$ nm (Figure \ref{fig:4}{\bf b}). An exponential distribution of unpeeling lengths is known to correspond to the distribution of intervals expected in random partitioning a given contour length, in agreement with our hypothesis that DNA binds KF aggregates at hydrophilic spots in a random fashion.

A force {\em vs.} contour length representation (Figure \ref{fig:4}{\bf c}) emphasizes the release of DNA segments in a stepwise manner during the unpeeling process. The mechanical work performed at each unpeeling event ($W=F\Delta x$) was then inferred from the rupture force value and released extension (see Methods). A histogram of the dissipated work (Figure S5, Supporting Information) shows an exponential distribution with average unpeeling energy of $13.5\pm5$ kcal/mol. This value sets an upper limit to the free energy of binding of KF to DNA.

To gain a better understanding of the affinity of DNA binding to KF particles, we used a simple theoretical model that reproduces the experimental force-extension curves. We considered a model previously used to characterize the DNA-dendrimer condensation transition (see Methods).\cite{Ritort:2006fk} The model reproduces the essential features of the experimental curves (Figure \ref{fig:4}{\bf d}) over a wide range of pulling speeds (30-500 nm/s). Despite the apparent large number of free parameters, only certain values in very specific ranges can reproduce these features (Section S3, Supporting Information). In brief, the experimental force-extension curves could be well described by assuming: (a) a low binding energy of DNA to KF aggregates  ($\Delta G\sim6\pm2$ kcal/mol); (b) a brittle unpeeling of the DNA segments ($x^{\dagger}_{i}= 2\pm$1 nm, the barrier lying close to the formed conformation); and (c) a broad right-tailed distribution $p\left(B\right)$ of high energy activation barriers given by $p\left(B\right)=\left(1/w^{\prime}\right) \exp{\left[-\left(B-B_{0}\right)/w^{\prime}\right]}$ with $B\ge B_{0}=89$ $k_{B}T$ and $w^{\prime}=5$ $k_{B}T$.

\subsection{DNA binds to KF aggregates at forces lower than 1pN}
\label{sec:dsDNA3}

We followed the kinetics of DNA compaction by performing constant-force experiments at forces such that EB prevails (molecular extension $\ge$ 5.5 $\mu$m, Figure \ref{fig:2}{\bf d}). The DNA molecule was maintained at constant force using force-feedback, and we followed the time-evolution of the molecular extension while KF was flowed in.

\begin{figure}
\includegraphics{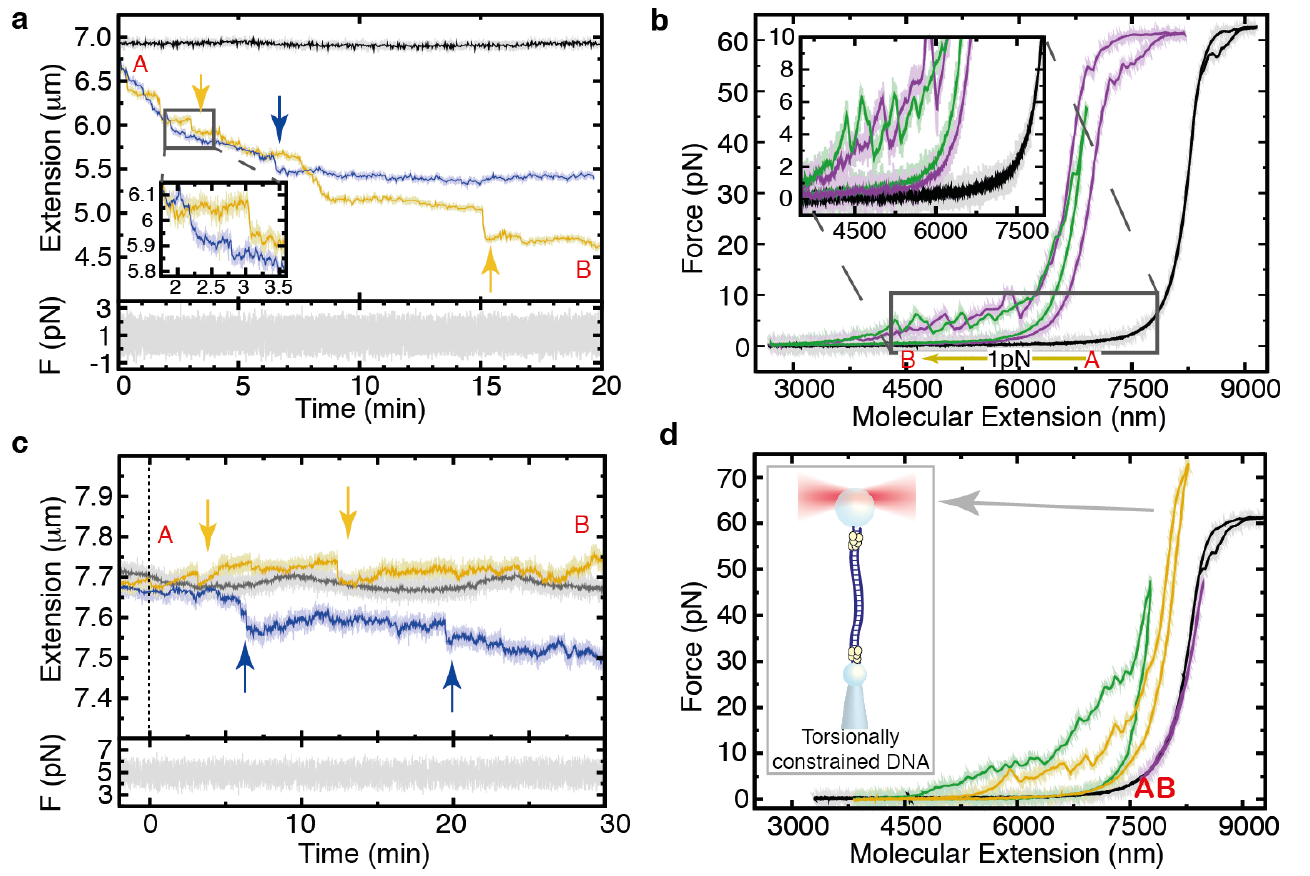}
\caption{{\bf Kinetics of DNA binding to KF particles at a constant force.} {\bf (a)} DNA compaction at 1 pN. A control without peptide (gray) and two equivalent experiments at 40 $\mu$M KF (blue and yellow) are shown. The molecule was maintained at an initial extension A that relaxed at constant force down to a final value B. {\bf (b)} Stretching of a DNA molecule before (black) and after (green, purple) the constant-force experiment at 1 pN.  {\bf (c)} KF does not compact DNA at 5 pN. A DNA molecule subjected at 5 pN is rinsed with KF and changes in the molecular extension are monitored. A control experiment without peptide (gray), and two independent experiments at 40 $\mu$M KF (blue and yellow) are shown. Large fluctuations indicative of individual binding events are observed (arrows) {\bf (d)} Stretching of a DNA molecule only at forces higher than 5 pN immediately after the constant-force experiment at 5pN (purple). If the force was relaxed below 5 pN (green and yellow), the characteristic sawtooth pattern was immediately recovered. The force-extension curve of that molecule before flowing KF is shown in black. For all plots, raw data (1 kHz) is shown in light colors, and filtered data (1 Hz bandwidth for kinetic experiments, 10 Hz bandwidth for pulling experiments) is presented in dark colors. Pulling speed is 500 nm/s. 
}
\label{fig:5}
\end{figure}

At 1pN, a fast compaction took place (Figure \ref{fig:5}{\bf a}). The extension was reduced up to $40\%$ in 20 min at a reproducible rate. This compaction is characterized by intermittent drops of extension that shorten the molecule by hundreds of nanometers in a few seconds (Figure \ref{fig:5}{\bf a}, arrows). Pulling curves performed after this experiment (Figure \ref{fig:5}{\bf b}) showed again the characteristic sawtooth pattern.

In contrast, at 5pN the molecular extension remained constant within 100 nm after flowing KF for more than 30 min (Figure \ref{fig:5}{\bf c}). Still, intermittent large fluctuations on the order of tens of nanometers were often detected (Figure \ref{fig:5}{\bf c}, arrows). These large fluctuations were never observed in controls without KF (Figure \ref{fig:5}{\bf c}, gray), and we attribute them to individual binding events. Pulling cycles performed between 5-40 pN immediately after the peptide flow show a slight decrease in the persistence length and weak hysteresis effects suggesting very weak binding of DNA to KF (Figure \ref{fig:5}{\bf d}). Only by further decreasing the extension and force of the molecule full binding events were observed (Figure \ref{fig:5}{\bf d}).

Interestingly an overstretching transition was not always observed (Figure \ref{fig:5}{\bf d}, yellow). We attribute this to the recruitment of DNA segments close to both ends of the tethered molecule by KF particles that induce a torsionally constrained fiber, inhibiting the overstretching transition (Figure \ref{fig:5}{\bf d}, Inset).\cite{leger1999structural} Otherwise, KF binding does not suppress or tilt the overstretching plateau as observed for DNA intercalators.\cite{vladescu2007quantifying} Moreover, the characteristic sawtooth pattern of KF remained visible after fully overstretching the DNA (Figure S6, Supporting Information).

\subsection{Unzipping experiments reveal different binding modes of KF to dsDNA and ssDNA}
\label{sec:unzip}

\begin{figure}
\includegraphics{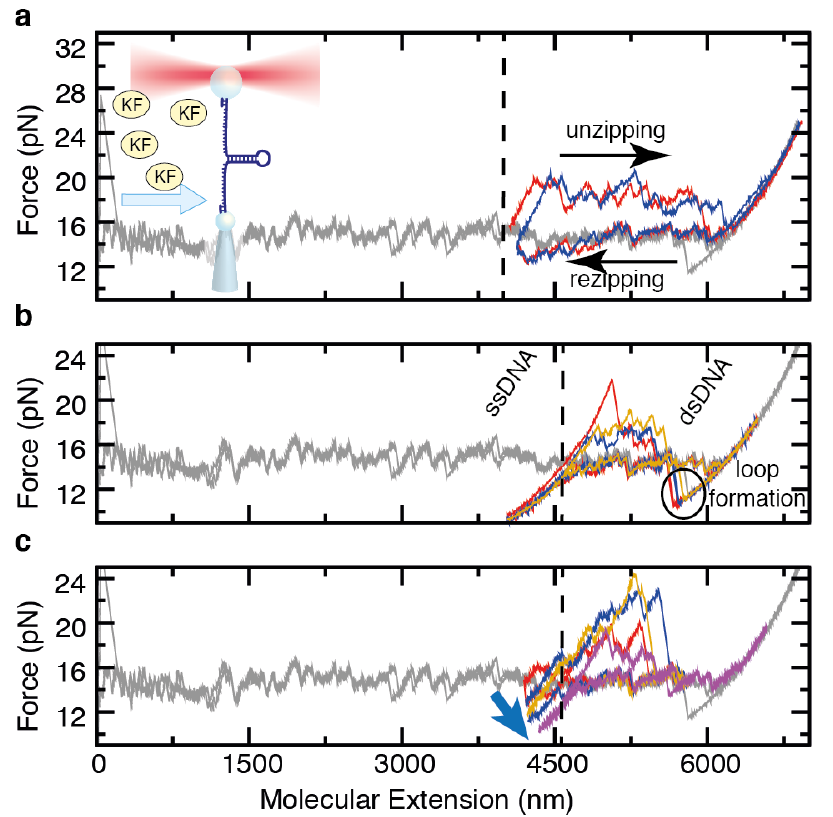}
\caption{{\bf Unzipping experiments show that KF binds both dsDNA and ssDNA.} {\bf (a)} An unzipping pattern of the DNA hairpin before incubation with KF is fully represented in gray in each panel. The dashed line represents the position at which the molecule remained unzipped during the peptide flow. Consecutive pulling cycles of the dsDNA stem region after incubation with KF (blue, red) show a strong distortion of the unzipping pattern. However, the re-zipping of the hairpin remains unaffected indicating that the molecule can hybridize again. {\bf (Inset)} A 6.8-kb hairpin is maintained partially unzipped leaving less than half of the dsDNA stem closed during the incubation with KF. This configuration prevents KF-DNA interactions mediated by the beads. {\bf (b)} The region of the hairpin that remained as ssDNA during the peptide flow (left of dashed line) cannot hybridize again in contrast to what happens without KF (gray curve) or in the region maintained as dsDNA during the peptide flow (right of dashed line). Three pulling cycles are plotted in blue, red and yellow. {\bf (c)} If the hairpin is rinsed with KF and then submitted to several consecutive pulling cycles, the non-hybridizing region increases with time. Four pulling cycles that reflect this trend are shown (red, blue, yellow and purple). Data is filtered at 10 Hz bandwidth, v=50 nm/s.}
\label{fig:6}
\end{figure}

In a different set-up (Figure \ref{fig:6}{\bf a}), a 6.8-kb DNA hairpin was tethered and partially unzipped, maintaining at least half of the dsDNA stem open (Figure \ref{fig:6}{\bf a}, dashed line) and then KF was flowed into the chamber. In this configuration the released ssDNA is long but rigid enough to severely restrict thermal fluctuations in the molecular extension ($\rm{r.m.s.d.}\sim$20 nm). The advantage of this set-up is that the long separation between the hairpin and the beads ($\sim$4 $\mu$m) inhibits any interaction between the beads and both the dsDNA region and linkers. The unzipping pattern of a DNA molecule is a fingerprint of its base-sequence \cite{huguet2010single} (Figure \ref{fig:6}{\bf a}, gray), and changes of that pattern indicate a direct interaction between the peptide and DNA. Moreover, with this set-up we could explore the effect of KF on a DNA molecule maintained at zero force and forming a random-coil (the force stretches the linkers but not the hairpin).

After flowing KF for 3 min, the unzipping pattern substantially changed (Figure \ref{fig:6}{\bf a}) and forces up to 22 pN were needed to unzip the DNA. We attribute this to the increased force required to simultaneously break the base-pairing interactions and unpeel DNA segments from the KF particles. Consecutive unzipping curves show that ssDNA remains bound to KF particles at the maximum forces (25 pN). Surprisingly enough, the re-zipping trajectories overlapped with the re-zipping curves of naked DNA over a wide range of extensions ($\ge$1500 $\rm{nm}$). This is in agreement with a re-annealing mechanism in which re-hybridization takes place first, followed by the formation of the complex. A similar phenomenology has been observed  in the formation of amyloid nucleic acid fibers, in which the binding of amyloid peptides to oligonucleotides promotes their hybridization.\cite{braun2011amyloid}

Note that in this experiment we only unzipped the region of the hairpin that remained in double-stranded form while KF is flowed (right of the dashed line in Figure \ref{fig:6}{\bf b}). However, when we tried to rezip the region of the hairpin that remained as ssDNA during the peptide flow (left of the dashed line in Figure \ref{fig:6}{\bf b}),  we could not recover the characteristic unzipping pattern of the molecule. This indicates that KF can bind ssDNA in a way that prevents re-hybridization of ssDNA strands. If the molecule was continuously submitted to unzipping/re-zipping cycles, the region of the DNA hairpin that previously re-zipped progressively looses that capability (Figure \ref{fig:6}{\bf c}) suggesting that KF is slowly binding to the stretched ssDNA (arrow in Figure \ref{fig:6}{\bf c}).

These results suggest that KF binds to the phosphate backbone in a configuration that does not interfere with base-pairing interactions when DNA is in its double-stranded form. However, when DNA is in its single-stranded form, KF can adopt configurations that interfere with the re-zipping of the molecule. These experiments also show that the interaction of the peptide with ssDNA is slow (in the order of minutes), as only the ssDNA regions of the hairpin that remained exposed for long times to the peptide were unable to re-hybridize.

\subsection{KF binds ssDNA} 
\label{sec:ssdna}

\begin{figure}
\includegraphics{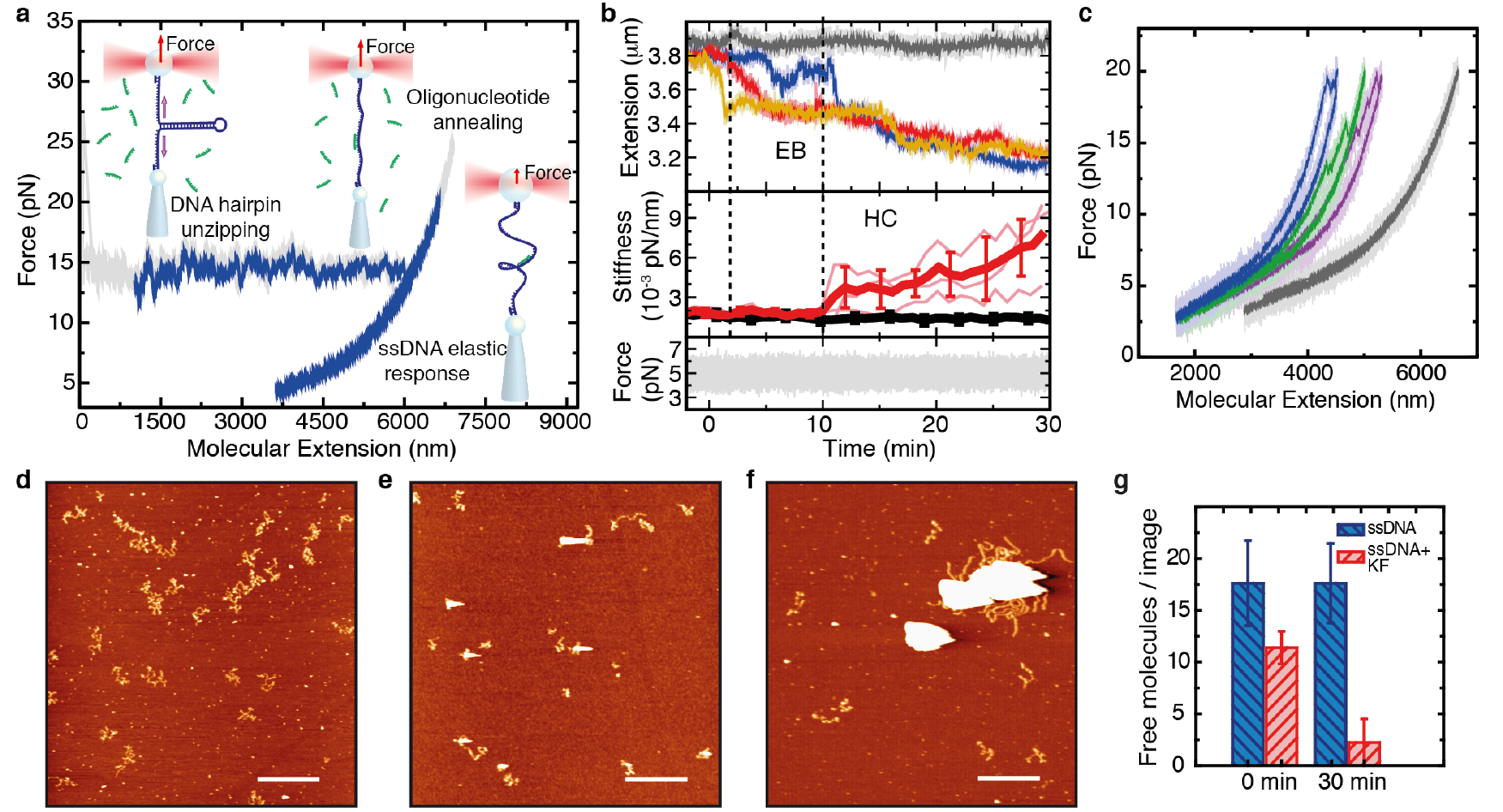}
\caption{{\bf Kinetics of ssDNA binding to KF particles.} {\bf (a)} Method used to generate a long ssDNA template from a DNA hairpin. The specific binding of an oligonucleotide to the hairpin inhibits the hybridization of the molecule (blue) at forces lower than the average unzipping force. The full force-extension curve of the hairpin is plotted as a reference (gray). {\bf (b)} Compaction and stiffening of ssDNA at 5 pN are representative of EB and HC respectively. (Top) Extension of a ssDNA molecule rinsed with KF at 5 pN. Three independent experiments at 40 $\mu$M KF (blue, red and yellow) and a control without peptide (gray) are shown (compaction starts at t=0). (Middle) Average stiffness of the ssDNA molecule at 5 pN during the peptide flow (red) and a control without peptide (black). Three individual experiments are shown in light red. The stiffness is measured from the fluctuations in the trap position. {\bf (c)} Stretching curves of a ssDNA molecule before (gray) and after incubation with 40 $\mu$M KF at 3 pN for 25 min (purple, green, blue). Data is collected at 1 kHz (light colors) and filtered to 1 Hz bandwidth (dark colors).  Pulling speed is 100 nm/s.  {\bf (d)} ssDNA molecules (1.8 nM molecules, 5 $\mu$M nucleotides) are adsorbed on a mica surface showing a much compact conformation than dsDNA due to its lower persistence length. {\bf (e)} Immediately after mixing KF (100 $\mu$M) with ssDNA molecules (5 $\mu$M nucleotides) we observe the formation of aggregation spots and a substantial decrease in the number of ssDNA molecules per image. {\bf (f)} After 30 min incubation at room temperature, these effects are more evident as big aggregates are seen. {\bf (g)} Histogram of ssDNA molecules at different incubation times (0 and 30 min) with and without KF, determined as the average number of free individual molecules per image of 4 $\mu$m$^2$ (mean$\pm$SD, N$\ge$6).}
\label{fig:7}
\end{figure}

To characterize the interaction of ssDNA with KF aggregates we developed a simple method to generate a long ssDNA template (13-kb) for optical tweezers experiments (see Methods). By using this setup we could therefore measure the elastic response of the ssDNA down to forces as low as 1-2 pN (Figure \ref{fig:7}{\bf a}). 

 We then followed their molecular extension in the presence of KF. At 5 pN we observed a slow compaction ($\sim$20-30 min) with an absolute reduction in extension close to 16$\%$ (Figure \ref{fig:7}{\bf b}, upper panel), demonstrating that ssDNA binds KF. This phenomenology was reproducible within different experiments, and the slow kinetics agree with the results from unzipping experiments (Figure \ref{fig:6}{\bf c}). We also measured the time evolution of the stiffness of the KF-ssDNA fiber by recording the magnitude of the thermally induced fluctuations in the molecular extension (Section S4, Supporting Information). At 5 pN the molecule stiffened with time at a rate of (13$\pm$5)$\cdot10^{-3}$ pN/(nm$\cdot$s) (Figure \ref{fig:7}{\bf b}, middle panel). However this change was only observed 10-15 min after compaction of the fiber started, suggesting that stiffness changes are mostly due to the hydrophobic collapse of the KF-ssDNA aggregate rather than electrostatic binding of ssDNA to KF particles. 

Pulling curves obtained after the peptide flow (Figure \ref{fig:7}{\bf c}) also show force rips in the stretching curves. However the sawtooth pattern was smoother than for dsDNA suggesting the occurrence of fewer events and higher unpeeling forces (Figure \ref{fig:7}{\bf c}). We attribute this to the increased hydrophobic forces that stabilize the KF-ssDNA complex, that also lead to a systematic shortening of the effective contour length of the ssDNA. At a higher stretching force of 10 pN, KF did not induce compaction of ssDNA though, but intermittent jumps in the extension were observed, indicative of individual binding events (Figure S7{\bf a}, Supporting Information). At this higher force, the stiffness of the molecule remained constant within the resolution of measurements (Figure S7{\bf b}, Supporting Information). However, KF-ssDNA compaction could be induced by lowering the force down to 5 pN (Figure S7{\bf a}, Supporting Information) reproducing the phenomenology reported in Figure \ref{fig:7}{\bf b}.

Binding of ssDNA to KF aggregates was further corroborated by AFM. ssDNA molecules (2743-bp) were generated by heat denaturation and fast cooling down to 4$^\circ$C (Figure \ref{fig:7}{\bf d}). Addition of KF to the ssDNA preparation triggered the formation of aggregation spots immediately after mixing (Figure \ref{fig:7}{\bf e}). Interestingly enough, longer incubation times yielded a reduction of free ssDNA molecules, together with the formation of large aggregates of KF surrounded by double stranded DNA (Figure \ref{fig:7}{\bf f-g}). This observation suggests that KF induces the rehybridization of ssDNA, as ssDNA molecules do not anneal after 30 min incubation at room temperature in the absence of KF. This is in agreement with the unzipping results (see previous section), and is likely due to the fact that KF traps and maintains close in space different ssDNA molecules that are occasionally able to rehybridize.  

\section{Discussion}
\label{sec:Discussion}

\begin{figure}
\includegraphics{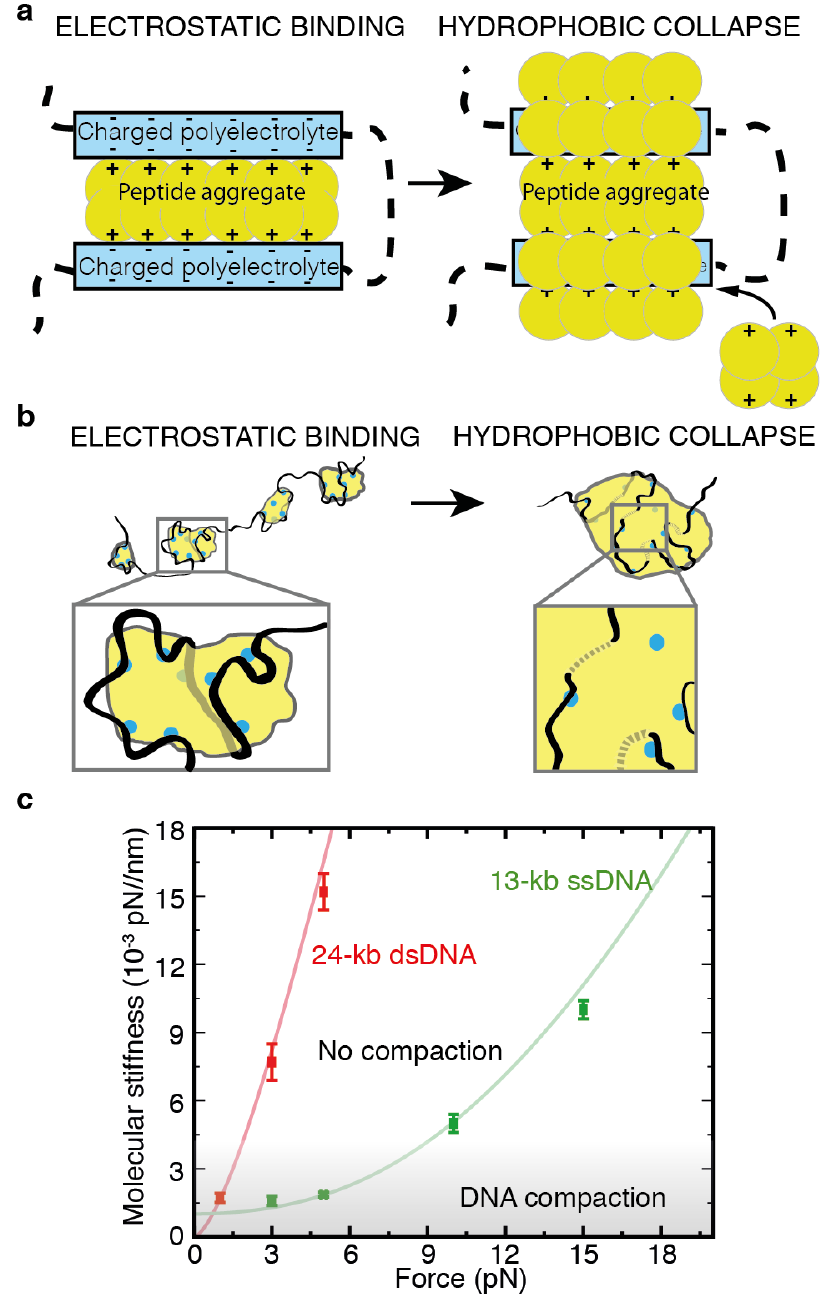}
\caption{{\bf Model of KF-DNA complex formation.} {\bf (a)} Scheme of the two kinetic steps during the formation of the KF-DNA complex: Electrostatic binding (EB) and Hydrophobic collapse (HC). At low forces DNA (blue) binds to KF particles (yellow) due to the electrostatic attraction of the phosphate backbone to positive residues exposed on the aggregate surface (EB). This process is followed by the formation of new hydrophobic contacts between aggregates that form a larger collapsed structure (HC) {\bf (b)} Pictorial representation of how bending fluctuations induce the EB of DNA (black) to electrophilic spots (blue) on the aggregate surface, and how hydrophobic interactions between peptides drive to a collapsed structure in which DNA becomes entrapped within the aggregated complex. {\bf (c)} Phase diagram showing conditions that trigger DNA compaction due to EB of KF in constant force experiments. Average stiffness of the dsDNA (red) and ssDNA (green) molecules before flowing KF in the constant force experiments ($2\le N\le9$, mean$\pm$SD). We only observe DNA compaction in the region highlighted in gray, suggesting that compaction depends directly on DNA bending fluctuations {\em via} template stiffness. A WLC model (red line) and FJC model (green line) with the parameters determined in the main text for dsDNA and ssDNA respectively are plotted as a reference.}
\label{fig:8}
\end{figure}

By combining single molecule techniques and bulk measurements we showed that KF forms particles that bind and compact DNA. Our measurements reveal that this process is characterized by two distinct phases controlled by the molecular extension of the DNA. First, there is a fast and weak binding regime determined by electrostatic binding (EB) to positive residues exposed on the surface of the KF particles (Figure \ref{fig:8}{\bf a}). This binding is triggered by spontaneous
bending fluctuations along DNA. Upon reduction of the molecular
extension a slow remodeling of the KF-DNA complex takes place; we propose that this new regime is led by
the formation of new hydrophobic contacts that stabilize a hydrophobically
collapsed (HC) structure. A capture distance separating both regimes is identified, corresponding to a relative extension of 66\%. Remarkably enough, theoretical studies between spherical charged aggregates and oppositely charged polymeric chains predicted a capture distance leading to the irreversible adsorption of the chain to the aggregates.\cite{podgornik1993stretching} The recruitment of DNA segments along the
surface of the aggregate (EB) is driven by the electrostatic attraction
between the negative charge of DNA and the positive charge of L-Orn
residues that are most likely exposed on the surface of the particle
forming hydrophilic spots (Figure \ref{fig:8}{\bf b}). This interpretation is supported by the following facts: (i) The persistence length of DNA is reduced during the initial binding of the peptide, suggesting a charge compensation that reduces self-repulsion of the DNA phosphate backbone (ii) A KF analog without a positive charge does not bind to DNA (Figure S2, Supporting Information), providing evidence that the positive charge is essential for binding (iii) A salt titration shows that binding of KF to DNA is inhibited at high salt condition, and that the strength of the interaction increases with decreasing ionic strength (Figure S3, Supporting Information) (iv) The zeta-potential value of KF, DNA, and KF-DNA complexes (Table S1, Supporting Information) indicate different surface charge densities depending on whether DNA is complexed with KF or not. In addition, the phenomenology observed during the EB regime cannot be understood from the isolated action of KF peptides (each bearing a single positive charge). It must be due instead to the concurrent action of several positive charges that are contained in each peptide aggregate, which are then able to electrostatically interact with DNA in a similar fashion that dendrimers or other polycationic agents do. Modeling of the
experimental results shows that EB of DNA to KF particles is consistent with an exponential distribution of unpeeling lengths (average value $\Delta l_{0}=31\pm6$ nm), a low binding energy ($\Delta G\sim6\pm2$ kcal/mol), a brittle unpeeling process ($x^{\dagger}_{i}= 2\pm$1 nm), and a disordered collection of activation barriers higher than $B_{0}=89 k_{B}T$ with an exponential right tail of width $w^{\prime}=5 k_{B}T$. This is in contrast to results obtained for nucleosomal particles that show a single characteristic unpeeling length of 26 nm and a single energy barrier of 36-38 $k_{B}T$.\cite{Brower-Toland:2002fk} The force at which we observe dsDNA compaction (1pN) is in the same range of forces that has been reported for other DNA compacting molecules such as histone-like FIS and HU proteins,\cite{skoko2005low, van2004dual} or polycationic condensing agents.\cite{todd2008interplay, Ritort:2006fk, Hormeno:2011} 

Unzipping experiments show that KF also forms complexes with ssDNA. Interestingly, unzipping experiments together with AFM show that KF does not inhibit dsDNA hybridization. These results agree with previous studies on the formation of amyloid nucleic acid fibers, which showed how charged surfaces of peptide complexes recruit oligonucleotides and promote their hybridization.\cite{braun2011amyloid} The two distinct phases observed in the formation of KF-dsDNA complexes (EB and HC) were also seen for ssDNA, showing that the mechanism of complexation is similar in both polyelectrolytes. Yet, our experiments suggest a stronger stabilization of the complex for ssDNA, which is likely due to its increased hydrophobicity and lower persistence length. Indeed, we have found that KF can compact ssDNA at higher forces than dsDNA does, indicating the role of spontaneous bending fluctuations to initiate EB. Finally, constant-force experiments with both DNA substrates show a correlation between the degree of compaction of the molecule and the effective rigidity of the tether (Figure \ref{fig:8}{\bf c}).

In relation to its biological activity, and as previously showed for amyloid fibrils,\cite{calamai2006nature, di2012binding, cohlberg2002heparin} the enhanced positive charge of KF aggregates could allow them to interact more effectively with other polyanionic molecules such as polysaccharides ({\em e.g.} glycosaminoglycans). Whether or not the cytotoxic effects of KF are related to its interaction with polyanions, we hypothesize that these interactions could play a role in modulating the activity of the peptide either intracellularly or in the extracellular matrix. In particular KF might also interact with the phospholipids of the plasma membrane inducing the formation of pores and cell necrosis.\cite{molina2011irvalec}

\section{Conclusion}
\label{sec:Conclusion}

This study represents the first attempt to extract quantitative information about the binding affinity and kinetic steps involved in the interaction between a nucleic acid (DNA) and an anticancer self-aggregating peptide (KF) at the single molecule level. To date most studies of aggregation kinetics have been performed using ensemble techniques where the individual behavior of molecules cannot be distinguished. Using optical tweezers, we have shown that KF binds DNA in two kinetic steps (an initial electrostatic binding that is followed by an hydrophobic collapse of the peptide-DNA complex), and characterized the spectrum of binding affinities, kinetics barriers and lengths of DNA segments sequestered within the KF-DNA complex. The proposed methodology is not limited to the characterization of amorphous aggregates.\cite{yoshimura2012distinguishing} Protein aggregation, a topic of major interest due to the role of the aggregation of misfolded proteins in neurodegenerative diseases,\cite{ross2005role} might be well addressed using single molecule force spectroscopy. As well, AFM images of the nucleoid-associated proteins Dps \cite{ceci2004dna, ceci2007neutrophil} and of the drug cisplatin \cite{hou2009cisplatin} are very similar to those we found here for KF-DNA, suggesting that many biochemical studies of protein complexes that face similar aggregation or compaction phenomena are susceptible of being studied with this approach. For instance, the aggregation of the splicing factor MBNL1 by mutant mRNA hairpins is at the core of Myotonic dystrophy type I.\cite{dansithong2008cytoplasmic, dickson2010repeat} Research on the formation of these RNA-protein aggregates, and of peptides that disrupt this interaction \cite{garcia2011vivo}  could greatly benefit from the aforementioned approaches. Our study confirms force spectroscopy studies of single aggregates as potentially very useful to characterize the thermodynamic and mechanical properties of nucleic acid-peptide complexes. On a longer term, the study of the mechanical response of aggregates related to many relevant neurodegenerative diseases could also be approached by using a peptide template instead of a DNA molecule.

\section{Materials and Methods}
\label{sec:Methods}

\subsection{Optical tweezers set-up}
\label{sec:instrument}

A miniaturized dual beam optical tweezers instrument described in \cite{huguet2010single} has been used for the single molecule experiments. A single optical trap is created by focusing two counter-propagating laser beams ($\lambda = 845$ nm, $P=200$ mW) into the center of a fluidics chamber mounted on a motorized stage. The optical trap can be displaced in a range of 12 $\mu$m by using a pair of piezoelectric actuators mechanically coupled to the laser optical fibers. For manipulation with the optical tweezers, DNA molecules are differentially end-labeled with digoxigenins and biotins so each end of the molecule can specifically bind to antidigoxigenin and streptavidin-coated beads, respectively (Supporting Information). A single DNA molecule is then tethered between two polystyrene beads. One is subjected on the tip of a micropipette, whereas the other one is confined in the optical trap (Figure \ref{fig:2}{\bf a}, Inset). The molecule can be stretched by moving the trap relatively to the micropipette, and both the extension and force applied to it are determined in real time. 

Force applied to the optically trapped bead is directly determined from the change in light momentum by measuring the deflection of the laser beams with position sensing detectors (PSD).\cite{smith2002optical} To measure the position of the optical trap, about $8\%$ of the lasers light is split before entering the objective lenses and redirected to PSD's. In this way, trap displacements can be followed with sub-nanometer resolution. The relative molecular extension is inferred by subtracting the trap compliance (F/k) to the absolute displacement of the optical trap (trap stiffness = 70 pN/$\mu$m). The extension and force applied to the molecule can be recorded at 1 kHz rate, and a resolution of 0.1 pN is achieved. 

\subsection{Stretching experiments of dsDNA and ssDNA}
\label{sec:stretchdna}
The persistence length ($l_p$) and stretch modulus ($S$) of each dsDNA molecule were determined before flowing KF with a fit to the Worm-Like Chain (WLC) model. Average values of $l_p=44.7\pm2.0$ nm and $S=1419\pm240$ pN were obtained (N=10), in good accordance with the generally accepted parameters.\cite{smith1996overstretching, wang1997stretching, baumann1997ionic} A compatible value of $l_p=43.0\pm2.0$ nm was also found using the inextensible WLC model. The stretching curves also showed the characteristic overstretching plateau at a force of $62.5\pm0.5$ pN with an extension $\sim$70$\%$ of the contour length.  Molecules that showed an abnormally high hysteresis on the overstretching transition (generally attributed to highly nicked DNA molecules)  were discarded. For the fits to the WLC model, the correction to the Marko-Siggia interpolation formula suggested in \cite{bouchiat1999estimating} was used, and a Levenberg-Marquadt algorithm was used for both minimizations.\cite{marquardt1963algorithm} Details of the experimental set-up and molecular synthesis are found in the Supporting Information.

To generate a 13-kb ssDNA molecule, a 6.8-kb DNA hairpin was fully unzipped in a buffer containing a 30-base oligonucleotide that binds to the loop and its flanking region due to base-pair complementarity. The high bending rigidity of the duplex at the loop region strongly stabilizes the ssDNA form over the dsDNA form at forces lower than the average unzipping force (14.5pN). Stretching curves of the ssDNA molecule were fitted to the Freely-Jointed Chain Model (Section S5, Supporting Information), finding a Kuhn length of $b=1.57\pm0.05$ nm (N=5) in agreement with previous results.\cite{smith1996overstretching, huguet2010single}

\subsection{Statistical analysis and simulations}
\label{sec:dataanalysis}

For every data point ($x_{\rm exp},f_{\rm exp}$) of a force-distance curve, we determined its most probable apparent contour length ($l_{0}$) by finding the theoretical WLC \cite{bouchiat1999estimating} that passes closest to that point at the  force $f_{\rm exp}$:
\begin{equation}
 \left| x_{\rm exp}-x_{WLC}\left( l_{0}, f_{\rm  exp} \right) \right|=\min_{l} \left(\left| x_{\rm  exp}-x_{WLC}\left( l, f_{\rm exp} \right) \right|\right).
\end{equation}

The theoretical extension ($x_{WLC}\left( l_{0}, f_{\rm exp} \right)$) was determined using the elastic parameters ($l_p$, $S$) obtained from a WLC fit to the relaxation curve.

To determine the mechanical work ($W=F\Delta x$) performed to disrupt each KF-DNA contact in DNA stretching experiments, we determined the average rupture force and the extension of DNA released at every unpeeling event. This work is partially used to stretch the released DNA up to the rupture force. The rest of work is dissipated into the solvent in the form of heat:
\begin{equation}
 W_{\rm dissipated}=\frac{F\Delta F}{k}-\Delta G_{\rm stretching}
\end{equation}
\begin{equation}
 \Delta G_{\rm stretching}=\frac{\Delta L}{L}\int_{0}^{x_{\rm rup}}{F_{WLC}\left(x\right)dx}
\end{equation}

where $x_{\rm rup}$ is the molecular extension of the DNA fiber at the rupture force. In the above expressions we use the fact that the force-extension curve of the WLC model is a sole function of $x/L$.

 The model considered to reproduce the KF-DNA stretching curves, simulates the contacts made between KF aggregates and DNA segments as a set of $N$ non-interacting two-level systems. When force is applied to the molecule, each segment can yield an extension $x_{i}$  in a thermally activated process characterized by a critical force $f_c$, and a dissociation rate $k_{c}$. Each segment is described by its free energy of formation $\Delta G_{i}$, activation barrier $B_i$, and distance to the transition state $x_{i}^{\dagger}$. The released extension $x_{i}$ was assumed to follow the experimental distribution (Figure \ref{fig:4}{\bf b}), and we introduced some structural disorder by assuming that $\Delta G$ and $B$ are also exponentially distributed. Simulation parameters that best describe the experimental curves are: $p\left(\Delta G\right)=\left(1/w\right) e^{-\left(\Delta G-\Delta G_{0}\right)/w}$ with $\Delta G\ge \Delta G_{0}=$10 $k_{B}T$ and $w=$1 $k_{B}T$; $x^{\dagger}_{i}= 2\pm$1 nm; $f_c=5\pm3$ pN; $k_{c}$=0.5 s$^{-1}$; $p\left(B\right)=\left(1/w^{\prime}\right)e^{-\left(B-B_{0}\right)/w^{\prime}}$ with $B\ge B_{0}=$89 $k_{B}T$ and $w^{\prime}=5$ $k_{B}T$. The contact-length distribution ($x_i$) was assumed to follow the experimental distribution: $p\left(x_{i}\right)=\left(1/w\right) e^{-\left(x_{i}-x_{i,0}\right)/w}$ for $x_{i}\ge x_{i,0}$ ($p\left(x_{i}\right)$=0 otherwise) with $x_{i,0}$=8 nm and $w$=24 nm. Other specific parameters for the simulations shown in Figure \ref{fig:4}{\bf d}: $k_{\rm trap}$=0.07 pN/nm, v=500 nm/s, $l_{p}$=35 nm, DNA slack=6500 nm. Errors are an estimation of the range in which the features of the process are well reproduced by the model when each parameter is independently modified.

\subsection{Sample flow set-up}
\label{sec:flow}

A syringe pump (PicoPump, KDScientific) and a glass syringe have been used for KF sample infusion. Polyethylene PE-10 tubing (BD Intramedic) is used to connect the syringe to the fluidics chamber. For the experiments performed with a constant force protocol a buffer flow-rate of 3 $\mu$l/min is used to keep a low drag force on the bead. For the other experiments a higher flow-rate of 9 $\mu$l/min was preferred (Supporting Information). The arrival of the KF solution into the experimental area can be monitored due to the slight change in the refractive index of the medium caused by the $2\%$ DMSO content of the peptide buffer.

We were not able to establish DNA tethers between polystyrene beads with KF in the buffer, suggesting that strong condensation effects appear on untethered molecules. Therefore, experiments were always performed by flowing KF to DNA molecules that had been tethered in peptide-free buffer. Flow experiments performed with polystyrene beads without DNA do not show aggregation of KF onto the bead surfaces, nor an increase in stickiness between beads, ruling out aggregation on bead surfaces.
 
 \subsection{AFM sample preparation and imaging}
KF-DNA reactions included 1.65 ng of DNA (0.9 nM dsDNA molecules, pGEM3Z, 2743 bp) (Promega) linearized with {\it Bam}HI and 100 $\mu$M of KF in 20 mM Tris-HCl (pH 7.5) and 100 mM NaCl. To facilitate adsorption of DNA in a buffer devoid of Mg$^{2+}$ ions, we pre-treated the mica surface with 100mM spermine tetrahydrochloride (S85610, Fluka, Sigma) dissolved in 10 mM Tris-HCl (pH 7.5). Pre-treatment with spermine consisted in deposition of 20 $\mu$l of 100 mM spermine tetrahydrochloride on a freshly cleaved mica surface, one minute adsorption, washing with Milli-Q water, and drying with nitrogen gas. Use of spermine at low concentrations allowed uniform adsorption of DNA molecules. Immediately after the spermine treatment, the mixture of KF and DNA incubated at room temperature and for the stated time was deposited on the mica. After 30 s, the mica surface was washed with filtered-MilliQ water and blown dry in a gentle stream of nitrogen gas. To study KF-ssDNA interactions, a stock of ssDNA was produced by heat denaturing the linearized pGEM plasmid at 95$^{\circ}$C for 5 minutes and placing the tube quickly after on ice. $\sim$3 kb DNA molecules remain stable in its ssDNA form following this procedure as long as they remain at 4$^{\circ}$C for at least one week. KF and ssDNA were mixed at the same proportions as for dsDNA and followed the sample preparation procedure described above. Samples were imaged in air at room temperature at identical conditions as previously described.\cite{Fuentes:2012} Standard image processing consisted of plane subtraction and flattening using WSxM freeware.\cite{Horcas:2007}

\acknowledgement
\label{sec:Acknow}

The authors thank Pharmamar SA and G. Acosta for providing the peptides Kahalalide F and its analog. JCS acknowledges a grant associated to ICREA Academia 2008; FR is supported by the Human Frontier Science Program [Grant No. RGP55-2008], the Spanish Ministry of Economy and Competitiveness [Grant No. FIS2010-19342] and an ICREA Academia award;
FMH acknowledges a grant from the European Research Council [Starting Grant number 206117], and the Spanish Ministry of Economy and Competitiveness [grant number FIS2011-24638]; MEFP was supported by a contract from CSIC [contract number 200920I123, associated to the ERC grant]. FA acknowledges CICYT [grant number: CTQ2012-30930], and SBD a invited professor position from AGAUR (Generalitat de Catalunya).

\section{Supporting Information Available}
Additional methods, optical tweezers results, DLS measurements and AFM images. This material is available free of charge {\em via} the Internet at \texttt{http://pubs.acs.org/.}

\providecommand*\mcitethebibliography{\thebibliography}
\csname @ifundefined\endcsname{endmcitethebibliography}
  {\let\endmcitethebibliography\endthebibliography}{}

\begin{tocentry}
\includegraphics{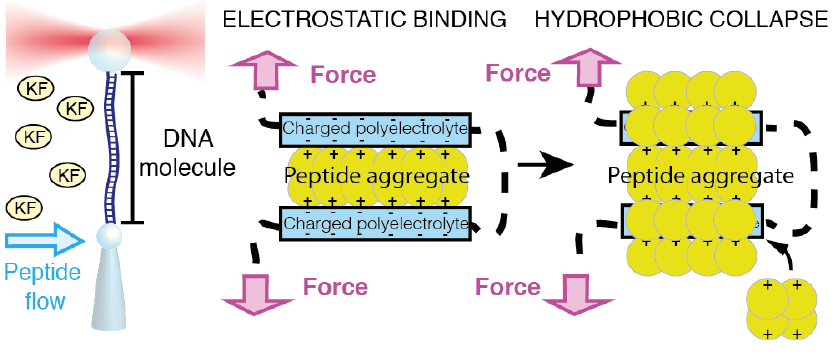}
\end{tocentry}

\pagebreak

\begin{center}
\textbf{\large Supporting Information: Electrostatic Binding and Hydrophobic Collapse of Peptide-Nucleic Acid Aggregates Quantified Using Force Spectroscopy}
\end{center}
\setcounter{equation}{0}
\setcounter{figure}{0}
\setcounter{table}{0}
\setcounter{page}{1}
\makeatletter

\renewcommand{\thesection}{S\arabic{section}}
\renewcommand{\thesubsection}{\thesection.\arabic{subsection}}
 
%
\makeatletter 
\def\tagform@#1{\maketag@@@{(S\ignorespaces#1\unskip\@@italiccorr)}}
\makeatother
 
\makeatletter
\makeatletter \renewcommand{\fnum@figure}
{\figurename~S\thefigure}
\makeatother

\renewcommand{\thetable}{\arabic{table}}
\makeatletter
\makeatletter \renewcommand{\fnum@table}
{\tablename~S\thetable}
\makeatother
 
\renewcommand{\bibnumfmt}[1]{[S#1]}
\renewcommand{\citenumfont}[1]{\textit{S#1}}


\section{S1 Supplementary Methods}
\label{sec:submeth}

\subsection{S1.1 Optical tweezers experiments}
\label{sec:exp set-up}

\subsubsection{S1.1.1 Stretching experiments with dsDNA}
\label{sec:dsDNA}

For the dsDNA stretching experiments,  1 $\mu$l of a $\sim0.5$ pmol/ml dilution of a 24-kb dsDNA stock solution (see Section S1.2) was mixed and incubated for 30' with 5 $\mu$l antidigoxigenin-coated beads (0.5\% w/v, 3.15 $\mu$m diameter) and 14 $\mu$l TE (Tris 10 mM, EDTA 1 mM, 0.01\% $NaN_3$, pH7.5) 500 mM NaCl. The sample was then diluted to 1 ml in TE 100 mM NaCl, 0.1 mg/ml BSA (New England Biolabs). The sample was incubated at least 30 min with BSA before starting the experiments to passivate the bead surface. A sample of streptavidin-coated beads was also prepared by diluting 1 $\mu$l of beads (0.5\% w/v, 1.87 $\mu$m diameter, Kisker Biotech) in 1 ml TE 100 mM NaCl, 0.1 mg/ml BSA. Beads were differentially flowed into the fluidics chamber through lateral channels, and a dsDNA tether between an optically trapped bead and a pipette-subjected bead was created as previously described. 

The molecule was submitted to several stretch/relaxation cycles up to a maximum pulling force of 45 pN and a minimum molecular extension of $\sim 4.5$ $\mu$m to characterize its elastic properties. At least three consecutive pulling cycles were fitted to the Inextensible Worm-Like Chain (WLC) in the low-force regime (F$\le$5 pN), and to the extensible WLC in the high-force regime (F$\le$40 pN). 

After flowing KF, the low-force region (F$\le$5 pN) of the release cycles was still well described by the inextensible WLC model. The contour length and persistence length of the molecule were then obtained as an average of at least four pulling cycles. Higher force-data was not used for the fits, as data is not well described by the extensible WLC model after flowing KF. 

For the experiments in which KF was flowed without keeping the molecule at a constant force, the DNA was maintained at an end-to-end distance of 6 $\mu$m (flow-rate: 9 $\mu$l/min). This corresponds to a drag force on the trapped bead of $\sim$14 pN perpendicularly to the stretching direction. In these experiments the flow was stopped after 5, 15 and 30 min and stretching curves were collected at a pulling speed of 500 nm/s. 

For the constant force experiments a lower flow-rate of 3 $\mu$l/min was used to maintain the overall tension on the DNA molecule similar to the force on the stretching direction. This buffer flow corresponds to a drag force of $\sim$4 pN perpendicular to the stretching direction. To accurately perform the constant force experiments at 1 pN, we flowed KF keeping the DNA molecule at a higher tension of 3 pN. Once the chamber was fully equilibrated the flow was stopped and the force set at 1 pN.

\subsubsection{S1.1.2 DNA unzipping experiments}
\label{sec:unzip}

For the DNA unzipping experiments a 6.8-kb DNA hairpin sample (see Section S1.2) was tethered by bringing the bead in the optical trap in close contact to the bead on the micropipette due to the short length of the hairpin handles. The DNA and beads incubations were prepared following the same protocol as for the DNA stretching experiments, the only difference being the specific DNA concentration required to optimize the tethering of a single DNA molecule between beads. The DNA hairpin was then fully unzipped in TE 100 mM NaCl at a pulling rate of 50 nm/s, and maintained at a position in which at least half of the hairpin stem remained unzipped. The molecule was then rinsed with KF for at least 3 min (flow-rate: 9 $\mu$l/min). The peptide flow was then stopped and unzipping/rezipping curves recorded.

\subsubsection{S1.1.3 Experiments with ssDNA}
\label{sec:ssDNA}

For the ssDNA experiments a 6.8-kb DNA hairpin was prepared and tethered as for the unzipping experiments. The tethered hairpin was fully unzipped in a buffer (TE 100 mM NaCl) that contains a 250 nM concentration of a 30-base oligonucleotide complementary to the loop and its flanking region (see Table S2 for oligonucleotide sequence, oligo name: Blockloop30). The annealing of an oligonucleotide to the tethered hairpin stabilizes the ssDNA form over the dsDNA form down to forces close to 1 pN.  The elastic response of the ssDNA molecules was then recorded at a pulling rate of 100 nm/s in the force range 5 pN $<$ F $<$ 20 pN and characterized by means of the FJC model (see Section S5). Then, the molecule was maintained at a constant force by means of the force-feedback while a freshly prepared 40 $\mu$M KF solution was flowed (flow-rate: 3 $\mu$l/min).

\subsection{S1.2 DNA substrates preparation}
\label{sec:dnasynth}

For the DNA stretching experiments we prepared a 24508-bp DNA molecule with biotin and digoxigenin tags at the 3'-ends of the molecule. The DNA template was prepared by cleaving N6-methyladenine free $\lambda$-DNA (New England Biolabs) with {\it Xba}I restriction enzyme, and purified using Wizard DNA clean-up system kit (Promega). The digoxigenin tag was prepared by annealing  to the cosL end of $\lambda$-DNA an oligonucleotide tailed at its 3'-end with digoxigenin-labeled dUTP's using terminal transferase (Roche). The biotin tag was prepared by annealing two complementary oligonucleotides designed to create an {\it Xba}I cohesive end at one side. One of the oligonucleotides was tailed at its 3'-end with multiple biotins with biotin-labeled dUTP's using terminal transferase (Roche). Oligonucleotides were purified after tailing steps using the QIAquick Nucleotide Removal Kit (Qiagen). The digoxigenin tag was annealed to the half $\lambda$-DNA molecule by incubation for 10 min at $68^{\circ}C$ in a 10-fold excess oligonucleotide. Then, the biotin tag was annealed by incubation for 1h at $42^{\circ}C$ (using 20-fold excess oligonucleotides) followed by cooling down to room temperature. Ligation was performed as an overnight reaction at $16^{\circ}C$ using T4 DNA ligase (New England Biolabs).

For the DNA unzipping experiments a 6838-bp DNA hairpin with a tetraloop at one end and two 29-bp dsDNA handles at the other end was prepared. The synthesis is based on a previously described method\cite{huguet2010single} but was modified to introduce double digoxigenin and biotin tags at each handle to enhance tether lifetimes (Figure S8). Briefly, N6-methyladenine free $\lambda$-DNA (New England Biolabs) was digested with {\it Bam}HI, phosphorylated at its 5'-ends with T4 polynucleotide kinase (New England Biolabs), and purified using Wizard SV Gel and PCR clean-up system (Promega). The 6770-bp restriction fragment contained between positions 41733 and 48502 (cosR end) was used as the stem of the DNA hairpin. To create the end-loop of the molecule, the stem was annealed to an oligonucleotide (BamHI-loop2) that self-assembles in a hairpin structure with a tetraloop at one end, and an overhang complementary to the {\it Bam}HI restriction site on the other end. To create the dsDNA handles,  we used two partially complementary oligonucleotides (cosRlong and Bio-cosRshort3) that hybridize forming a protruding end complementary to the cosR end. The Bio-cosRshort3 oligonucleotide was purchased 5'-biotinylated, and the cosRlong oligonucleotide was tailed with multiple digoxigenins at its 3'-end as previously explained. To create the doubly biotynilated dsDNA handle, a third oligonucleotide (splint3) complementary to the unpaired regions of BIO-cosRshort3 was purchased and tailed at its 3' end with multiple biotins with biotin-labeled dUTP's using terminal transferase (Roche). To create the digoxigenin dsDNA handle we used a modified oligonucleotide (inverted-splint, Thermo Scientific) complementary to the unpaired region of cosRlong. This oligonucleotide contains two modifications: a C3 spacer at its 3' end to block this end in tailing reactions, and a polarity inversion at its 5' end using a 5'-5' linkage. This end can therefore be tailed with digoxigenin-labeled dUTP's using terminal transferase (Roche). In this way, both ends of the handle could be tailed with multiple digoxigenins (Figure S8). Oligonucleotides were purified after tailing steps using the Qiaquick Nucleotide Removal Kit (Qiagen). The oligonucleotides were annealed to the 6770-bp stem by incubation for 10 min at $70^{\circ}C$, followed by incubation for 10 min at $55^{\circ}C$ and cooling down to room temperature. Ligation was performed as an overnight reaction at $16^{\circ}C$ using T4 DNA ligase (New England Biolabs).

The sequence of the oligonucleotides used for the hairpin synthesis, together with the 30-base oligonucleotide used to generate the ssDNA template are specified in Table S2.

\subsection{S1.3 Kahalalide F sample preparation}
\label{sec:sampleprep}

Stock solutions containing 2 mM KF were prepared by dissolving a $\sim$1-2 mg sample of pure lyophylized Kahalalide F (gently provided by Pharmamar) in 100\% DMSO (Sigma, Molecular Biology Grade). Stock solutions were then vortexed at low speed for 20 min and filtered with a 0.2 $\mu$m pore PTFE membrane filter (Millipore) that was pre-rinsed with 300 $\mu l$ DMSO to reduce filter extractables. Sample concentration after stock filtration was verified by means of HPLC: An aliquot before and after filtering was collected and stored in ACN:H20 (50:50). These samples were sequentially processeÄd in an HPLC using a 5\%-100\% acetonitrile gradient. The elution peak was monitored at 220 nm, and the integrated area was compared between both samples. No sample loss happened during filtration (Figure S9). Ready-to-use Kahalalide F stock solutions were stored at -20$^\circ$C as 20-60 $\mu$l aliquots. To prepare diluted working solutions, a stock solution aliquot was thawed and sequentially diluted in TE 100 mM NaCl ({\em e.g.} 40 $\mu$M). The final DMSO concentration was always adjusted to 2\%. Once thawed, stock solutions were not stored again to avoid the problems related to freeze-thaw cycles.\cite{borchardt2005pharmaceutical, kozikowski2003effect} Samples were always prepared in glass vials, and a glass syringe used for sample filtering. Use of glass material was preferred to avoid material leaching from disposable plasticware,\cite{mcdonald2008bioactive} and the good performance of glass infusion devices for low concentration samples of KF in clinical practice.\cite{nuijen2001compatibility}

A mass spectrum of a 40 $\mu$M KF sample in 100 mM NH4OAc buffer pH7.0 (2\% DMSO) is shown in Figure S10. The concentration of NH4OAc was set to obtain an ionic strength comparable to that used in the optical tweezers experiments.

\subsubsection{S1.3.1 Mass Spectrometry Methods}
\label{sec:MSmethod}
Sample are introduced using an Automated Nanoelectrospray. Triversa NanoMate (Advion BioSciences, Ithaca, NY, USA) sequentially aspirated the sample from a 384-well plate with disposable, conductive pipette tips, and infused it through the nanoESI Chip, which consists of 400 nozzles in a 20x20 array. Spray voltage was 1.75 kV and delivery pressure was 0.5 psi. Mass Spectrometer:  Synapt HDMS (Waters, Manchester, UK). Samples were acquired with Masslynx software v.4 SCN 639 (Waters). 
{\bf MS Conditions for TOF results:} NanoESI. Positive mode TOF. V mode. Sampling cone: 20 V. Source temperature: 20$^\circ$C. Trap Collision Energy: 10. Transfer Collision Energy: 10. Trap Gas Flow: 8 ml/min. Vacuum Backing pressure: 5.89 mbar. m/z range: 300 to 5000, 500-15000. Instrument calibrated with CsI (external calibration).
{\bf MS Conditions for Ion Mobility results:} NanoESI. Positive mode Ion Mobility mode. V mode. Sampling cone: 20 V. Source temperature: 20$^\circ$C. Trap Collision Energy: 10. Transfer Collision Energy: 10. Trap Gas Flow: 8 ml/min. IMS Wave Velocity: 300 m/s. IMS Wave Height: 9.5 V. Transfer Wave Velocity: 200 m/s. Transfer Wave Height: 8 V. Vacuum Backing pressure: 5.89 mbar. m/z range: 300 to 5000, 500-15000. Instrument calibrated with CsI (external calibration).

 \subsection{S1.4 Dynamic Light Scattering Methods}
 A Photon Correlation Spectrometer (PCS) 3D from LS INSTRUMENTS was used for DLS measurements. The instrument is equipped with a He-Ne laser (632.8 nm). Measurements of at least 90 s were recorded at an angle of $90^{\circ}$. The hydrodynamic radius was calculated by a manual exponential fitting of the first cumulant parameter. Standard deviations were calculated from the second cumulant. The measurement temperature of $25^{\circ}C$ was maintained by a decaline bath, which matches the refractive index of glass and does not therefore interfere with the measurement. The evolution of the hydrodynamic radius was observed during 1 hour. The time indicated corresponds to the minutes past after preparation of the sample and the beginning of the measurement. 

Zeta potential measurements were carried out at $25^{\circ}C$ with a Malvern Instrument Zetasizer Nano Z by laser Doppler electrophoresis. Disposable polystyrene cells were used. Solutions of peptide, DNA and the mixture were measured at the same concentration as the light scattering measurements were performed. Commercially available $\lambda$-DNA (New England Biolabs) was used for the experiments.
 
\section{S2 DLS measurements of KF and KF-DNA complexes}
\label{sec:Aggregation}

To characterize the size of KF aggregates we performed DLS measurements. The hydrodynamic radius of KF particles one minute after sample preparation was 170$\pm$20 nm (Figure S11{\bf a}, red) and showed a high polydispersity. The hydrodynamic radius increased linearly with time with a growing rate of 3.2$\pm$0.6 nm/min, indicative of a rate of aggregation proportional to the surface of the aggregate. We attribute such growth to the hydrophobic interactions between peptides. However, the hydrodynamic radius of KF-DNA mixtures remained constant within experimental errors during 60 min after sample preparation (Figure S11{\bf a}, black). This may be explained by the stabilizing effect induced by the added DNA. In fact, the average size of the particles could be stabilized by adding DNA at a latter time-stage (Figure S11{\bf b}, arrow).  DNA as a strongly charged polyelectrolyte interacts with the positive charge of KF and might form an anionic, water-soluble shell around the peptide. We measured the zeta potential of KF particles finding a low positive value (Table S1) in agreement with the single positively charged residue of the peptide. A higher negative value was obtained for KF-DNA mixtures. KF oligomerization was also observed with size-exclusion HPLC (Figure S12).

\section{S3 Simulation of KF-DNA stretching experiments}
\label{sec:simul}

As described in the main text, KF-DNA stretching curves are simulated using a set of N non-interacting two-states systems (Figure 4{\bf d}, Inset) that describe the contacts made between KF and DNA. Each contact can be in two conformations: formed or dissociated. The formed and dissociated states of contact $i^{th}$ have extensions corresponding to 0 and $x_i$ respectively. Initially the N contacts are found in the formed conformation. As the DNA is stretched and the force increased, the tilting of the free-energy landscape towards larger extensions favors the dissociated conformation, releasing an extension $x_i$.

The parameters that best reproduce the experimental force-extension curves are described in the main text. A set of figures in which a different parameter is modified in each panel (the others being kept at the optimal value) is presented in Figure S13{\bf a-d}. In each panel, the optimal simulation is presented in black, whereas simulations with varying values of the modified parameters are presented in colors.

For $\Delta G_0 \le$ 5 $k_{B}T$ most of the contacts remain in the dissociated conformation ($60\%$ for $\Delta G_0=$ 5 $k_{B}T$) at the end of a simulated pulling cycle (minimum extension of $\sim$ 4.5 $\mu$m). This high fraction of dissociated contacts cannot explain the sawtooth pattern observed in the force-extension curve (FEC) in subsequent pulls. This suggests that a value of $\Delta G_0 \le$ 5 $k_{B}T$  is too low to reproduce the experimental curves. On the other hand, for $\Delta G_0\sim$ 10 $k_{B}T$ more than 90$\%$ of the contacts are formed again at the end of a pulling cycle (minimum extension of $\sim$ 4.5 $\mu$m), in agreement with the experimental results both at low (50 nm/s) and high pulling speeds (500 nm/s). Finally, for free energies greater than 15 $k_{B}T$ a sawtooth pattern is also observed in the relaxation curve, in disagreement with the releasing part of the observed experimental FEC for the same range of pulling speeds (Figure S13{\bf a}).

Experimental force-extension curves show a broad distribution of rupture forces with most rupture events occurring at forces lower than 20 pN, but with a large rightmost tail (Figure S5{\bf a}). This phenomenology can be well reproduced by considering: (i) that the transition state is close to the formed conformation (Figure S13{\bf b}); and (ii) that the process is characterized by a disordered ensemble of barriers rather than a single valued barrier. This structural disorder is introduced in the form of an exponential distribution of barriers with a rightmost exponential tail of width $w^{\prime}\ge$ 1 $k_{B}T$ (Figure S13{\bf c}).

To model the experimental force-extension curves, we have assumed that the distribution of released lengths ($x_i$) follows the experimental distribution shown in Figure 4{\bf b}. In Figure S13{\bf d} a simulation using the experimental distribution is compared to simulations in which the system is characterized by a single contact length ($x_i$) rather than an exponential distribution. Finally the optimal distributions for the different parameters of the simulation are shown in Figure S13{\bf e}.

\section{S4 Measurement of tether stiffness from trap-distance fluctuations}
\label{sec:ssDNAstiff}

A bead confined in an optical trap fluctuates around its equilibrium position due to thermal fluctuations. By the equipartition theorem the fluctuations of the bead position along the stretching direction are directly related to the effective stiffness of the system formed by the tethered molecule and the optical trap:\cite{forns2011improving}

\begin{equation}
<\delta y^2>=<y^2>-<y>^2=\frac{k_B T}{k_{\rm trap} + k_{\rm mol}}~~~.
\end{equation}

In an ideal force-feedback the stiffness of the trap vanishes and the bead fluctuations are only determined by the stiffness of the tether:

\begin{equation}
<\delta y^2>=\frac{k_B T}{k_{\rm mol}}~~~.
\end{equation}

In this case the fluctuations of the position of the optical trap are expected to match those of the bead ($<\delta y^2>$). However, in our experimental set-up a finite frequency feedback of 1 kHz is used ({\em i.e.} the force-feedback corrects the position of the optical trap by moving the piezoelectric actuators at a 1 kHz rate), and Eq. (S2)  is not satisfied. To extract the value of $k_{\rm mol}$ we followed a phenomenological approach that uses a modified version of Eq. (S2) containing a proportionality constant {\it c}:

\begin{equation}
<\delta y^2>=c \frac{k_B T}{k_{\rm mol}}~~~.
\end{equation}

The constant {\it c} includes all effects of the finite frequency of the feedback. The value of {\it c} has been obtained by fitting a set of measurements of ssDNA and dsDNA tethers at different average forces. The fluctuations in the position of the optical trap remain inversely proportional to $k_{\rm mol}$ in the investigated range of stiffness ($1-10\cdot10^{-3}$ pN/nm) (Figure S14{\bf a}), verifying the validity of the method based on Eq. (S3). This calibration method has been used to estimate the changes in stiffness that KF induces in ssDNA. The stiffness of the molecule during the peptide flow is measured in short time-windows to ensure reliable measurements of $<\delta y^2>$.

Data is recorded at 1 kHz and low frequencies are filtered out to remove instrumental drift as experiments run for long times (30 min). Filtering low frequencies is also important to correct the changes in extension due to the compaction of the DNA with KF. A time-window of 3 s has been found to be the optimal value to reduce drift without affecting the measurements (vertical dashed line in Figure S14{\bf b}). Using a shorter time-window removes fluctuations that are relevant to determine the stiffness of the tether due to the large autocorrelation time shown by the data (Figure S14{\bf c}).

\section{S5 Elastic properties of ssDNA before incubation with KF}
\label{sec:ssDNAchar}

Before flowing KF, pulling curves of the ssDNA molecule were recorded. The relative molecular extension was determined by subtracting the trap compliance (F/k) to the absolute displacement of the optical trap (trap stiffness: 70 pN/$\mu$m). To obtain the absolute molecular extension, the elastic response of the tethered molecule was aligned to a reference unzipping trajectory in the range of forces from 15 to 20 pN (ssDNA response of the molecule). This method allowed to accurately calibrate the absolute extension of the molecule within less than 50 nm. 

The method used to prepare the ssDNA template leaves a very short dsDNA stretch (88-bp) in relation to the length of the ssDNA chain (13650 bases). As the rigidity of dsDNA is $\sim$70-fold that of ssDNA, pulling curves mainly reflect the elastic response of the ssDNA region. At a given salt concentration, pulling curves of ssDNA can be modeled with the Freely-Jointed Chain (FJC) model:\cite{huguet2010single}

\begin{equation}
x(F)=d_0N_b \left[\frac{1}{\tanh 
(\frac{Fb}{k_bT})}-\frac{k_bT}{Fb} \right] 
\end{equation}
where: $b$: Kuhn length , $d_0$: Interphosphate distance, $N_b$: Number of bases.
\newline

Pulling curves were fitted to the FJC model finding average values of $b=1.57\pm$0.05 nm and $d_0$=$0.57\pm0.03$ nm (N=5), compatible with previously reported results for the same ionic condition.\cite{smith1996overstretching, huguet2010single} Data was fitted on the force-range 5-25 pN, as experiments were performed at a salt concentration ([NaCl]=100 mM) at which secondary structure formation is not observed at low forces.
\newpage

\section{S6 Supplementary Figures}
\begin{figure} [ht]
\addtolength{\abovecaptionskip}{-10pt}
\begin{center}
\includegraphics[width=16cm]{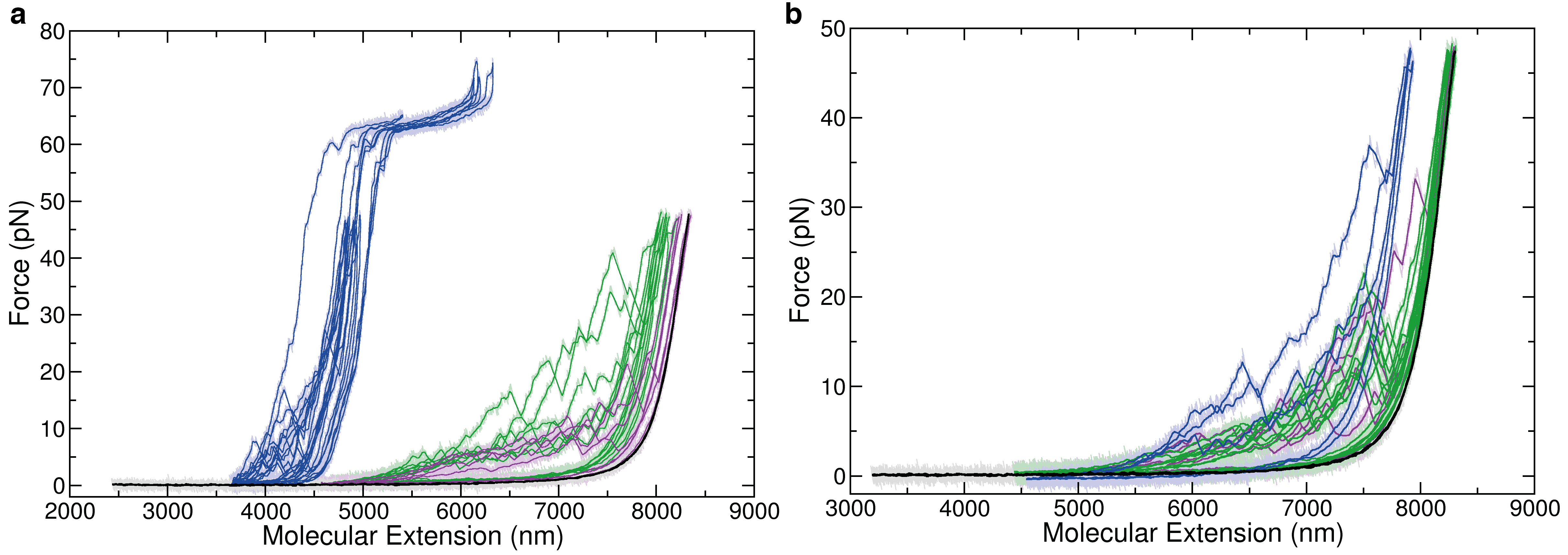}
\end{center}
\caption{{\bf Reproducibility of KF-DNA force-extension curves.} ({\bf a}) Pulling cycles of a 24-kb DNA molecule before (black) and after flowing 40 $\mu$M KF at different waiting times: 5 min (purple), 15 min (green), 30 min (blue). ({\bf b}) Pulling cycles of a 24-kb DNA molecule before (black) and after incubation with 40 $\mu$M KF at different waiting times of the interaction: 5 min (purple), 15 min (green), 30 min (blue). Similar trends to those reported in Figure 2{\bf a} are seen. Raw data was obtained at 1 kHz acquisition rate (light colors) and filtered to 10 Hz bandwidth (dark colors). Pulling speed is 500 nm/s.}
\label{fig:supp1}
\end{figure}

\begin{figure} [h!]
\addtolength{\abovecaptionskip}{-10pt}
\begin{center}
\includegraphics[width=9cm]{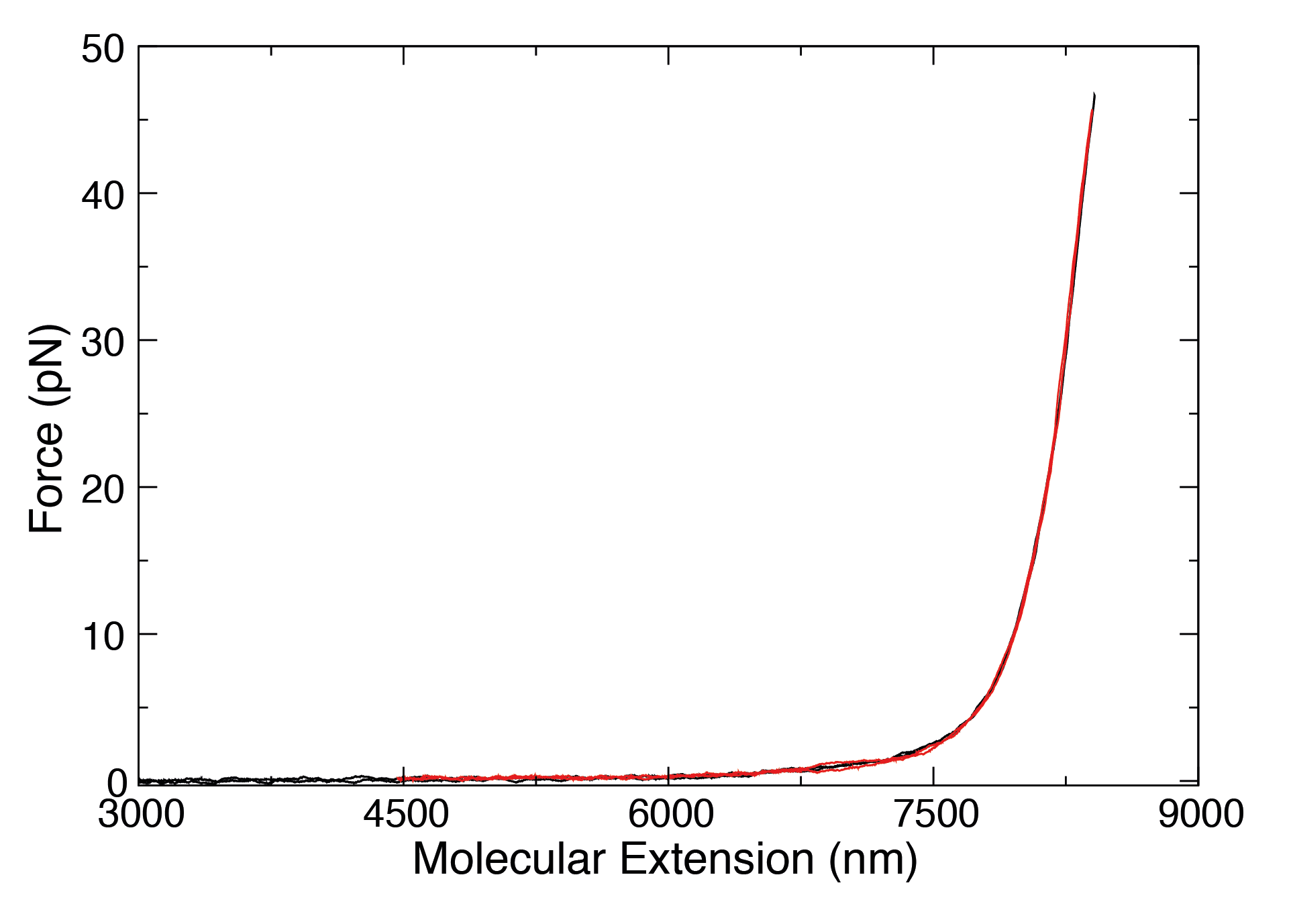}
\end{center}
\caption{{\bf The ornithine residue is essential for DNA binding.} Stretching curves of a 24-kb DNA molecule before (black) and after (red) flowing a KF analog (40 $\mu$M) in which the ornithine residue has been replaced by a glutamic acid. The characteristic sawtooth pattern induced by KF is not observed, and compatible values for the elastic parameters are found if force-extension curves are fitted to the WLC model before and after flowing the analog. Data is filtered at 10Hz bandwidth.}
\label{fig:supp2}
\end{figure}

\begin{figure} [h]
\addtolength{\abovecaptionskip}{-10pt}
\begin{center}
\includegraphics[width=14cm]{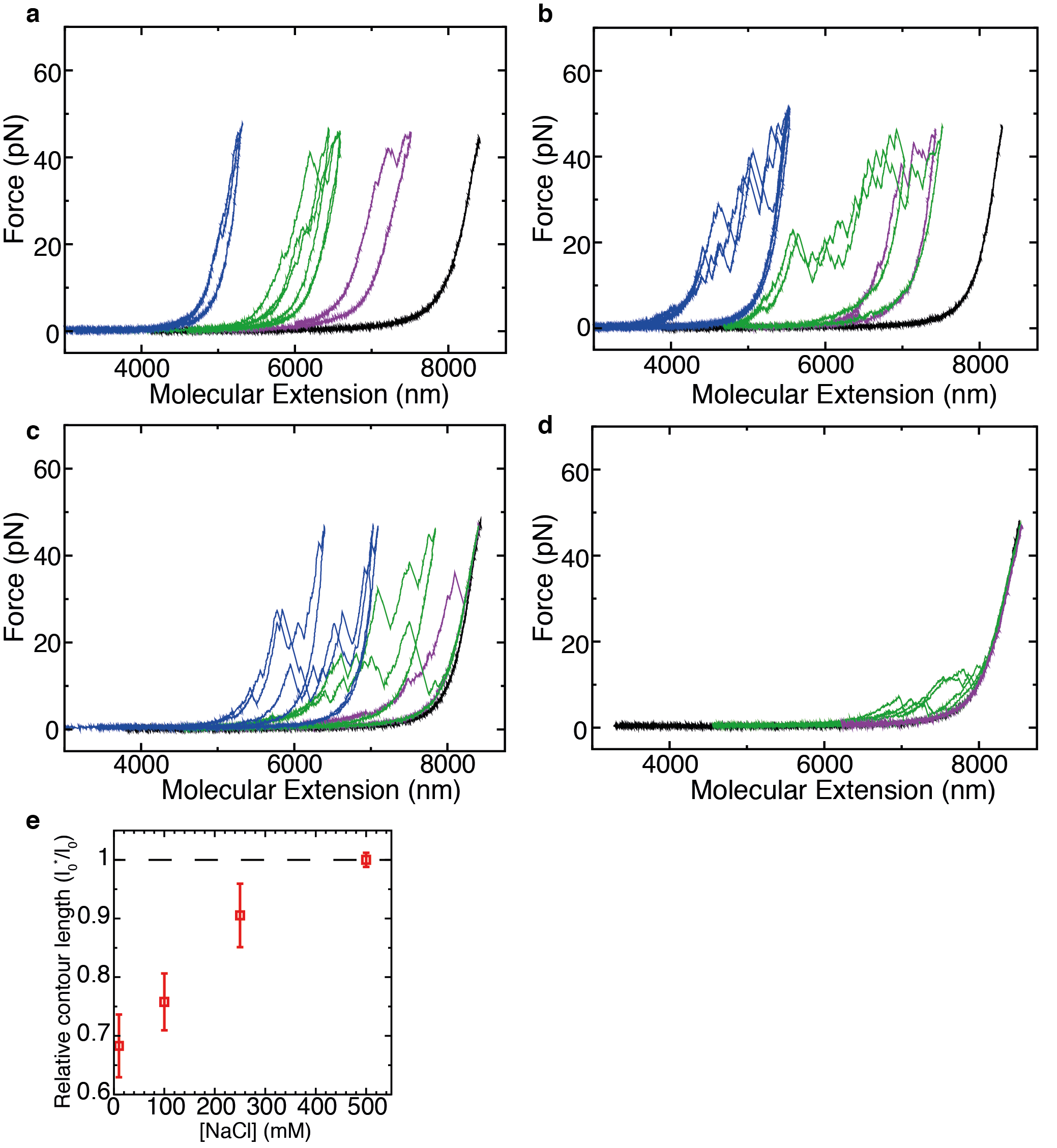}
\end{center}
\caption{{\bf Effect of ionic strength on KF-DNA interaction. (a) 10 mM NaCl. (b) 100 mM NaCl. (c) 250 mM NaCl. (d) 500 mM NaCl.} In panels (a-d) we show representative experiments at each salt condition in which a DNA molecule is pulled before (black) and after flowing 10 $\mu$M KF (color). After flowing the peptide, the DNA molecule was repeatedly pulled between a maximum force of 45 pN and a minimum extension of 6 $\mu$m (purple), 4 $\mu$m (green) and 3 $\mu$m (blue). At the highest ionic strength (500 mM NaCl) we did not observe binding of the peptide to DNA, except for one experiment in which we obtained the results shown in panel d (green curve). Data is filtered at 50 Hz bandwidth, pulling speed v=500 nm/s. We performed at least 5 experiments at each condition. (e) Apparent contour length of the DNA molecule (${l_0}^{*}$) relative to its original extension ($l_0$=8.3 $\mu$m), after being repeatedly pulled between a maximum force of 45 pN and a minimum extension of 4 $\mu$m. The degree of compaction of the DNA molecule increases with decreasing ionic strength (mean$\pm$SD, N=5, except for 500 mM in which we only observed binding in one experiment).}
\label{fig:supp3}
\end{figure}

\begin{figure} [h]
\addtolength{\abovecaptionskip}{-10pt}
\begin{center}
\includegraphics[width=11cm]{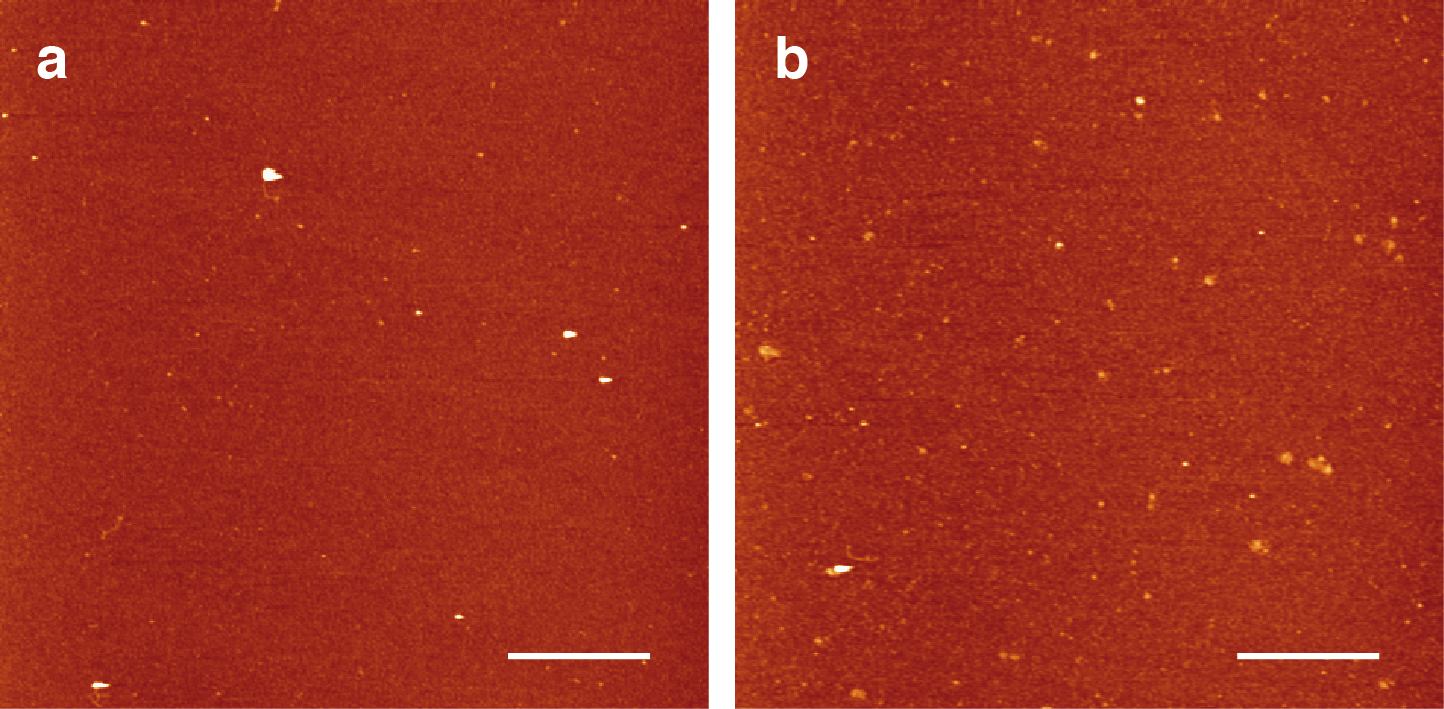}
\end{center}
\caption{{\bf AFM imaging of KF in the absence of DNA.} ({\bf a}) 100 $\mu$M KF immediately after dilution in aqueous buffer. ({\bf b}) 100 $\mu$M KF incubated for 30 min at room temperature. Aggregation spots are occasionally observed on the surface at both incubation times. Bar scale is 600 nm. Color scale (from dark to bright) is 0-2 nm in all AFM images.}
\label{fig:supp3}
\end{figure}

\begin{figure} [h]
\addtolength{\abovecaptionskip}{-10pt}
\begin{center}
\includegraphics[width=16cm]{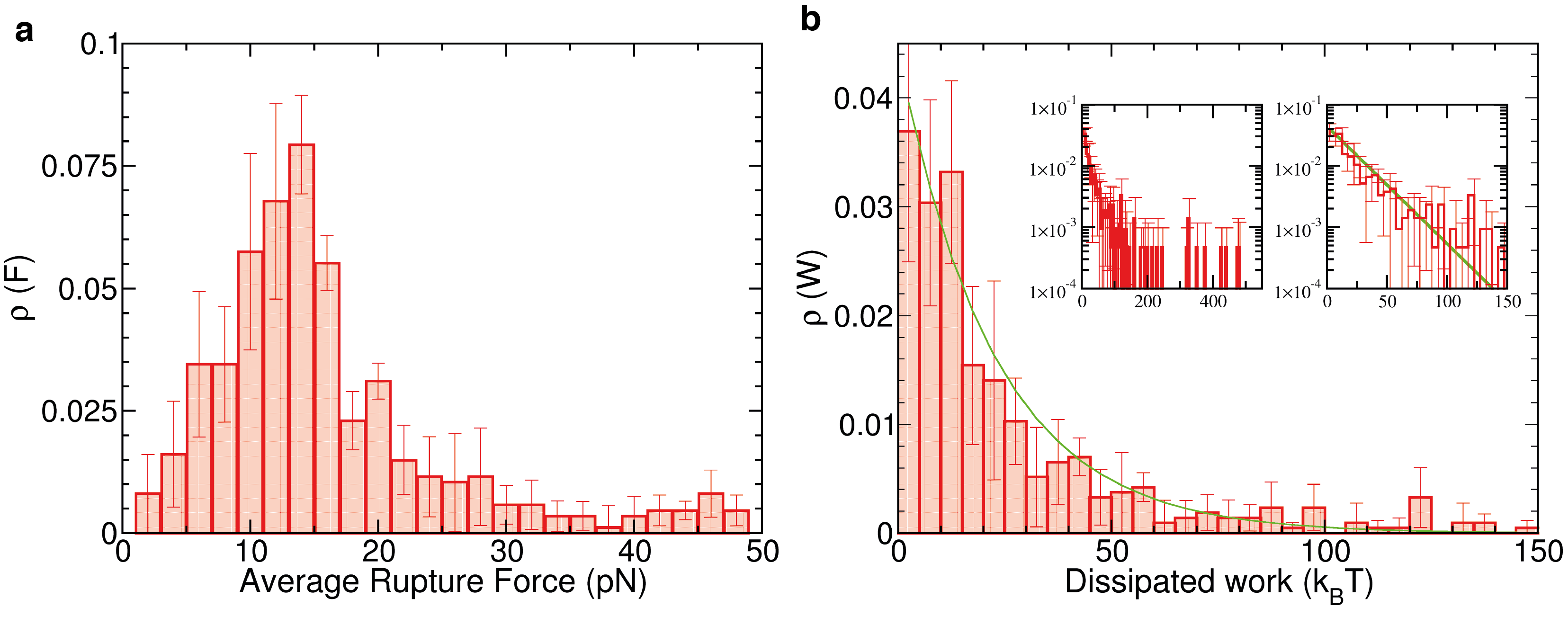}
\end{center}
\caption{{\bf Analysis of KF-DNA unpeeling events} ({\bf a}) Average rupture force of KF-DNA contacts. The average rupture force is calculated as the mean of the average force immediately before and after an unpeeling event ({\bf b}) Histogram of dissipated work in individual unpeeling events (calculated using equation (2) in main text). The histogram follows an exponential distribution of mean $23\pm8$ $k_{B}T$ (green). A rightmost tail corresponding to individual unpeeling events with $W_{\rm dissipated}\ge$ 150 $k_{B}T$ is observed. Insets show a log-normal plot (right) and the same plot with an enlarged range of $W_{\rm dissipated}$ values (left). Unpeeling events with dissipated work as large as 400 $k_{B}T$ are observed. For both figures N=435 events from 3 molecules. Error bars are the statistical error measured between different molecules.}
\label{fig:supp4}
\end{figure}

\begin{figure} [h]
\addtolength{\abovecaptionskip}{-10pt}
\begin{center}
\includegraphics[width=11cm]{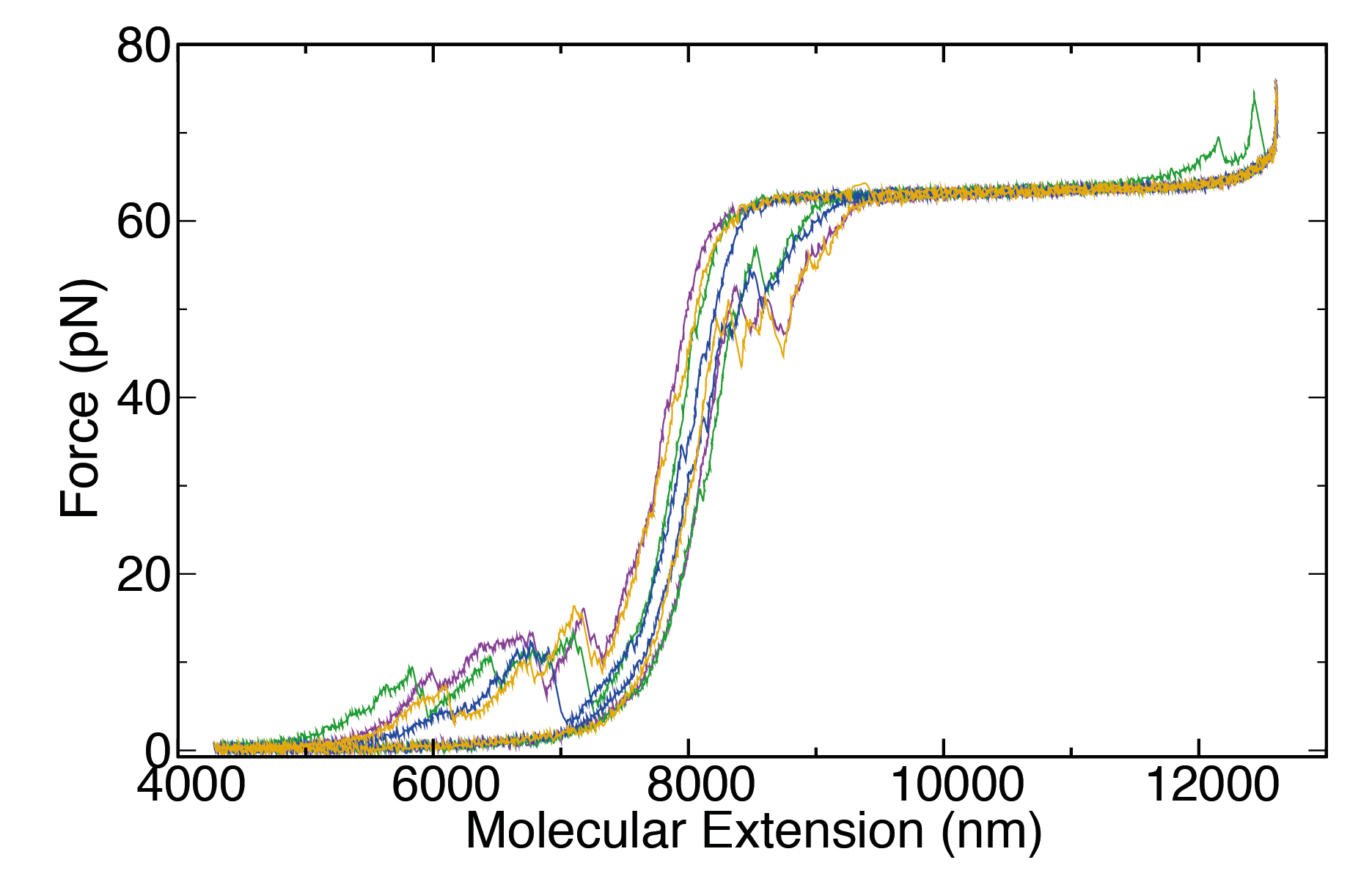}
\end{center}
\caption{{\bf KF binding does not change the overstretching transition.} Four consecutive stretching curves (purple, green, blue and yellow) of a 24-kb DNA molecule after incubation with 50 $\mu$M KF. The molecule is fully overstretched at each pulling cycle. The sawtooth pattern at low extensions is clearly visible at each cycle. Data was filtered at 100 Hz bandwidth, v=1000 nm/s.}
\label{fig:supp5}
\end{figure}

\begin{figure} [h]
\addtolength{\abovecaptionskip}{-10pt}
\begin{center}
\includegraphics[width=14cm]{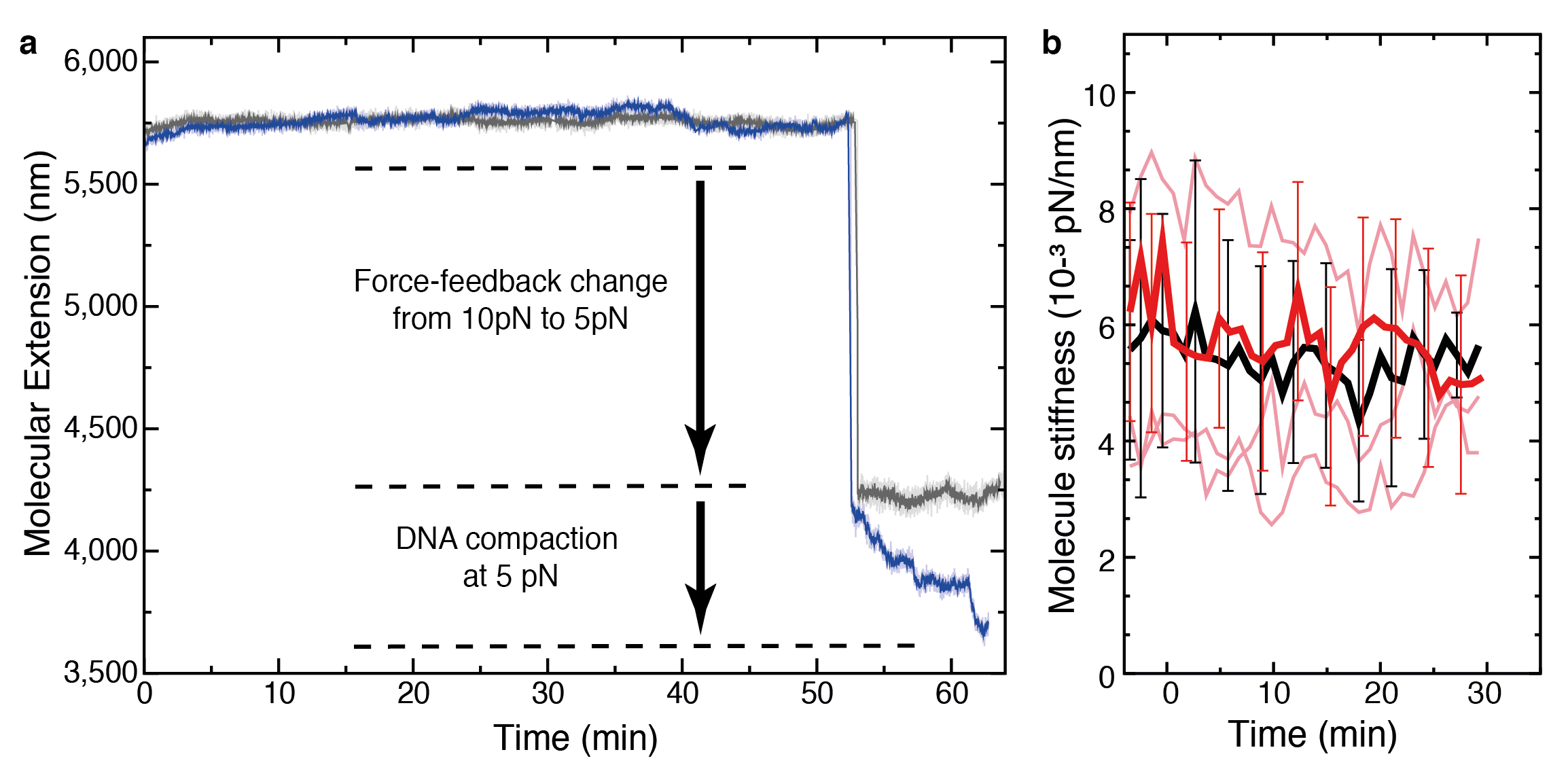}
\end{center}
\caption{{\bf KF compacts ssDNA at 5 pN but not at 10 pN} ({\bf a}) A ssDNA molecule is maintained for more than 50 min at a constant force of 10 pN with a flow of KF, without observing a significant decrease in extension (blue). However, when the force is lowered to 5 pN (top arrow) a compaction equivalent to that reported in Figure 7{\bf b} is seen (bottom arrow). A negative control  without peptide in the flowed buffer does not show DNA compaction (gray). Raw data is obtained at 1 kHz (light colors) and filtered at 1 Hz bandwidth (dark colors).  ({\bf b})  Average stiffness of ssDNA at 10 pN during the first 30 min of the peptide flow (red) compared to a negative control without peptide (black). Three individual experiments are shown in light red. In contrast to the results shown in Figure 7{\bf b}, the molecular stiffness remains constant in time.}
\label{fig:supp6}
\end{figure}

\begin{figure} [h]
\addtolength{\abovecaptionskip}{-10pt}
\begin{center}
\includegraphics[width=9.5cm]{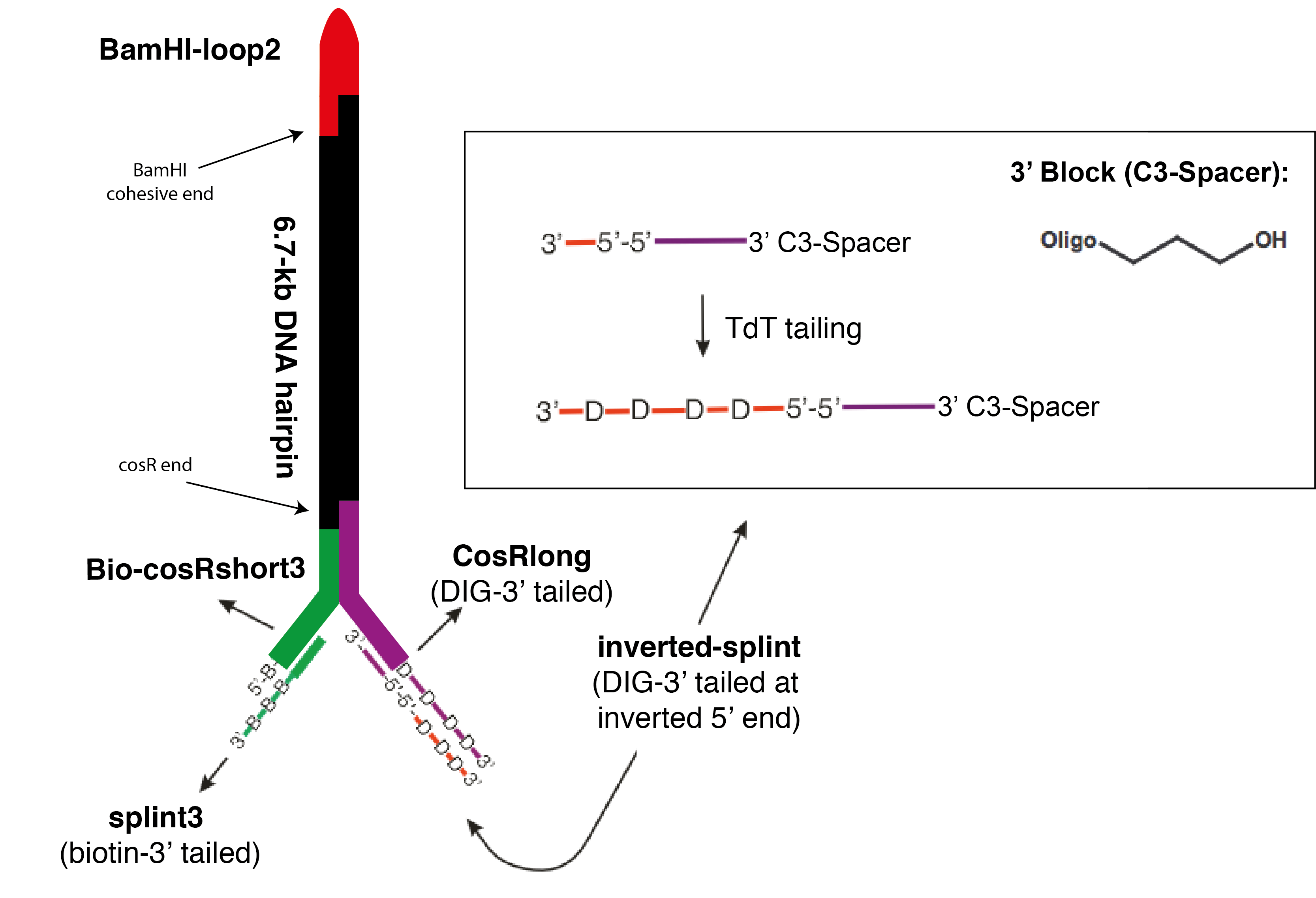}
\end{center}
\caption{{\bf Scheme of the DNA hairpin synthesis.} The DNA hairpin is created by ligating a set of oligonucleotides (red, green, purple) to a 6.7-kb restriction fragment of $\lambda$-DNA (black). The cosR end, and an {\it Xba}I cohesive end were respectively used to anneal the dsDNA handles and the end-loop to the $\lambda$-DNA fragment. To create the biotinylated handle (green) one oligonucleotide was purchased 5'-biotinlylated (Bio-cosRshort3) and the other one was tailed with multiple biotins at its 3' end (splint3). To create a dsDNA handle with multiple digoxigenins at each strand we used an oligonucleotide containing a 5'-5' inversion and a blocked 3' end (inverted-splint). In this way the two oligonucleotides that create this handle (cosRlong, inverted-splint) could be  tailed with multiple digoxigenins at the appropriate end.}
\label{fig:supp7}
\end{figure}

\begin{figure} [h]
\addtolength{\abovecaptionskip}{-10pt}
\begin{center}
\includegraphics[width=16.5cm]{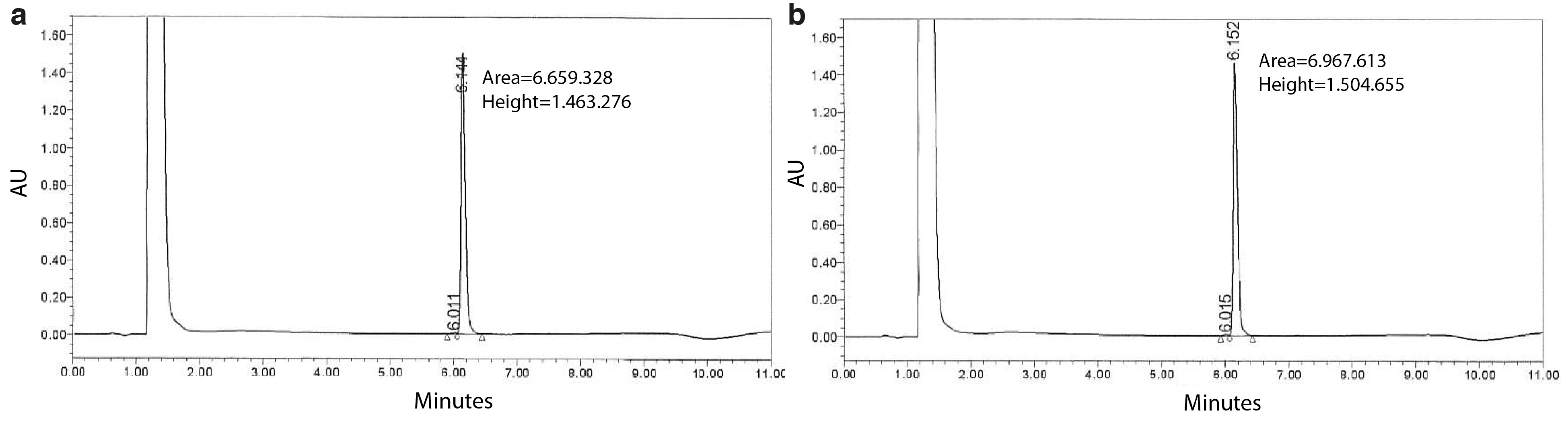}
\end{center}
\caption{{\bf KF stock filtration does not reduce sample concentration} ({\bf a}) Chromatogram of a freshly dissolved 2 mM KF stock solution before filtration (wavelength: 220 nm). The elution peak of KF is seen at t=6.4 min. ({\bf b}) Chromatogram of a freshly dissolved 2 mM KF stock solution after filtration. The elution peak of KF is seen at t=6.5 min. The area and height of the filtered and non-filtered samples are comparable, indicating that KF concentration is not reduced due to filtration. The high peak at t=1 min corresponds to DMSO.}
\label{fig:supp8}
\end{figure}

\begin{figure} [h]
\addtolength{\abovecaptionskip}{-10pt}
\begin{center}
\includegraphics[width=9cm]{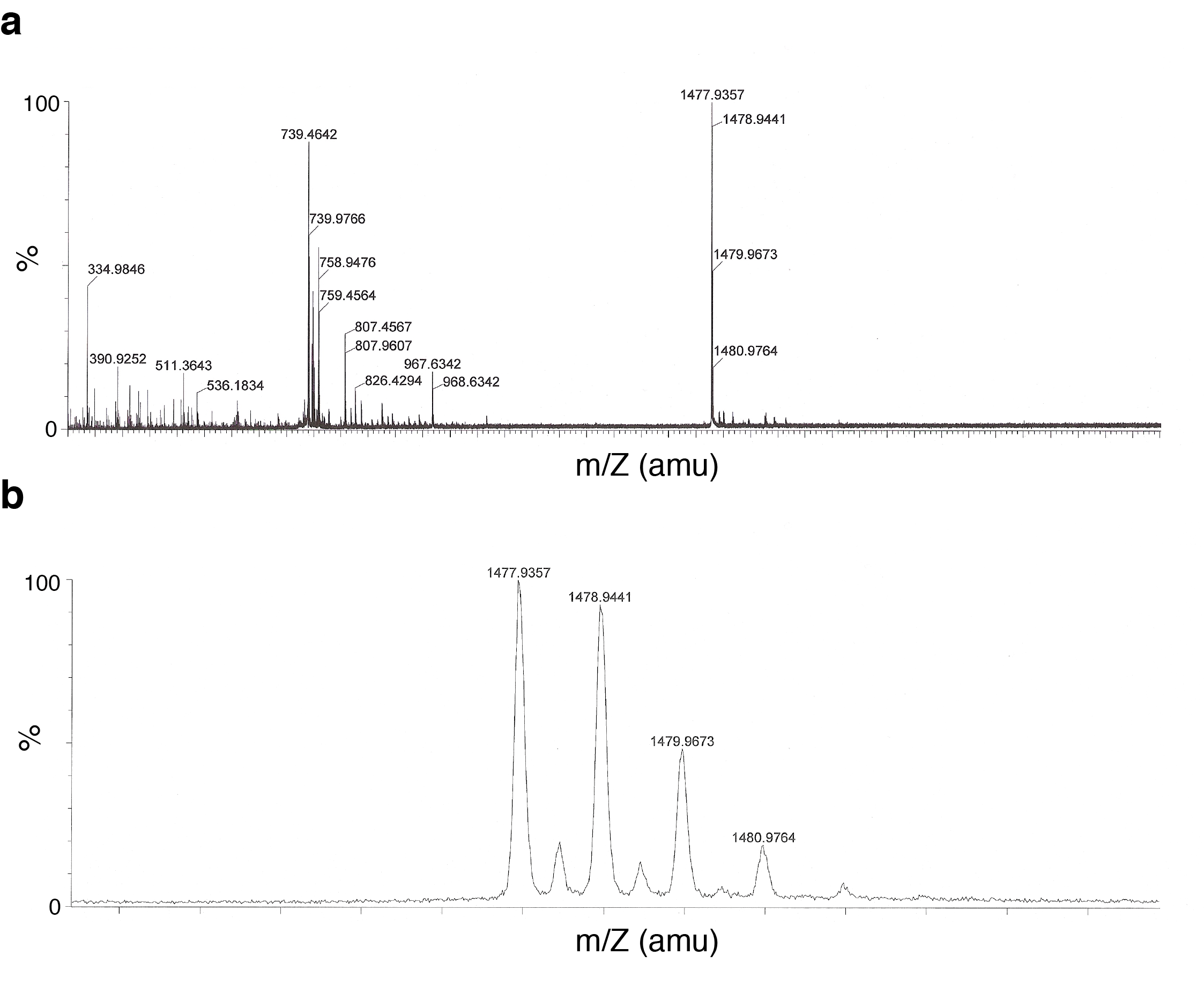}
\end{center}
\caption{{\bf Mass spectrometry of Kahalalide F} ({\bf a}) Mass spectrum of a 40 $\mu$M KF sample. Peaks found at 1478 m/Z correspond to the singly charged ion and doubly charged dimer. The peaks at 967 m/Z and 511 m/Z correspond to a fragmentation reaction of the peptide at D-val/D-Pro.\cite{stokvis2002quantitative} ({\bf b}) Zoom of the main peak of the mass spectrum from (a).}
\label{fig:supp9}
\end{figure}

\begin{figure} [h]
\addtolength{\abovecaptionskip}{-10pt}
\begin{center}
\includegraphics[width=10cm]{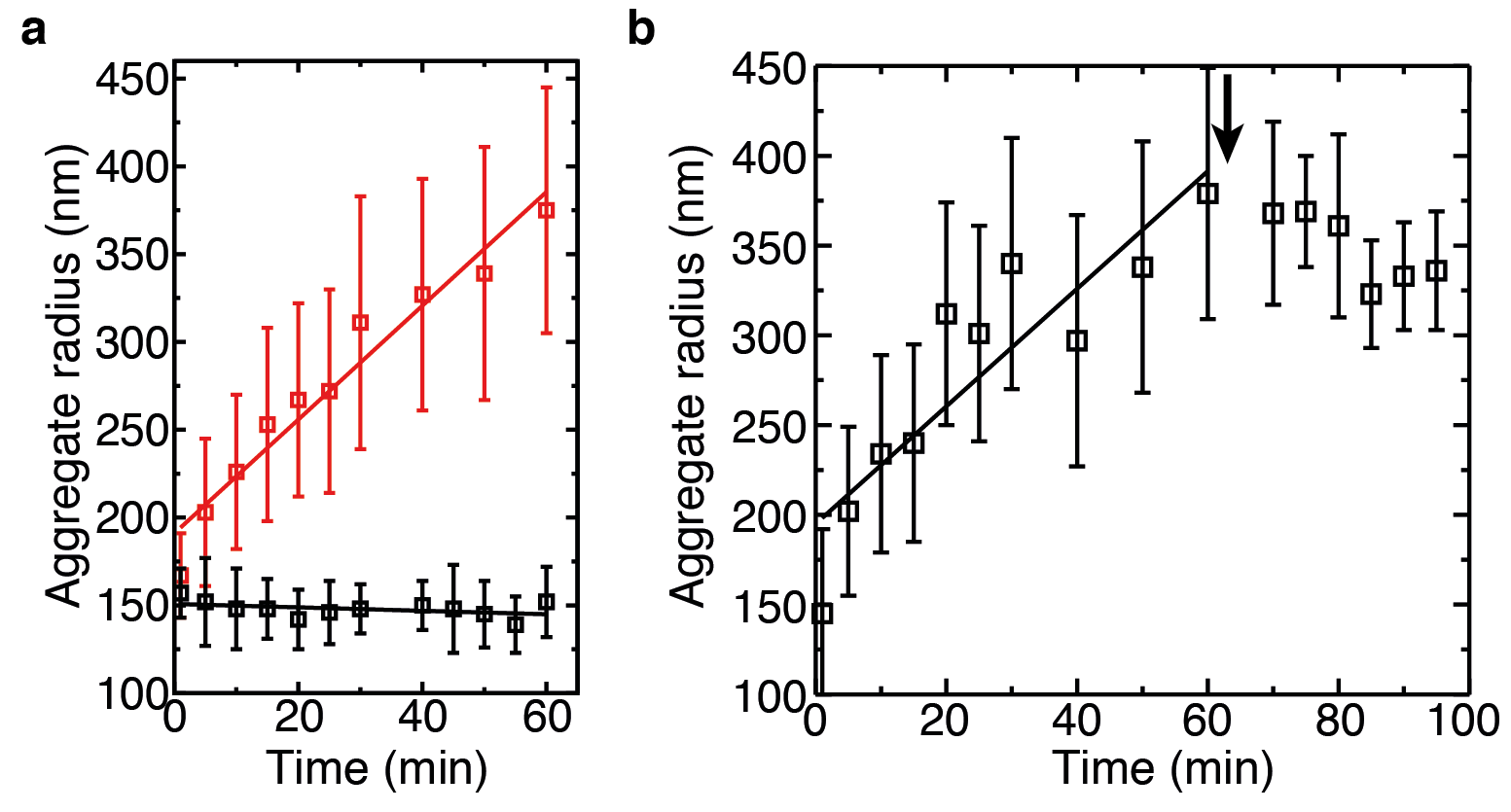}
\end{center}
\caption{{\bf DLS measurements of KF and KF-DNA complexes.} {\bf(a)} {\bf (red)} Hydrodynamic radius of KF particles in the buffer used for optical tweezers experiments (40 $\mu$M KF, $25^{\circ}C$, mean$\pm$SD, N=9). KF forms nanometer-sized aggregates whose size grows linearly with time. The aggregation rate is obtained from a linear fit (red line). Significant differences are not seen between KF stock aliquots stored at $-20^{\circ}C$ or solutions freshly prepared from lyophilized KF. {\bf(black)} Hydrodynamic radius of KF aggregates when DNA is added to the sample immediately after dilution (KF 40 $\mu$M, 48-kb $\lambda$-DNA 6.25 $\mu$g/ml, $25^{\circ}C$, mean$\pm$SD, N=3). The size of the KF aggregates remains constant up to 1 hour after dilution, showing that DNA has a stabilizing effect on the size of the aggregates. A linear fit is shown in black. {\bf(b)} The size of the KF aggregates (black) can also be stabilized by adding DNA (arrow) after 60 min (mean$\pm$SD, N$\ge$3). A linear fit is shown to highlight the stabilization of the particle size after adding DNA.}
\label{fig:supp10}
\end{figure}

\begin{figure} [h]
\addtolength{\abovecaptionskip}{-10pt}
\begin{center}
\includegraphics[width=8cm]{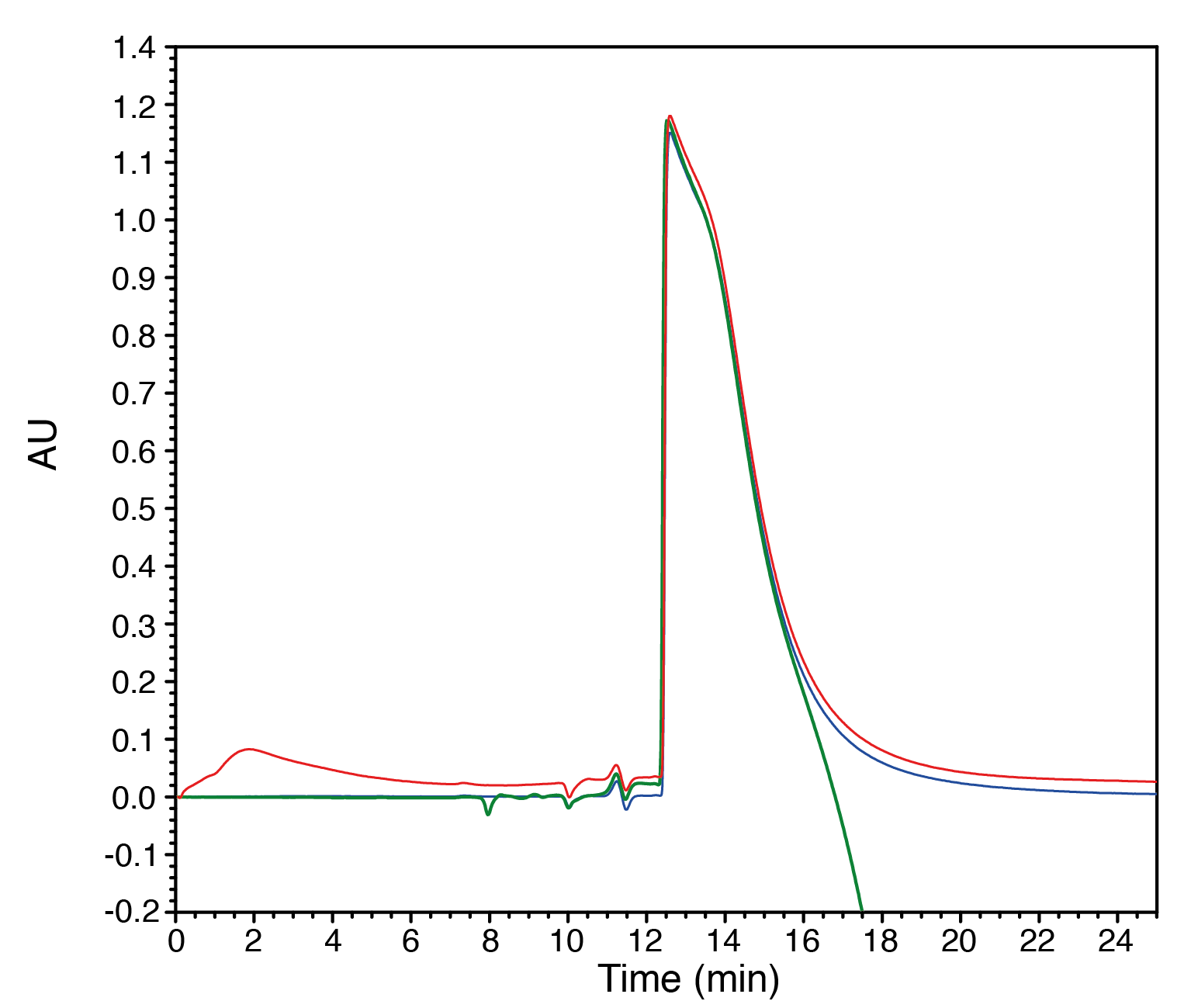}
\end{center}
\caption{{\bf Aggregation of KF dilutions.} Size-exclusion chromatography of 40$\mu$M KF dilutions at different waiting times: immediately after dilution (blue), after 30 min (green) and after 3 h (red). The width of the elution peak indicates peptide aggregation at all times.}
\label{fig:supp11}
\end{figure}

\begin{figure} [h]
\addtolength{\abovecaptionskip}{-10pt}
\begin{center}
\includegraphics[width=13.2cm]{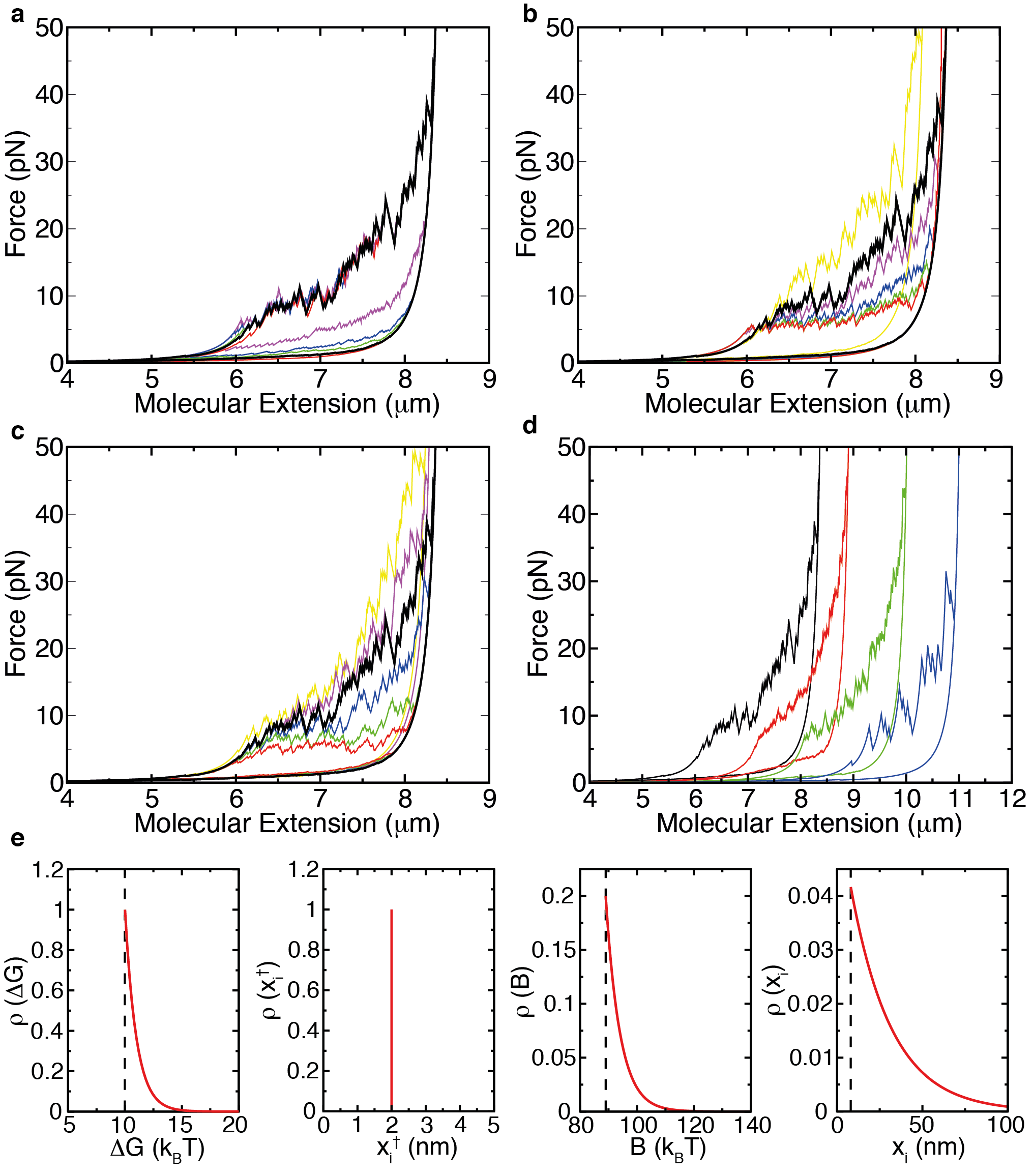}
\end{center}
\caption{{\bf Simulations of KF-DNA stretching experiments.} ({\bf a}) Simulations varying $\Delta G_0$ (minimum value of the exponential distribution of $\Delta G$ values, width $w=$1 $k_{B}T)$: $\Delta G_0$=5 $k_BT$ (red), $\Delta G_0$=10 $k_BT$ (black, optimal value), $\Delta G_0$=15 $k_BT$ (green), $\Delta G_0$=20 $k_BT$ (blue), $\Delta G_0$=40 $k_BT$ (purple). ({\bf b}) Simulations varying $x_i^{\dagger}$: $x_i^{\dagger}$=10 nm (red), $x_i^{\dagger}$=8 nm (green), $x_i^{\dagger}$=5 nm (blue), $x_i^{\dagger}$=3 nm (purple), $x_i^{\dagger}$=2 nm (black, optimal value), $x_i^{\dagger}$=1 nm (yellow). ({\bf c}) Simulations varying $w^{\prime}$ (exponential tail of the barrier B): $w^{\prime}$=0 $k_BT$ (red), $w^{\prime}$=1 $k_BT$ (green), $w^{\prime}$=3 $k_BT$ (blue), $w^{\prime}$=5 $k_BT$ (black, optimal value), $w^{\prime}$=7 $k_BT$ (purple), $w^{\prime}$=10 $k_BT$ (yellow) ({\bf d}) Simulations varying $x_{i}$ (contact-length distribution): experimental distribution $p\left(x_{i}\right)=\left(1/w\right) \exp{\left[-\left(x_{i}-x_{i,0}\right)/w\right]}$ for $x_{i}\ge x_{i,0}$ ($p\left(x_{i}\right)$=0 otherwise) with $x_{i,0}$=8 nm and $w$=24 nm (black), $x_{i}$= 10 nm (red), $x_{i}$=30 nm (green), $x_{i}$=100 nm (blue). Simulations are shifted by 1 $\mu$m for clarity. ({\bf e}) Optimal distributions for the different parameters of the simulation (corresponding to the black lines in panels a-d).}
\label{fig:supp12}
\end{figure}
\newpage

\begin{figure} [p]
\addtolength{\abovecaptionskip}{-10pt}
\begin{center}
\includegraphics[width=16cm]{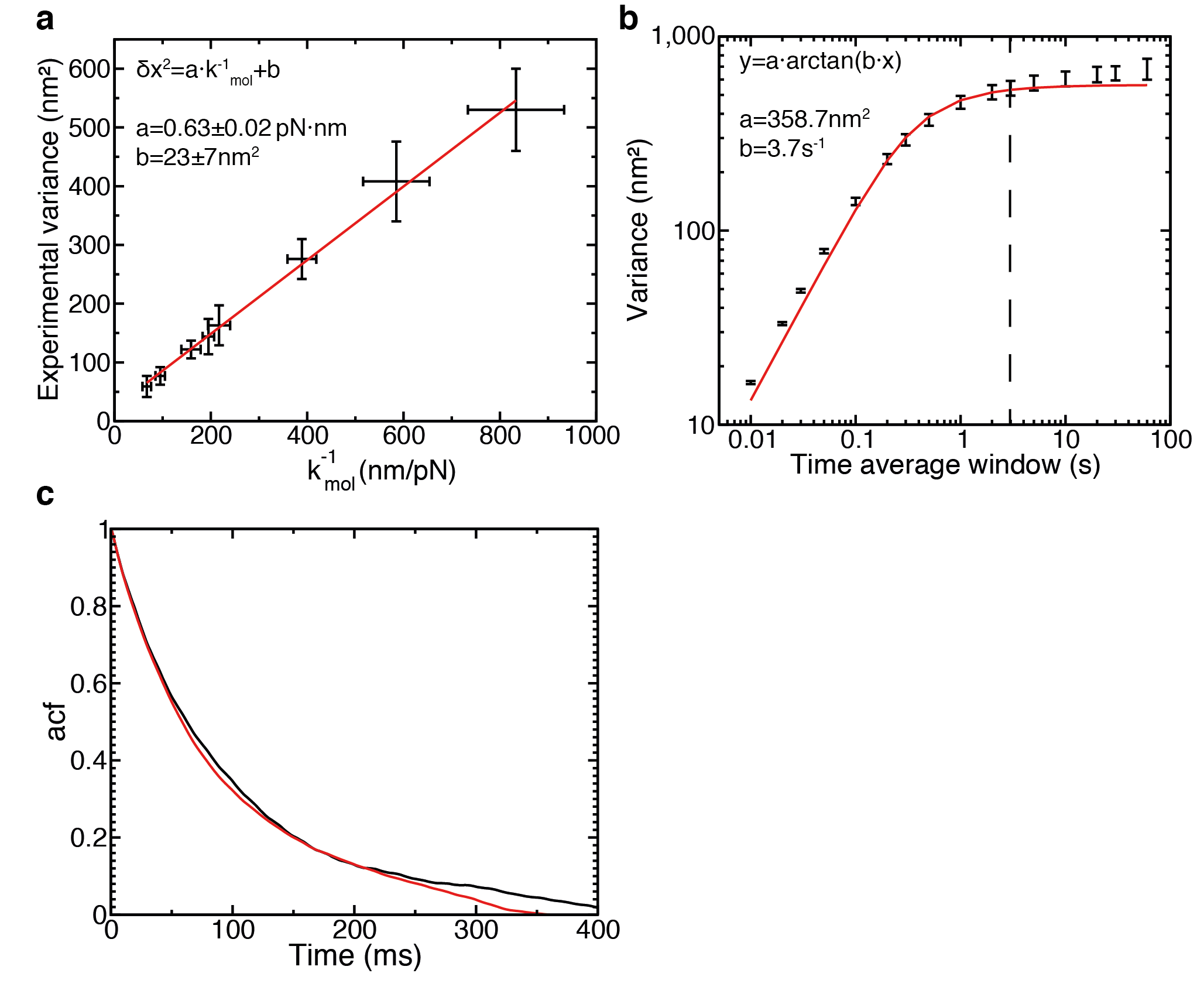}
\end{center}
\caption{{\bf Stiffness determination from trap-position fluctuations.} ({\bf a}) Phenomenological calibration of the molecular stiffness. The experimental variance of the trap position for a set of ssDNA and dsDNA molecules at different average forces was measured with the force-feedback protocol. The stiffness of the molecule (${\rm k_{mol}}$) was determined as the derivative of the force-extension curve at each force. ({\bf b}) Drift correction of the experimental data. The position of the trap is recorded at 1 kHz, and drift is removed by subtracting the average position using time-windows in the range 0.01-60 s (black). A window of 3 s (dashed line) is the optimal value to remove drift without distorting the measurement of the fluctuations. Data is fitted to an arctangent function (red). Measurements correspond to a 13-kb ssDNA molecule at 5 pN (N=5). ({\bf c}) Autocorrelation function of the distance during two constant force experiments at 8 pN for a ssDNA molecule after subtracting the drift.  Data with (red) and without (black) a buffer flow are shown.}
\label{fig:supp13}
\end{figure}

\clearpage

\section{S7 Supplementary Tables}

\begin{table}[ht] 
\begin{tabular}{c c c}
\hline
 & Zeta Potential & Electrophoretic  \\
 & (mV) & Mobility ($\mu m$$\cdot$cm$\cdot$V$\cdot$$s^{-1}$) \\
\hline
KF & $7.3\pm0.3$ & $0.6\pm0.1$\\
DNA & $-46.1\pm5.5$ & $-3.4\pm0.3$ \\
DNA/KF & $-30.1\pm1.2$ & $-2.4\pm0.1$ \\
\hline
\end{tabular}
\caption{{\bf Zeta potential measurements.} Zeta pontential of KF (40 $\mu$M) and $\lambda$-DNA (6.25 $\mu$g/ml, 0.2 nM) solutions as well as their mixtures in water at $25^{\circ}C$ (mean$\pm$SD, N$\ge$3). KF aggregates in water have a slighlty positive zeta potential -too low to prevent aggregation-, but suggesting that peptide aggregates might preferentially expose the positively charged residues on the surface. The addition of DNA to the sample shifts the zeta potential to negative values, further supporting that DNA binds to the peptide aggregates preventing their growth.}
\label{table:zetapot} 
\end{table}

\begin{table}[ht] 
\centering 
\begin{tabular}{cc}
\hline\hline 
Oligonucleotide name & Oligonucleotide sequence \\ [0.5ex] 
\hline 
BamHI-loop II & 5'-GATCGCCAGTTCGCGTTCGCCAGCATCCG \\ 
 & ACTACGGATGCTGGCGAACGCGAACTGGC-3' \\ 
cosRlong & 5'-Pho-GGGCGGCGACCTAAGATCTATTATATATGTG \\
& TCTCTATTAGTTAGTGGTGGAAACACAGTGCCAGCGC-3' \\
BIO-cosRshort3 &5'-Bio-GACTTCACTAATACGACTCACTATAGGG\\
&AAATAGAGACACATATATAATAGATCTT-3'\\
splint3 & 5'-TCCCTATAGTGAGTCGTATTAGTGAAGTC-3' \\
inverted-splint &  3'-AAAAA-5'-5'-GCGCTGGCACTGTGTTTCCACCACTAAC(SpC3)-3' \\ [1ex]   
\hline 
Blockloop30 & 5'-TAGTCGGATGCTGGCGAACGCGAACTGGCG-3' \\ [1ex]   
\hline 
\end{tabular} 
\caption{{\bf Oligonucleotides used for the DNA hairpin synthesis and to generate the ssDNA template.}} 
\label{table:oligos} 
\end{table}

\providecommand*\mcitethebibliography{\thebibliography}
\csname @ifundefined\endcsname{endmcitethebibliography}
  {\let\endmcitethebibliography\endthebibliography}{}


\begin{mcitethebibliography}{49}
\providecommand*\natexlab[1]{#1}
\providecommand*\mciteSetBstSublistMode[1]{}
\providecommand*\mciteSetBstMaxWidthForm[2]{}
\providecommand*\mciteBstWouldAddEndPuncttrue
  {\def\EndOfBibitem{\unskip.}}
\providecommand*\mciteBstWouldAddEndPunctfalse
  {\let\EndOfBibitem\relax}
\providecommand*\mciteSetBstMidEndSepPunct[3]{}
\providecommand*\mciteSetBstSublistLabelBeginEnd[3]{}
\providecommand*\EndOfBibitem{}
\mciteSetBstSublistMode{f}
\mciteSetBstMaxWidthForm{subitem}{(\alph{mcitesubitemcount})}
\mciteSetBstSublistLabelBeginEnd
  {\mcitemaxwidthsubitemform\space}
  {\relax}
  {\relax}

\bibitem[Calamai {\em et~al.}(2006)Calamai, Kumita, Mifsud, Parrini, Ramazzotti,
  Ramponi, Taddei, Chiti, and Dobson]{calamai2006nature}
Calamai,~M.; Kumita,~J.; Mifsud,~J.; Parrini,~C.; Ramazzotti,~M.; Ramponi,~G.;
  Taddei,~N.; Chiti,~F.; Dobson,~C. Nature and Significance of the Interactions
  between Amyloid Fibrils and Biological Polyelectrolytes. \emph{Biochemistry}
  \textbf{2006}, \emph{45}, 12806--12815\relax
\mciteBstWouldAddEndPuncttrue
\mciteSetBstMidEndSepPunct{\mcitedefaultmidpunct}
{\mcitedefaultendpunct}{\mcitedefaultseppunct}\relax
\EndOfBibitem
\bibitem[Bucciantini {\em et~al.}(2002)Bucciantini, Giannoni, Chiti, Baroni,
  Formigli, Zurdo, Taddei, Ramponi, Dobson, and
  Stefani]{bucciantini2002inherent}
Bucciantini,~M.; Giannoni,~E.; Chiti,~F.; Baroni,~F.; Formigli,~L.; Zurdo,~J.;
  Taddei,~N.; Ramponi,~G.; Dobson,~C.; Stefani,~M. Inherent Toxicity of
  Aggregates Implies a Common Mechanism for Protein Misfolding Diseases.
  \emph{Nature} \textbf{2002}, \emph{416}, 507--511\relax
\mciteBstWouldAddEndPuncttrue
\mciteSetBstMidEndSepPunct{\mcitedefaultmidpunct}
{\mcitedefaultendpunct}{\mcitedefaultseppunct}\relax
\EndOfBibitem
\bibitem[Gsponer and Vendruscolo(2006)Gsponer, and
  Vendruscolo]{gsponer2006theoretical}
Gsponer,~J.; Vendruscolo,~M. Theoretical Approaches to Protein Aggregation.
  \emph{Protein Pept. Lett.} \textbf{2006}, \emph{13}, 287--293\relax
\mciteBstWouldAddEndPuncttrue
\mciteSetBstMidEndSepPunct{\mcitedefaultmidpunct}
{\mcitedefaultendpunct}{\mcitedefaultseppunct}\relax
\EndOfBibitem
\bibitem[Coan and Shoichet(2008)Coan, and Shoichet]{coan2008stoichiometry}
Coan,~K.; Shoichet,~B. Stoichiometry and Physical Chemistry of Promiscuous
  Aggregate-Based Inhibitors. \emph{J. Am. Chem. Soc.} \textbf{2008},
  \emph{130}, 9606--9612\relax
\mciteBstWouldAddEndPuncttrue
\mciteSetBstMidEndSepPunct{\mcitedefaultmidpunct}
{\mcitedefaultendpunct}{\mcitedefaultseppunct}\relax
\EndOfBibitem
\bibitem[Puchalla {\em et~al.}(2008)Puchalla, Krantz, Austin, and
  Rye]{puchalla2008burst}
Puchalla,~J.; Krantz,~K.; Austin,~R.; Rye,~H. Burst Analysis Spectroscopy: A
  Versatile Single-Particle Approach for Studying Distributions of Protein
  Aggregates and Fluorescent Assemblies. \emph{Proc. Natl. Acad. Sci. U.S.A.}
  \textbf{2008}, \emph{105}, 14400--14405\relax
\mciteBstWouldAddEndPuncttrue
\mciteSetBstMidEndSepPunct{\mcitedefaultmidpunct}
{\mcitedefaultendpunct}{\mcitedefaultseppunct}\relax
\EndOfBibitem
\bibitem[Feng {\em et~al.}(2008)Feng, Toyama, Wille, Colby, Collins, May, Prusiner,
  Weissman, and Shoichet]{feng2008small}
Feng,~B.; Toyama,~B.; Wille,~H.; Colby,~D.; Collins,~S.; May,~B.; Prusiner,~S.;
  Weissman,~J.; Shoichet,~B. Small-Molecule Aggregates Inhibit Amyloid
  Polymerization. \emph{Nat. Chem. Biol.} \textbf{2008}, \emph{4},
  197--199\relax
\mciteBstWouldAddEndPuncttrue
\mciteSetBstMidEndSepPunct{\mcitedefaultmidpunct}
{\mcitedefaultendpunct}{\mcitedefaultseppunct}\relax
\EndOfBibitem
\bibitem[Coan {\em et~al.}(2009)Coan, Maltby, Burlingame, and
  Shoichet]{coan2009promiscuous}
Coan,~K.; Maltby,~D.; Burlingame,~A.; Shoichet,~B. Promiscuous Aggregate-Based
  Inhibitors Promote Enzyme Unfolding. \emph{J. Med. Chem.} \textbf{2009},
  \emph{52}, 2067--2075\relax
\mciteBstWouldAddEndPuncttrue
\mciteSetBstMidEndSepPunct{\mcitedefaultmidpunct}
{\mcitedefaultendpunct}{\mcitedefaultseppunct}\relax
\EndOfBibitem
\bibitem[Braun {\em et~al.}(2011)Braun, Humphreys, Fraser, Brancale, Bochtler, and
  Dale]{braun2011amyloid}
Braun,~S.; Humphreys,~C.; Fraser,~E.; Brancale,~A.; Bochtler,~M.; Dale,~T.
  Amyloid-Asssociated Nucleic Acid Hybridisation. \emph{PLoS One}
  \textbf{2011}, \emph{6}, e19125\relax
\mciteBstWouldAddEndPuncttrue
\mciteSetBstMidEndSepPunct{\mcitedefaultmidpunct}
{\mcitedefaultendpunct}{\mcitedefaultseppunct}\relax
\EndOfBibitem
\bibitem[Di~Domizio {\em et~al.}(2012)Di~Domizio, Zhang, Stagg, Gagea, Zhuo, Ladbury,
  and Cao]{di2012binding}
Di~Domizio,~J.; Zhang,~R.; Stagg,~L.; Gagea,~M.; Zhuo,~M.; Ladbury,~J.; Cao,~W.
  Binding with Nucleic Acids or Glycosaminoglycans Converts Soluble Protein
  Oligomers to Amyloid. \emph{J. Biol. Chem.} \textbf{2012}, \emph{287},
  736--747\relax
\mciteBstWouldAddEndPuncttrue
\mciteSetBstMidEndSepPunct{\mcitedefaultmidpunct}
{\mcitedefaultendpunct}{\mcitedefaultseppunct}\relax
\EndOfBibitem
\bibitem[Macedo {\em et~al.}(2012)Macedo, Millen, Braga, Gomes, Ferreira, Kraineva,
  Winter, Silva, and Cordeiro]{macedo2012nonspecific}
Macedo,~B.; Millen,~T.; Braga,~C.; Gomes,~M.; Ferreira,~P.; Kraineva,~J.;
  Winter,~R.; Silva,~J.; Cordeiro,~Y. Nonspecific Prion Protein--Nucleic Acid
  Interactions Lead to Different Aggregates and Cytotoxic Species.
  \emph{Biochemistry} \textbf{2012}, \emph{51}, 5402--5413\relax
\mciteBstWouldAddEndPuncttrue
\mciteSetBstMidEndSepPunct{\mcitedefaultmidpunct}
{\mcitedefaultendpunct}{\mcitedefaultseppunct}\relax
\EndOfBibitem
\bibitem[Motamedi-Shad {\em et~al.}(2012)Motamedi-Shad, Garfagnini, Penco, Relini,
  Fogolari, Corazza, Esposito, Bemporad, and Chiti]{motamedi2012rapid}
Motamedi-Shad,~N.; Garfagnini,~T.; Penco,~A.; Relini,~A.; Fogolari,~F.;
  Corazza,~A.; Esposito,~G.; Bemporad,~F.; Chiti,~F. Rapid Oligomer Formation
  of Human Muscle Acylphosphatase Induced by Heparan Sulfate. \emph{Nat.
  Struct. Mol. Biol.} \textbf{2012}, \emph{19}, 547--554\relax
\mciteBstWouldAddEndPuncttrue
\mciteSetBstMidEndSepPunct{\mcitedefaultmidpunct}
{\mcitedefaultendpunct}{\mcitedefaultseppunct}\relax
\EndOfBibitem
\bibitem[Cherny {\em et~al.}(2004)Cherny, Hoyer, Subramaniam, and
  Jovin]{cherny2004double}
Cherny,~D.; Hoyer,~W.; Subramaniam,~V.; Jovin,~T. Double-Stranded DNA
  Stimulates the Fibrillation of $\alpha$-Synuclein \textit{in Vitro} and Is
  Associated with the Mature Fibrils: an Electron Microscopy Study. \emph{J.
  Mol. Biol.} \textbf{2004}, \emph{344}, 929--938\relax
\mciteBstWouldAddEndPuncttrue
\mciteSetBstMidEndSepPunct{\mcitedefaultmidpunct}
{\mcitedefaultendpunct}{\mcitedefaultseppunct}\relax
\EndOfBibitem
\bibitem[Cohlberg {\em et~al.}(2002)Cohlberg, Li, Uversky, and
  Fink]{cohlberg2002heparin}
Cohlberg,~J.; Li,~J.; Uversky,~V.; Fink,~A. Heparin and Other
  Glycosaminoglycans Stimulate the Formation of Amyloid Fibrils from
  $\alpha$-Synuclein \textit{in Vitro}. \emph{Biochemistry} \textbf{2002},
  \emph{41}, 1502--1511\relax
\mciteBstWouldAddEndPuncttrue
\mciteSetBstMidEndSepPunct{\mcitedefaultmidpunct}
{\mcitedefaultendpunct}{\mcitedefaultseppunct}\relax
\EndOfBibitem
\bibitem[Dale(2006)]{dale2006protein}
Dale,~T. Protein and Nucleic Acid Together: a Mechanism for the Emergence of
  Biological Selection. \emph{J. Theor. Biol.} \textbf{2006}, \emph{240},
  337--342\relax
\mciteBstWouldAddEndPuncttrue
\mciteSetBstMidEndSepPunct{\mcitedefaultmidpunct}
{\mcitedefaultendpunct}{\mcitedefaultseppunct}\relax
\EndOfBibitem
\bibitem[Hamann and Scheuer(1993)Hamann, and Scheuer]{hamann1993kahalalide}
Hamann,~M.~T.; Scheuer,~P.~J. {K}ahalalide {F}: a Bioactive Depsipeptide from
  the Sacoglossan Mollusk \textit{Elysia Rufescens} and the Green Alga
  \textit{Bryopsis sp}. \emph{J. Am. Chem. Soc.} \textbf{1993}, \emph{115},
  5825--5826\relax
\mciteBstWouldAddEndPuncttrue
\mciteSetBstMidEndSepPunct{\mcitedefaultmidpunct}
{\mcitedefaultendpunct}{\mcitedefaultseppunct}\relax
\EndOfBibitem
\bibitem[Lopez-Macia {\em et~al.}(2001)Lopez-Macia, Jimenez, Royo, Giralt, and
  Albericio]{lopez2001synthesis}
Lopez-Macia,~A.; Jimenez,~J.~C.; Royo,~M.; Giralt,~E.; Albericio,~F. Synthesis
  and Structure Determination of {K}ahalalide {F}. \emph{J. Am. Chem. Soc.}
  \textbf{2001}, \emph{123}, 11398--11401\relax
\mciteBstWouldAddEndPuncttrue
\mciteSetBstMidEndSepPunct{\mcitedefaultmidpunct}
{\mcitedefaultendpunct}{\mcitedefaultseppunct}\relax
\EndOfBibitem
\bibitem[Garcia-Rocha {\em et~al.}(1996)Garcia-Rocha, Bonay, and
  Avila]{garcia1996antitumoral}
Garcia-Rocha,~M.; Bonay,~P.; Avila,~J. The Antitumoral Compound {K}ahalalide
  {F} Acts on Cell Lysosomes. \emph{Cancer Lett.} \textbf{1996}, \emph{99},
  43--50\relax
\mciteBstWouldAddEndPuncttrue
\mciteSetBstMidEndSepPunct{\mcitedefaultmidpunct}
{\mcitedefaultendpunct}{\mcitedefaultseppunct}\relax
\EndOfBibitem
\bibitem[Suarez {\em et~al.}(2003)Suarez, Gonzalez, Cuadrado, Berciano, Lafarga, and
  Munoz]{suarez2003kahalalide}
Suarez,~Y.; Gonzalez,~L.; Cuadrado,~A.; Berciano,~M.; Lafarga,~M.; Munoz,~A.
  {K}ahalalide {F}, a New Marine-Derived Compound, Induces Oncosis in Human
  Prostate and Breast Cancer Cells. \emph{Mol. Cancer Ther.} \textbf{2003},
  \emph{2}, 863--872\relax
\mciteBstWouldAddEndPuncttrue
\mciteSetBstMidEndSepPunct{\mcitedefaultmidpunct}
{\mcitedefaultendpunct}{\mcitedefaultseppunct}\relax
\EndOfBibitem
\bibitem[Sewell {\em et~al.}(2005)Sewell, Mayer, Langdon, Smyth, Jodrell, and
  Guichard]{sewell2005mechanism}
Sewell,~J.~M.; Mayer,~I.; Langdon,~S.~P.; Smyth,~J.~F.; Jodrell,~D.~I.;
  Guichard,~S.~M. The Mechanism of Action of {K}ahalalide {F}: Variable Cell
  Permeability in Human Hepatoma Cell Lines. \emph{Eur. J. Cancer}
  \textbf{2005}, \emph{41}, 1637--1644\relax
\mciteBstWouldAddEndPuncttrue
\mciteSetBstMidEndSepPunct{\mcitedefaultmidpunct}
{\mcitedefaultendpunct}{\mcitedefaultseppunct}\relax
\EndOfBibitem
\bibitem[Molina-Guijarro {\em et~al.}(2011)Molina-Guijarro, Mac{\'\i}as, Garc{\'\i}a,
  Mu{\~n}oz, Garc{\'\i}a-Fern{\'a}ndez, David, N{\'u}{\~n}ez,
  Mart{\'\i}nez-Leal, Moneo, Cuevas, Lillo, Villalobos~Jorge, Valenzuela, and
  Galmarini]{molina2011irvalec}
Molina-Guijarro,~J.; Mac{\'\i}as,~{\'A}.; Garc{\'\i}a,~C.; Mu{\~n}oz,~E.;
  Garc{\'\i}a-Fern{\'a}ndez,~L.; David,~M.; N{\'u}{\~n}ez,~L.;
  Mart{\'\i}nez-Leal,~J.; Moneo,~V.; Cuevas,~C. {\em et~al.}  Irvalec Inserts into
  the Plasma Membrane Causing Rapid Loss of Integrity and Necrotic Cell Death
  in Tumor Cells. \emph{PLoS One} \textbf{2011}, \emph{6}, e19042\relax
\mciteBstWouldAddEndPuncttrue
\mciteSetBstMidEndSepPunct{\mcitedefaultmidpunct}
{\mcitedefaultendpunct}{\mcitedefaultseppunct}\relax
\EndOfBibitem
\bibitem[Li and Walker(2011)Li, and Walker]{li2011signature}
Li,~I.; Walker,~G. Signature of Hydrophobic Hydration in a Single Polymer.
  \emph{Proc. Natl. Acad. Sci. U.S.A.} \textbf{2011}, \emph{108},
  16527--16532\relax
\mciteBstWouldAddEndPuncttrue
\mciteSetBstMidEndSepPunct{\mcitedefaultmidpunct}
{\mcitedefaultendpunct}{\mcitedefaultseppunct}\relax
\EndOfBibitem
\bibitem[Huguet {\em et~al.}(2009)Huguet, Forns, and Ritort]{huguet2009statistical}
Huguet,~J.; Forns,~N.; Ritort,~F. Statistical Properties of Metastable
  Intermediates in DNA Unzipping. \emph{Phys. Rev. Lett.} \textbf{2009},
  \emph{103}, 248106\relax
\mciteBstWouldAddEndPuncttrue
\mciteSetBstMidEndSepPunct{\mcitedefaultmidpunct}
{\mcitedefaultendpunct}{\mcitedefaultseppunct}\relax
\EndOfBibitem
\bibitem[Ritort {\em et~al.}(2006)Ritort, Mihardja, Smith, and
  Bustamante]{Ritort:2006fk}
Ritort,~F.; Mihardja,~S.; Smith,~S.~B.; Bustamante,~C. Condensation Transition
  in {DNA}-Polyaminoamide Dendrimer Fibers Studied Using Optical Tweezers.
  \emph{Phys. Rev. Lett.} \textbf{2006}, \emph{96}, 118301\relax
\mciteBstWouldAddEndPuncttrue
\mciteSetBstMidEndSepPunct{\mcitedefaultmidpunct}
{\mcitedefaultendpunct}{\mcitedefaultseppunct}\relax
\EndOfBibitem
\bibitem[Leger {\em et~al.}(1999)Leger, Romano, Sarkar, Robert, Bourdieu, Chatenay,
  and Marko]{leger1999structural}
Leger,~J.~F.; Romano,~G.; Sarkar,~A.; Robert,~J.; Bourdieu,~L.; Chatenay,~D.;
  Marko,~J.~F. Structural Transitions of a Twisted and Stretched {DNA}
  Molecule. \emph{Phys. Rev. Lett.} \textbf{1999}, \emph{83}, 1066--1069\relax
\mciteBstWouldAddEndPuncttrue
\mciteSetBstMidEndSepPunct{\mcitedefaultmidpunct}
{\mcitedefaultendpunct}{\mcitedefaultseppunct}\relax
\EndOfBibitem
\bibitem[Vladescu {\em et~al.}(2007)Vladescu, McCauley, Nu{\~n}ez, Rouzina, and
  Williams]{vladescu2007quantifying}
Vladescu,~I.~D.; McCauley,~M.~J.; Nu{\~n}ez,~M.~E.; Rouzina,~I.;
  Williams,~M.~C. Quantifying Force-Dependent and Zero-Force {DNA}
  Intercalation by Single-Molecule Stretching. \emph{Nat. Methods}
  \textbf{2007}, \emph{4}, 517--522\relax
\mciteBstWouldAddEndPuncttrue
\mciteSetBstMidEndSepPunct{\mcitedefaultmidpunct}
{\mcitedefaultendpunct}{\mcitedefaultseppunct}\relax
\EndOfBibitem
\bibitem[Huguet {\em et~al.}(2010)Huguet, Bizarro, Forns, Smith, Bustamante, and
  Ritort]{huguet2010single}
Huguet,~J.~M.; Bizarro,~C.~V.; Forns,~N.; Smith,~S.~B.; Bustamante,~C.;
  Ritort,~F. Single-Molecule Derivation of Salt Dependent Base-Pair Free
  Energies in {DNA}. \emph{Proc. Natl. Acad. Sci. U.S.A.} \textbf{2010},
  \emph{107}, 15431--15436\relax
\mciteBstWouldAddEndPuncttrue
\mciteSetBstMidEndSepPunct{\mcitedefaultmidpunct}
{\mcitedefaultendpunct}{\mcitedefaultseppunct}\relax
\EndOfBibitem
\bibitem[Podgornik and J{\"o}nsson(1993)Podgornik, and
  J{\"o}nsson]{podgornik1993stretching}
Podgornik,~R.; J{\"o}nsson,~B. Stretching of Polyelectrolyte Chains by
  Oppositely Charged Aggregates. \emph{Europhys. Lett.} \textbf{1993},
  \emph{24}, 501--506\relax
\mciteBstWouldAddEndPuncttrue
\mciteSetBstMidEndSepPunct{\mcitedefaultmidpunct}
{\mcitedefaultendpunct}{\mcitedefaultseppunct}\relax
\EndOfBibitem
\bibitem[Brower-Toland {\em et~al.}(2002)Brower-Toland, Smith, Yeh, Lis, Peterson,
  and Wang]{Brower-Toland:2002fk}
Brower-Toland,~B.; Smith,~C.; Yeh,~R.; Lis,~J.; Peterson,~C.; Wang,~M.
  Mechanical Disruption of Individual Nucleosomes Reveals a Reversible
  Multistage Release of DNA. \emph{Proc. Natl. Acad. Sci. U.S.A.}
  \textbf{2002}, \emph{99}, 1960--1965\relax
\mciteBstWouldAddEndPuncttrue
\mciteSetBstMidEndSepPunct{\mcitedefaultmidpunct}
{\mcitedefaultendpunct}{\mcitedefaultseppunct}\relax
\EndOfBibitem
\bibitem[Skoko {\em et~al.}(2005)Skoko, Yan, Johnson, and Marko]{skoko2005low}
Skoko,~D.; Yan,~J.; Johnson,~R.; Marko,~J. Low-Force DNA Condensation and
  Discontinuous High-Force Decondensation Reveal a Loop-Stabilizing Function of
  the Protein Fis. \emph{Phys. Rev. Lett.} \textbf{2005}, \emph{95},
  208101\relax
\mciteBstWouldAddEndPuncttrue
\mciteSetBstMidEndSepPunct{\mcitedefaultmidpunct}
{\mcitedefaultendpunct}{\mcitedefaultseppunct}\relax
\EndOfBibitem
\bibitem[Van~Noort {\em et~al.}(2004)Van~Noort, Verbrugge, Goosen, Dekker, and
  Dame]{van2004dual}
Van~Noort,~J.; Verbrugge,~S.; Goosen,~N.; Dekker,~C.; Dame,~R.~T. Dual
  Architectural Roles of {HU}: Formation of Flexible Hinges and Rigid
  Filaments. \emph{Proc. Natl. Acad. Sci. U.S.A.} \textbf{2004}, \emph{101},
  6969--6974\relax
\mciteBstWouldAddEndPuncttrue
\mciteSetBstMidEndSepPunct{\mcitedefaultmidpunct}
{\mcitedefaultendpunct}{\mcitedefaultseppunct}\relax
\EndOfBibitem
\bibitem[Todd and Rau(2008)Todd, and Rau]{todd2008interplay}
Todd,~B.; Rau,~D. Interplay of Ion Binding and Attraction in DNA Condensed by
  Multivalent Cations. \emph{Nucleic Acids Res.} \textbf{2008}, \emph{36},
  501--510\relax
\mciteBstWouldAddEndPuncttrue
\mciteSetBstMidEndSepPunct{\mcitedefaultmidpunct}
{\mcitedefaultendpunct}{\mcitedefaultseppunct}\relax
\EndOfBibitem
\bibitem[Horme{\~n}o {\em et~al.}(2011)Horme{\~n}o, Moreno-Herrero, Ibarra,
  Carrascosa, Valpuesta, and Arias-Gonzalez]{Hormeno:2011}
Horme{\~n}o,~S.; Moreno-Herrero,~F.; Ibarra,~B.; Carrascosa,~J.~L.;
  Valpuesta,~J.~M.; Arias-Gonzalez,~J.~R. Condensation Prevails over B--A
  Transition in the Structure of DNA at Low Humidity. \emph{Biophys. J.}
  \textbf{2011}, \emph{100}, 2006--2015\relax
\mciteBstWouldAddEndPuncttrue
\mciteSetBstMidEndSepPunct{\mcitedefaultmidpunct}
{\mcitedefaultendpunct}{\mcitedefaultseppunct}\relax
\EndOfBibitem
\bibitem[Yoshimura {\em et~al.}(2012)Yoshimura, Lin, Yagi, Lee, Kitayama, Sakurai,
  So, Ogi, Naiki, and Goto]{yoshimura2012distinguishing}
Yoshimura,~Y.; Lin,~Y.; Yagi,~H.; Lee,~Y.; Kitayama,~H.; Sakurai,~K.; So,~M.;
  Ogi,~H.; Naiki,~H.; Goto,~Y. Distinguishing Crystal-Like Amyloid Fibrils and
  Glass-Like Amorphous Aggregates from their Kinetics of Formation. \emph{Proc.
  Natl. Acad. Sci. U.S.A.} \textbf{2012}, \emph{109}, 14446--14451\relax
\mciteBstWouldAddEndPuncttrue
\mciteSetBstMidEndSepPunct{\mcitedefaultmidpunct}
{\mcitedefaultendpunct}{\mcitedefaultseppunct}\relax
\EndOfBibitem
\bibitem[Ross and Poirier(2005)Ross, and Poirier]{ross2005role}
Ross,~C.; Poirier,~M. What is the Role of Protein Aggregation in
  Neurodegeneration? \emph{Nat. Rev. Mol. Cell Biol.} \textbf{2005}, \emph{6},
  891--898\relax
\mciteBstWouldAddEndPuncttrue
\mciteSetBstMidEndSepPunct{\mcitedefaultmidpunct}
{\mcitedefaultendpunct}{\mcitedefaultseppunct}\relax
\EndOfBibitem
\bibitem[Ceci {\em et~al.}(2004)Ceci, Cellai, Falvo, Rivetti, Rossi, and
  Chiancone]{ceci2004dna}
Ceci,~P.; Cellai,~S.; Falvo,~E.; Rivetti,~C.; Rossi,~G.; Chiancone,~E. DNA
  Condensation and Self-Aggregation of \textit{Escherichia Coli} Dps are
  Coupled Phenomena Related to the Properties of the N-Terminus. \emph{Nucleic
  Acids Res.} \textbf{2004}, \emph{32}, 5935--5944\relax
\mciteBstWouldAddEndPuncttrue
\mciteSetBstMidEndSepPunct{\mcitedefaultmidpunct}
{\mcitedefaultendpunct}{\mcitedefaultseppunct}\relax
\EndOfBibitem
\bibitem[Ceci {\em et~al.}(2007)Ceci, Mangiarotti, Rivetti, and
  Chiancone]{ceci2007neutrophil}
Ceci,~P.; Mangiarotti,~L.; Rivetti,~C.; Chiancone,~E. The Neutrophil-Activating
  Dps Protein of \textit{Helicobacter Pylori}, HP-NAP, Adopts a Mechanism
  Different from \textit{Escherichia Coli} Dps to Bind and Condense DNA.
  \emph{Nucleic Acids Res.} \textbf{2007}, \emph{35}, 2247--2256\relax
\mciteBstWouldAddEndPuncttrue
\mciteSetBstMidEndSepPunct{\mcitedefaultmidpunct}
{\mcitedefaultendpunct}{\mcitedefaultseppunct}\relax
\EndOfBibitem
\bibitem[Hou {\em et~al.}(2009)Hou, Zhang, Wei, Ji, Dou, Wang, Li, and
  Wang]{hou2009cisplatin}
Hou,~X.; Zhang,~X.; Wei,~K.; Ji,~C.; Dou,~S.; Wang,~W.; Li,~M.; Wang,~P.
  Cisplatin Induces Loop Structures and Condensation of Single DNA Molecules.
  \emph{Nucleic Acids Res.} \textbf{2009}, \emph{37}, 1400--1410\relax
\mciteBstWouldAddEndPuncttrue
\mciteSetBstMidEndSepPunct{\mcitedefaultmidpunct}
{\mcitedefaultendpunct}{\mcitedefaultseppunct}\relax
\EndOfBibitem
\bibitem[Dansithong {\em et~al.}(2008)Dansithong, Wolf, Sarkar, Paul, Chiang, Holt,
  Morris, Branco, Sherwood, Comai, Berul, and Reddy]{dansithong2008cytoplasmic}
Dansithong,~W.; Wolf,~C.; Sarkar,~P.; Paul,~S.; Chiang,~A.; Holt,~I.;
  Morris,~G.; Branco,~D.; Sherwood,~M.; Comai,~L. {\em et~al.}  Cytoplasmic CUG RNA
  \textit{Foci} Are Insufficient to Elicit Key DM1 Features. \emph{PLoS One}
  \textbf{2008}, \emph{3}, e3968\relax
\mciteBstWouldAddEndPuncttrue
\mciteSetBstMidEndSepPunct{\mcitedefaultmidpunct}
{\mcitedefaultendpunct}{\mcitedefaultseppunct}\relax
\EndOfBibitem
\bibitem[Dickson and Wilusz(2010)Dickson, and Wilusz]{dickson2010repeat}
Dickson,~A.~M.; Wilusz,~C.~J. Repeat Expansion Diseases: when a Good RNA Turns
  Bad. \emph{Wiley Interdiscip. Rev.: RNA} \textbf{2010}, \emph{1},
  173--192\relax
\mciteBstWouldAddEndPuncttrue
\mciteSetBstMidEndSepPunct{\mcitedefaultmidpunct}
{\mcitedefaultendpunct}{\mcitedefaultseppunct}\relax
\EndOfBibitem
\bibitem[Garc{\'\i}a-L{\'o}pez {\em et~al.}(2011)Garc{\'\i}a-L{\'o}pez, Llamus{\'\i},
  Orz{\'a}ez, P{\'e}rez-Pay{\'a}, and Artero]{garcia2011vivo}
Garc{\'\i}a-L{\'o}pez,~A.; Llamus{\'\i},~B.; Orz{\'a}ez,~M.;
  P{\'e}rez-Pay{\'a},~E.; Artero,~R. \textit{In Vivo} Discovery of a Peptide
  that Prevents CUG--RNA Hairpin Formation and Reverses RNA Toxicity in
  Myotonic Dystrophy Models. \emph{Proc. Natl. Acad. Sci. U.S.A.}
  \textbf{2011}, \emph{108}, 11866--11871\relax
\mciteBstWouldAddEndPuncttrue
\mciteSetBstMidEndSepPunct{\mcitedefaultmidpunct}
{\mcitedefaultendpunct}{\mcitedefaultseppunct}\relax
\EndOfBibitem
\bibitem[Smith {\em et~al.}(2002)Smith, Cui, and Bustamante]{smith2002optical}
Smith,~S.~B.; Cui,~Y.; Bustamante,~C. Optical-Trap Force Transducer that
  Operates by Direct Measurement of Light Momentum. \emph{Methods Enzymol.}
  \textbf{2002}, \emph{361}, 134--162\relax
\mciteBstWouldAddEndPuncttrue
\mciteSetBstMidEndSepPunct{\mcitedefaultmidpunct}
{\mcitedefaultendpunct}{\mcitedefaultseppunct}\relax
\EndOfBibitem
\bibitem[Smith {\em et~al.}(1996)Smith, Cui, and Bustamante]{smith1996overstretching}
Smith,~S.; Cui,~Y.; Bustamante,~C. Overstretching {B}-{DNA}: the Elastic
  Response of Individual Double-Stranded and Single-Stranded {DNA} Molecules.
  \emph{Science} \textbf{1996}, \emph{271}, 795--799\relax
\mciteBstWouldAddEndPuncttrue
\mciteSetBstMidEndSepPunct{\mcitedefaultmidpunct}
{\mcitedefaultendpunct}{\mcitedefaultseppunct}\relax
\EndOfBibitem
\bibitem[Wang {\em et~al.}(1997)Wang, Yin, Landick, Gelles, and
  Block]{wang1997stretching}
Wang,~M.~D.; Yin,~H.; Landick,~R.; Gelles,~J.; Block,~S.~M. Stretching {DNA}
  with Optical Tweezers. \emph{Biophys. J.} \textbf{1997}, \emph{72},
  1335--1346\relax
\mciteBstWouldAddEndPuncttrue
\mciteSetBstMidEndSepPunct{\mcitedefaultmidpunct}
{\mcitedefaultendpunct}{\mcitedefaultseppunct}\relax
\EndOfBibitem
\bibitem[Baumann {\em et~al.}(1997)Baumann, Smith, Bloomfield, and
  Bustamante]{baumann1997ionic}
Baumann,~C.~G.; Smith,~S.~B.; Bloomfield,~V.~A.; Bustamante,~C. Ionic Effects
  on the Elasticity of Single {DNA} Molecules. \emph{Proc. Natl. Acad. Sci.
  U.S.A.} \textbf{1997}, \emph{94}, 6185--6190\relax
\mciteBstWouldAddEndPuncttrue
\mciteSetBstMidEndSepPunct{\mcitedefaultmidpunct}
{\mcitedefaultendpunct}{\mcitedefaultseppunct}\relax
\EndOfBibitem
\bibitem[Bouchiat {\em et~al.}(1999)Bouchiat, Wang, Allemand, Strick, Block, and
  Croquette]{bouchiat1999estimating}
Bouchiat,~C.; Wang,~M.; Allemand,~J.; Strick,~T.; Block,~S.; Croquette,~V.
  Estimating the Persistence Length of a Worm-Like Chain Molecule from
  Force-Extension Measurements. \emph{Biophys. J.} \textbf{1999}, \emph{76},
  409--413\relax
\mciteBstWouldAddEndPuncttrue
\mciteSetBstMidEndSepPunct{\mcitedefaultmidpunct}
{\mcitedefaultendpunct}{\mcitedefaultseppunct}\relax
\EndOfBibitem
\bibitem[Marquardt(1963)]{marquardt1963algorithm}
Marquardt,~D.~W. An Algorithm for Least-Squares Estimation of Nonlinear
  Parameters. \emph{J. Soc. Ind. Appl. Math.} \textbf{1963}, \emph{11},
  431--441\relax
\mciteBstWouldAddEndPuncttrue
\mciteSetBstMidEndSepPunct{\mcitedefaultmidpunct}
{\mcitedefaultendpunct}{\mcitedefaultseppunct}\relax
\EndOfBibitem
\bibitem[Fuentes-Perez {\em et~al.}(2012)Fuentes-Perez, Gwynn, Dillingham, and
  Moreno-Herrero]{Fuentes:2012}
Fuentes-Perez,~M.~E.; Gwynn,~E.~J.; Dillingham,~M.~S.; Moreno-Herrero,~F. Using
  DNA as a Fiducial Marker to Study SMC Complex Interactions with the Atomic
  Force Microscope. \emph{Biophys. J.} \textbf{2012}, \emph{102},
  839--848\relax
\mciteBstWouldAddEndPuncttrue
\mciteSetBstMidEndSepPunct{\mcitedefaultmidpunct}
{\mcitedefaultendpunct}{\mcitedefaultseppunct}\relax
\EndOfBibitem
\bibitem[Horcas {\em et~al.}(2007)Horcas, Fern{\'a}ndez, G{\'o}mez-Rodr{\'\i}guez,
  Colchero, G{\'o}mez-Herrero, and Baro]{Horcas:2007}
Horcas,~I.; Fern{\'a}ndez,~R.; G{\'o}mez-Rodr{\'\i}guez,~J.~M.; Colchero,~J.;
  G{\'o}mez-Herrero,~J.; Baro,~A.~M. WSXM: a Software for Scanning Probe
  Microscopy and a Tool for Nanotechnology. \emph{Rev. Sci. Instrum.}
  \textbf{2007}, \emph{78}, 013705--013705\relax
\mciteBstWouldAddEndPuncttrue
\mciteSetBstMidEndSepPunct{\mcitedefaultmidpunct}
{\mcitedefaultendpunct}{\mcitedefaultseppunct}\relax
\EndOfBibitem
\end{mcitethebibliography}

\begin{mcitethebibliography}{9}
\providecommand*\natexlab[1]{#1}
\providecommand*\mciteSetBstSublistMode[1]{}
\providecommand*\mciteSetBstMaxWidthForm[2]{}
\providecommand*\mciteBstWouldAddEndPuncttrue
  {\def\EndOfBibitem{\unskip.}}
\providecommand*\mciteBstWouldAddEndPunctfalse
  {\let\EndOfBibitem\relax}
\providecommand*\mciteSetBstMidEndSepPunct[3]{}
\providecommand*\mciteSetBstSublistLabelBeginEnd[3]{}
\providecommand*\EndOfBibitem{}
\mciteSetBstSublistMode{f}
\mciteSetBstMaxWidthForm{subitem}{(\alph{mcitesubitemcount})}
\mciteSetBstSublistLabelBeginEnd
  {\mcitemaxwidthsubitemform\space}
  {\relax}
  {\relax}

\bibitem[Huguet et~al.(2010)Huguet, Bizarro, Forns, Smith, Bustamante, and
  Ritort]{huguet2010single}
Huguet,~J.~M.; Bizarro,~C.~V.; Forns,~N.; Smith,~S.~B.; Bustamante,~C.;
  Ritort,~F. Single-Molecule Derivation of Salt Dependent Base-Pair Free
  Energies in {DNA}. \emph{Proc. Natl. Acad. Sci. U.S.A.} \textbf{2010},
  \emph{107}, 15431--15436\relax
\mciteBstWouldAddEndPuncttrue
\mciteSetBstMidEndSepPunct{\mcitedefaultmidpunct}
{\mcitedefaultendpunct}{\mcitedefaultseppunct}\relax
\EndOfBibitem
\bibitem[Borchardt et~al.(2005)Borchardt, Kerns, Lipinski, Thakker, and
  Wang]{borchardt2005pharmaceutical}
Borchardt,~R.; Kerns,~E.; Lipinski,~C.; Thakker,~D.; Wang,~B.
  \emph{Pharmaceutical Profiling in Drug Discovery for Lead Selection};
  American Assoc. of Pharm. Scientists, 2005; Vol.~1\relax
\mciteBstWouldAddEndPuncttrue
\mciteSetBstMidEndSepPunct{\mcitedefaultmidpunct}
{\mcitedefaultendpunct}{\mcitedefaultseppunct}\relax
\EndOfBibitem
\bibitem[Kozikowski et~al.(2003)Kozikowski, Burt, Tirey, Williams, Kuzmak,
  Stanton, Morand, and Nelson]{kozikowski2003effect}
Kozikowski,~B.~A.; Burt,~T.~M.; Tirey,~D.~A.; Williams,~L.~E.; Kuzmak,~B.~R.;
  Stanton,~D.~T.; Morand,~K.~L.; Nelson,~S.~L. The Effect of Freeze/Thaw Cycles
  on the Stability of Compounds in {DMSO}. \emph{J. Biomol. Screening}
  \textbf{2003}, \emph{8}, 210--215\relax
\mciteBstWouldAddEndPuncttrue
\mciteSetBstMidEndSepPunct{\mcitedefaultmidpunct}
{\mcitedefaultendpunct}{\mcitedefaultseppunct}\relax
\EndOfBibitem
\bibitem[McDonald et~al.(2008)McDonald, Hudson, Dunn, You, Baker, Whittal,
  Martin, Jha, Edmondson, and Holt]{mcdonald2008bioactive}
McDonald,~G.~R.; Hudson,~A.~L.; Dunn,~S. M.~J.; You,~H.; Baker,~G.;
  Whittal,~R.; Martin,~J.; Jha,~A.; Edmondson,~D.~E.; Holt,~A. Bioactive
  Contaminants Leach from Disposable Laboratory Plasticware. \emph{Science}
  \textbf{2008}, \emph{322}, 917\relax
\mciteBstWouldAddEndPuncttrue
\mciteSetBstMidEndSepPunct{\mcitedefaultmidpunct}
{\mcitedefaultendpunct}{\mcitedefaultseppunct}\relax
\EndOfBibitem
\bibitem[Nuijen et~al.(2001)Nuijen, Bouma, Manada, Jimeno, Lazaro, Bult, and
  Beijnen]{nuijen2001compatibility}
Nuijen,~B.; Bouma,~M.; Manada,~C.; Jimeno,~J.~M.; Lazaro,~L.; Bult,~A.;
  Beijnen,~J. Compatibility and Stability of the Investigational Polypeptide
  Marine Anticancer Agent {K}ahalalide {F} in Infusion Devices. \emph{Invest.
  New Drugs} \textbf{2001}, \emph{19}, 273--281\relax
\mciteBstWouldAddEndPuncttrue
\mciteSetBstMidEndSepPunct{\mcitedefaultmidpunct}
{\mcitedefaultendpunct}{\mcitedefaultseppunct}\relax
\EndOfBibitem
\bibitem[Forns et~al.(2011)Forns, De~Lorenzo, Manosas, Hayashi, Huguet, and
  Ritort]{forns2011improving}
Forns,~N.; De~Lorenzo,~S.; Manosas,~M.; Hayashi,~K.; Huguet,~J.; Ritort,~F.
  Improving Signal/Noise Resolution in Single-Molecule Experiments Using
  Molecular Constructs with Short Handles. \emph{Biophys. J.} \textbf{2011},
  \emph{100}, 1765--1774\relax
\mciteBstWouldAddEndPuncttrue
\mciteSetBstMidEndSepPunct{\mcitedefaultmidpunct}
{\mcitedefaultendpunct}{\mcitedefaultseppunct}\relax
\EndOfBibitem
\bibitem[Smith et~al.(1996)Smith, Cui, and Bustamante]{smith1996overstretching}
Smith,~S.; Cui,~Y.; Bustamante,~C. Overstretching {B}-{DNA}: the Elastic
  Response of Individual Double-Stranded and Single-Stranded {DNA} Molecules.
  \emph{Science} \textbf{1996}, \emph{271}, 795--799\relax
\mciteBstWouldAddEndPuncttrue
\mciteSetBstMidEndSepPunct{\mcitedefaultmidpunct}
{\mcitedefaultendpunct}{\mcitedefaultseppunct}\relax
\EndOfBibitem
\bibitem[Stokvis et~al.(2002)Stokvis, Rosing, Lopez-Lazaro, Rodriguez, Jimeno,
  Supko, Schellens, and Beijnen]{stokvis2002quantitative}
Stokvis,~E.; Rosing,~H.; Lopez-Lazaro,~L.; Rodriguez,~I.; Jimeno,~J.;
  Supko,~J.; Schellens,~J. H.~M.; Beijnen,~J.~H. Quantitative Analysis of the
  Novel Depsipeptide Anticancer Drug {K}ahalalide {F} in Human Plasma by
  High-Performance Liquid Chromatography under Basic Conditions Coupled to
  Electrospray Ionization Tandem Mass Spectrometry. \emph{J. Mass Spectrom.}
  \textbf{2002}, \emph{37}, 992--1000\relax
\mciteBstWouldAddEndPuncttrue
\mciteSetBstMidEndSepPunct{\mcitedefaultmidpunct}
{\mcitedefaultendpunct}{\mcitedefaultseppunct}\relax
\EndOfBibitem
\end{mcitethebibliography}
\end{document}